\documentclass[12pt]{article}

\usepackage{amsmath,graphicx}
\usepackage{amsfonts}
\usepackage{amssymb}

\def\hybrid{\topmargin -20pt    \oddsidemargin 0pt
        \headheight 0pt \headsep 0pt
        \textwidth 6.25in       
        \textheight 9.5in       
        \marginparwidth .875in
        \parskip 5pt plus 1pt   \jot = 1.5ex}

\hybrid

\numberwithin{equation}{section}
\numberwithin{table}{section}

\renewcommand{\Re}{\operatorname{Re}}
\renewcommand{\Im}{\operatorname{Im}}

\newcommand\e{\mathrm{e}}
\newcommand\iu{\operatorname{i}}
\newcommand\diff{\mathrm{d}}

\newcommand\M{{Y}}                      
\newcommand\ST{{\beta^{({\bf 2},{\bf 2})}}} 
\newcommand\G{{\alpha^{({\bf 1},{\bf 1})}}}  
\newcommand\B{{\beta^{({\bf 1},{\bf 1})}}}  
\newcommand\A{{\alpha^{({\bf 2},{\bf 2})}}}  

\newcommand\adM{M}                      
\newcommand\admi{\hat{m}}               
\newcommand\admA{m}                     
\newcommand\admspin{\tilde{m}}          
\newcommand\adN{N}                      
\newcommand\adni{\hat{n}}               
\newcommand\adnA{n}                     
\newcommand\adnspin{\tilde{n}}          

\newcommand\admu{\mu}                   
\newcommand\admui{\hat{\mu}}            
\newcommand\admuA{\mu}                  
\newcommand\admuspin{\mu}               

\newcommand\U{\mu}                      

\begin{document}
\begin{titlepage}

\begin{center}

\rightline{\small ZMP-HH/09-4}

\vskip 1.5cm

{\Large \bf Type II compactifications on manifolds }
\vskip 0.3cm
{\Large \bf  with $SU(2) \times SU(2)$ structure}

\vskip 1cm

{\bf Hagen Triendl$^{a}$ and Jan Louis$^{a,b}$}\\

\vskip 1cm

{}$^{a}${\em II. Institut f{\"u}r Theoretische Physik\\
Universit{\"a}t Hamburg\\
Luruper Chaussee 149\\
D-22761 Hamburg, Germany}\\

\vskip 0.4cm

{}$^{b}${\em Zentrum f\"ur Mathematische Physik, Universit\"at Hamburg,\\
Bundesstrasse 55, D-20146 Hamburg}

\vskip 1cm

{\tt hagen.triendl@desy.de, jan.louis@desy.de} \\

\end{center}

\vskip 1cm

\begin{center} {\bf ABSTRACT } \end{center}

We study compactifications of type II theories on $SU(2) \times SU(2)$
structure manifolds to six, five and four spacetime dimensions. We use the
framework of generalized geometry to describe the NS-NS sector of such
compactifications and derive the structure of
their moduli spaces. We show that in contrast to $SU(3) \times SU(3)$
structure compactifications, there is no dynamical $SU(2) \times
SU(2)$ structure interpolating between an $SU(2)$
structure and an identity structure.
Furthermore, we formulate type II compactifications on $SU(2) \times
SU(2)$ structures in the context of exceptional generalized geometry
which makes the U-duality group manifest and naturally incorporates the scalar degrees of freedom arising in
the Ramond-Ramond sector. Via this formalism we derive the structure of the
moduli spaces as it is expected from $N=4$ supergravity.
\vfill

April 2009

\end{titlepage}

\tableofcontents

\section{Introduction}
One class of  backgrounds which are commonly considered
in string theory are  of the form
\begin{equation}\label{stringbackground}
M_{1,d-1} \times \M_{10-d}\ ,
\end{equation}
where $M_{1,d-1}$ is a $d$-dimensional
Minkowskian space-time while $\M_{10-d}$ is a compact mani\-fold of
dimension $(10-d)$. In such backgrounds the Lorentz group decomposes
accordingly into $SO(1,d-1)\times SO(10-d)$.
The requirement that the
background preserves some amount of supersymmetry generically demands that $\M_{10-d}$
admits globally defined and nowhere vanishing
spinors. Generically this reduces the structure group $G$ from
$SO(10-d)$ to some subgroup $G\subset SO(10-d)$.\footnote{For exceptions to this situation see for example \cite{Tsimpis:2005kj}. We thank the referee of our paper for drawing our attention to this
possibility.} Such  manifolds are termed manifolds with $G$-structure in
the literature and they are  natural generalizations of
Calabi-Yau manifolds \cite{Hitchin}-\cite{Gauntlett:2003cy}.

More precisely,
Calabi-Yau manifolds form a subclass of manifolds with $G$-structure
in that the globally
defined spinors are also covariantly constant with respect to the
Levi-Civita connection. As a consequence the Ricci-tensor vanishes and
$G=SU(n)$ coincides with the holonomy group.
However, in the presence of fluxes and localized energy sources such as
D-branes and/or orientifold planes
the geometry back-reacts and is no longer necessarily Calabi-Yau \cite{Grana:2005jc}.

In type II theories a slightly more general set-up is possible.
The left and the right sector can preserve different ``structure
groups'' and backgrounds with $G \times G$ structure can be considered \cite{Gualtieri:2003dx}-\cite{Grana:2005ny}.
This situation is best described by a
formalism  -- termed generalized geometry -- that describes the metric
and the $B$-field of string theory as
components of the generalized metric of some bundle $E$ that locally
looks like the direct sum of the tangent and the cotangent
bundle. On this bundle, the T-duality group $SO(10-d,10-d)$ acts naturally,
and thus one has an $SO(10-d,10-d)$ covariant formulation of string
backgrounds. Within this formalism the notion of  classical geometry
can be relaxed in that $\M_{10-d}$ does not have to be a manifold.
Instead objects such as T-folds and generalizations thereof can be
considered \cite{{Hull:2004in}}-\cite{Shelton:2005cf}.\footnote{Throughout this paper we do not specify
if $\M_{10-d}$ is an honest manifold or some of the
generalizations. For example, we discuss backgrounds where only the
tangent space admits a
splitting of the form $TM_{1,9} = T_{1,d-1} \oplus F_{10-d}$.
Nevertheless we always call $\M$ the compactification
manifold and the analysis in this paper just carries over to this more general case.}

Recently the Ramond-Ramond (RR) sector has also been included in this
framework by embedding the T-duality group into the U-duality group
$E_{11-d(11-d)}$~\cite{Hull:2007zu,Pacheco:2008ps,Sim:2008,Grana:2008tbp}. The
resulting geometrical structure  has been termed `exceptional generalized
geometry' as
one has to replace the generalized tangent bundle by some exceptional
tangent bundle which admits a natural action of $E_{11-d(11-d)}$.
The metric of this bundle inherits all scalar degrees of freedom
of the low-energy type II theory, including the ones
coming from the RR sector.

So far, in type II theories mainly $SU(3) \times SU(3)$ backgrounds
generalizing Calabi-Yau threefold compactifications have been
discussed in the literature. The reason is that they preserve only
eight supercharges and therefore  are  a convenient starting point for
particle phenomenology. However, backgrounds with more supersymmetry
have also been of considerable interest due to their constraint low
energy couplings. In addition,  by appropriately orientifolding such
backgrounds,
one can also
construct theories with less supersymmetry.

In this paper we focus on type II backgrounds with 16 supercharges
(corresponding to $N=4$ in $d=4$) which we discuss in the Minkowskian
dimensions $d=6,5,4$.
These backgrounds feature the manifolds $\M_{4,5,6}$ which have
$SU(2) \times SU(2)$ structure and which are generalizations of
$K3, K3\times S^1$ and $K3\times T^2$ respectively.
Aspects of such manifolds were previously discussed for example
in refs.~\cite{Gauntlett:2003cy,Grana:2005sn,Bovy:2005qq,ReidEdwards:2008rd,
Lust:2009zb}, while
$N=4$ compactifications with background fluxes have been analyzed for
instance in refs.~\cite{N4SUGRA}.

One of our main interests in this paper
is the relation with $N=4$ supergravity, which
constrains the low energy couplings.
For example, the scalar fields of type IIA have to parameterize cosets
of the form \cite{de Roo:1984gd}
\begin{equation}\label{N=4coset}
\mathcal{M} = \frac{SO(10-d, n_V)}{SO(10-d)\times
  SO(n_V)}\times \mathbb{R}^+\ ,
\end{equation}
where $n_V$ counts the number of vector multiplets and the
$\mathbb{R}^+$ factor corresponds to the dilaton.\footnote{In $d=4$
  the $\mathbb{R}^+$ factor is enlarged to the coset $Sl(2,\mathbb{R})/SO(2)$
  since the antisymmetric tensor of the NS-sector is dual to an axion
  and contributes in the scalar couplings.}
In this paper we identify $\mathcal{M}$ as the deformation space of certain
$SU(2) \times SU(2)$ structure manifolds.
For a special class of $SU(2)$ structure backgrounds in $d=4$ this was already
discussed in \cite{ReidEdwards:2008rd}.
Here we analyze the
generic situation and concentrate on the scalar field space which
corresponds to the kinetic terms of the scalars in the low energy
effective Lagrangian.
A more detailed derivation of this  Lagrangian
including the possible gaugings and the potentials will be presented
elsewhere \cite{LST}. We choose a possible warp factor in \eqref{stringbackground} to be constant although most of our analysis carries over to the case of a non-trivial warp factor.

In order to analyze manifolds with  $SU(2) \times SU(2)$ structure
we use the pure spinor formalism within
the framework of generalized geometries \cite{Hitchin}.
We find that generic $SU(2) \times SU(2)$ structures do not exist but
instead only manifolds with a single $SU(2)$ structure or with an
identity structure can occur. The latter correspond to
backgrounds with 32 supercharges, which we do not study any further in this
paper. Instead we focus on backgrounds with an honest $SU(2)$
structure and thus 16 supercharges.

On the class of $SU(2)$ structure manifolds we impose the additional
constraint that at low energies no massive gravitino multiplets
survive as their presence would alter the scalar geometry
\eqref{N=4coset}.
Analogously to the analysis of $SU(3)\times SU(3)$ structures
in refs.~\cite{Grana:2005ny}
this constraint amounts to projecting out all $SU(2)$ doublets.
In this case we find that the deformations of the
$SU(2)$ structure span the Neveu-Schwarz (NS) subspace of the manifold
${\cal M}$ given in~\eqref{N=4coset}.\footnote{This NS subspace also
appears in corresponding compactifications of the heterotic string on
manifolds with $SU(2)$-structure \cite{LMM}.}

The metric on the scalar
field space or in other words the kinetic term of
the effective action is largely determined by algebraic properties of the pure
spinors.
In addition differential constraints can be imposed.
They appear in the scalar potential
and determine possible consistent backgrounds of the theory.
However, they also affect the metric in that they select a certain
subclass of manifolds with $SU(2)$ structure. For example, we will see
that projecting out the doublets can fix a certain component of $Y$ to
be related to $K3$.

The RR-sector can be included by using the
formalism of exceptional generalized geometry
\cite{Hull:2007zu,Pacheco:2008ps,Sim:2008,Grana:2008tbp}. The
$SO(10-d,10-d)$ pure spinors are appropriately embedded into representations of the U-duality
group $E_{11-d(11-d)}$, with new degrees of freedom corresponding to
the RR-scalars. Due to the RR-sector, type IIA differs from type IIB
and we analyze both cases separately. In each case we find agreement
with the supergravity results.\footnote{Note that there is an
intriguing relation between the supergravity moduli spaces and the
horizon geometry of extremal black hole attractors \cite{Ferrara:2008hwa}.}

The paper is organized as follows.
In section~\ref{section:compd6} we analyze four-dimensional
manifolds $\M_4$ with
$SU(2)\times SU(2)$ structure. Before we discuss geometrical and
topological issues, we recall in section~\ref{section:spectrumd6} the
supergravity field content for type II theories in ten dimensions in
an $N =4$ language. There we see that for ${N}=4$ compactifications to
be consistent, we have to project out all $SU(2)\times SU(2)$
doublets, which contain the extra gravitino multiplets that are
massive in the compactifications.
Then we turn to the geometrical description of $SU(2)\times SU(2)$ structures.
As a warm-up we start the discussion
with (geometric) $SU(2)$ structures in
section~\ref{section:SU2d6}. In
section~\ref{section:SU2SU2d6} we then use the pure spinor formalism to
describe generic $SU(2) \times SU(2)$ structure backgrounds. With this, we are able to determine the
moduli space in section~\ref{section:moduliSU2d6}.
In  section~\ref{section:RRd6} we include the RR-sector by
embedding the $SO(4,4)$ pure spinors into an $E_{5(5)}\equiv SO(5,5)$ covariant
formulation. This enables us to compute the complete moduli space for
type IIA and IIB strings and show the consistency with supergravity.

The more involved case of compactifications to $d=4$ is
discussed in section~\ref{section:compd4}. As in
section~\ref{section:compd6}, we start with an analysis of the
spectrum in section~\ref{section:spectrumd4}, before we turn to a
discussion of $SU(2)\times SU(2)$ structures in geometrical terms. After introducing $SU(2)$ structures in
section~\ref{section:SU2d4}, we use the pure spinor
formalism in section~\ref{section:SU2SU2d4} to describe the more general case of $SU(2) \times SU(2)$ structures for
six-dimensional manifolds $\M_6$. We find that it is convenient to
introduce a generalized almost
product structure which reduces the structure group from $SU(3) \times
SU(3)$ to $SU(2) \times SU(2)$. We show in
section~\ref{section:moduliSU2d4} that this generalized almost product structure cannot vary and that
the moduli space essentially reduces to the one of
section~\ref{section:moduliSU2d6}.
In section~\ref{section:RRd4} we embed the pure spinors in an
$E_{7(7)}$ covariant formalism, which we use in
section~\ref{section:moduliRRd4} to show that the moduli space of type
II theories is given by \eqref{N=4coset}.

Section~\ref{section:conclusions} contains our conclusions and some of
the more technical computations are assembled in 5 appendices.
In appendix~\ref{section:conventions} we briefly state our conventions and give explicitly the Fierz identities that are used in section~\ref{section:SU2d6} and~\ref{section:SU2d4}.
Appendix~\ref{SCFcone} discusses the different cones over the moduli space
which appear in the associated superconformal
supergravities. Specifically, appendix~\ref{section:Hitchind4} is
devoted to the hyperk\"ahler cone (or the Swann bundle) \cite{Swann:1990,deWit:1999fp} which exists
over the NS moduli space since it can be viewed as the scalar field
space of a theory with eight supercharges.
In particular we discuss the relationship of the
purity and compatibility conditions with an $SU(2)$ quotient of flat
space and express the hyperk\"ahler potential
in terms of geometric
quantities. In appendix~\ref{N4cone} we discuss the flat cones which
arise in the superconformal supergravity of theories with 16 supercharges.
In appendix~\ref{section:productstructured6} we give the proof that the generalized almost product structure which we introduced in section~\ref{section:SU2SU2d4} to define $SU(2)\times SU(2)$ structures on backgrounds of dimension six is rigid for $N =4$ compactifications.
In appendix~\ref{section:E7} we give the details on the
calculations in the $E_{7(7)}$ covariant formalism used in section~\ref{section:RRd4} and section~\ref{section:moduliRRd4}.
Finally, in
appendix~\ref{section:compd5} we work out the case of
compactifications on $SU(2) \times SU(2)$ structures in $d=5$, using the strategy of section~\ref{section:compd6} and section~\ref{section:compd4}. In appendix~\ref{section:spectrumd5} we discuss the spectrum of $N =2$ supergravities in five dimensions achieved by compactification of type II theories on backgrounds of $SU(2)\times SU(2)$ structure. In appendix~\ref{section:SU2d5}, we introduce $SU(2)$ structures in five dimensions, and generalize to $SU(2) \times SU(2)$ structures by use of the pure spinor formalism in appendix~\ref{section:SU2SU2d5}. There we also derive the NS moduli space. In appendix~\ref{section:RRd5} we use again the concept of exceptional generalized geometry to include the RR-sector and to compute the complete moduli space.
Appendix~\ref{section:compd5} has some overlap with the independent
work of~ref.~\cite{Sim:2008}.

\section{Compactifications on manifolds with $SU(2) \times SU(2)$ structure
 in $d=6$}
\label{section:compd6}

In this section we study backgrounds of the form
\eqref{stringbackground} with a six-dimensional Minkowskian space-time
$M_{1,5}$ times a four-dimensional manifold $\M_4$.
We focus  on the situation where these backgrounds
preserve $N=2$ in $d=6$. Since the spinor representation is
eight-dimensional in $d=6$ this corresponds to 16 supercharges or
$N =4$ supersymmetry in terms of a four-dimensional counting.

More precisely the Lorentz group for these backgrounds
is $SO(1,5)\times SO(4)$. The ten-dimensional spinor representation
decomposes accordingly as
\begin{equation}
{\bf 16} \to ({\bf 4}, {\bf 2}) \oplus ({\bf \bar 4},{\bf \bar 2})\ ,
\end{equation}
where the ${\bf 4}$ is a Weyl spinor of $SO(1,5)$ while the ${\bf
  2}$ denotes the spinor representation of $SO(4)$.

Preserving half of the supercharges amounts to choosing backgrounds
that admit one or two globally defined spinors, which
corresponds to manifolds $\M_4$ with a reduced structure group $SU(2)$
or $SU(2)\times SU(2)$, respectively.\footnote{In the following, whenever we refer to an object as being globally defined we mean globally defined and nowhere vanishing.} Let us now discuss both cases in turn. We will start with a general analysis of the spectrum of type II supergravities in such backgrounds, and  then introduce a proper geometrical description that allows us to determine the structure of the moduli space explicitly.

\subsection{Field decompositions and spectrum}\label{section:spectrumd6}
In this section we give the massless type II supergravity fields in
terms of their representations under the six-dimensional Lorentz group
and the structure group and analyze how they assemble in $N =2$
multiplets, in the spirit of~\cite{Grana:2005ny}.

Because we deal with massless particles, we can use the light-cone
gauge where we only have to consider representations of $SO(4)_{lc}$
instead of the whole $SO(1,5)$ Lorentz group. Since we treat the case
of an $SU(2) \times SU(2)$ structure group, we examine the decomposition of massless type II fields under the group $SO(4)_{lc} \times SU(2) \times SU(2)$. For this, let us recall the decomposition of the two Majorana-Weyl representations ${\bf 8}^s$ and ${\bf 8}^c$ and the vector representation ${\bf 8}^v$ under the breaking
\begin{equation}
SO(8) \to SO(4)_{lc} \times SO(4) \to SO(4)_{lc} \times SU(2) \ .
\end{equation}
We get
\begin{equation} \begin{aligned}
   {\bf 8}^s & \to {\bf 2}_{\bf 2} \oplus {\bf \bar{2}}_{\bf \bar{2}}
\to {\bf 2}_{\bf 2}  \oplus 2 \, {\bf 1}_{\bf \bar{2}} \ , \\
   {\bf 8}^c & \to {\bf 2}_{\bf \bar{2}} \oplus {\bf \bar{2}}_{\bf 2} \to {\bf 2}_{\bf \bar{2}}  \oplus 2\, {\bf 1}_{\bf 2} \ , \\
   {\bf 8}^v & \to {\bf 1}_{\bf 4} \oplus {\bf 4}_{\bf 1} \to {\bf
1}_{\bf 4}  \oplus 2 \, {\bf 2}_{\bf 1} \ ,
\end{aligned}\end{equation}
where the subscript denotes the representation under the group
$SO(4)_{lc}$.

In type IIA string theory the massless fermionic degrees of freedom
originate from the $({\bf 8}^s,{\bf 8}^v)$ and $({\bf 8}^v,{\bf 8}^c)$
representation of $SO(8)_L \times SO(8)_R$, while in type IIB they
originate from the $({\bf 8}^s,{\bf 8}^v)$ and $({\bf 8}^v,{\bf 8}^s)$ representation.
Respectively, under the decomposition $SO(8)_L \times SO(8)_R \to SO(4)_{lc} \times
SU(2)_L \times SU(2)_R$ we find
\begin{equation}\begin{aligned}\label{fermiondec}
  ({\bf 8}^s,{\bf 8}^v) & \to 2 ({\bf 1},{\bf 1})_{\bf 6} \oplus 2 ({\bf 1},{\bf 1})_{\bf 2} \oplus ({\bf 1},{\bf 2})_{\bf \bar{6}} \oplus ({\bf 1},{\bf 2})_{\bf \bar{2}} \oplus 4 ({\bf 2},{\bf 1})_{\bf \bar{2}} \oplus 2 ({\bf 2},{\bf 2})_{\bf 2} \ , \\
  ({\bf 8}^v,{\bf 8}^s) & \to 2 ({\bf 1},{\bf 1})_{\bf 6} \oplus 2 ({\bf 1},{\bf 1})_{\bf 2} \oplus 4 ({\bf 1},{\bf 2})_{\bf \bar{2}} \oplus ({\bf 2},{\bf 1})_{\bf \bar{6}} \oplus ({\bf 2},{\bf 1})_{\bf \bar{2}} \oplus 2 ({\bf 2},{\bf 2})_{\bf 2} \ , \\
  ({\bf 8}^v,{\bf 8}^c) & \to 2 ({\bf 1},{\bf 1})_{\bf \bar{6}} \oplus 2 ({\bf 1},{\bf 1})_{\bf \bar{2}} \oplus 4 ({\bf 1},{\bf 2})_{\bf 2} \oplus ({\bf 2},{\bf 1})_{\bf 6} \oplus ({\bf 2},{\bf 1})_{\bf 2} \oplus 2 ({\bf 2},{\bf 2})_{\bf \bar{2}} \ .
\end{aligned}\end{equation}
We see that half of the gravitinos, denoted by the subscript $\bf 6$
and $\bf \bar{6}$, come in the $({\bf 1},{\bf 1})$ representation
while the other half is in the doublet representations $({\bf 1},{\bf
2})$ and $({\bf 2},{\bf 1})$ of $SU(2)_L \times SU(2)_R$. We will see
below that the 6d graviton is in the $({\bf 1},{\bf 1})$
representation and thus this representation refers to the gravity
multiplet in six dimensions. Hence the $({\bf 1},{\bf 2})$ and $({\bf
2},{\bf 1})$ representations correspond to additional gravitino multiplets,
which acquire a mass at the Kaluza-Klein scale due to the $SU(2) \times SU(2)$ structure
background. We have to project out these representations to end up
with a standard $N =2$ supergravity in six
dimensions~\cite{Andrianopoli:2001zh}. This is
analogous to projecting out all $SU(3)\times SU(3)$ triplets
in~\cite{Grana:2005ny} to achieve a standard $N =2$
theory in four dimensions.
After this projection, the fermionic components in the $({\bf 1},{\bf
1})$ representation become part of the gravity multiplet, while the $({\bf 2},{\bf
2})$ components correspond to the fermionic degrees of freedom in the
$N =2$ vector  and tensor multiplets in type IIA and IIB,
respectively.\footnote{In type IIB, only the anti-self-dual part of the
antisymmetric two-tensor is part of the gravity multiplet \cite{Romans:1986er}. The
self-dual component forms a tensor multiplet together with scalars in
the RR-sector. This tensor multiplet is also in the $({\bf 1},{\bf 1})$ representation.}

The massless bosonic fields of type II supergravity can be decomposed
analogously. For the NS-NS-sector, it is convenient to consider
the combination
\begin{equation}
E_{MN} = g_{MN} + B_{MN} + \phi \, \eta_{MN} \ ,
\end{equation}
which decomposes as
\begin{equation}\begin{aligned}\label{NSdec}
 E_{\mu \nu} & : ({\bf 1},{\bf 1})_{{\bf 9}} \oplus ({\bf 1},{\bf 1})_{{\bf 1}} \oplus ({\bf 1},{\bf 1})_{{\bf 3} \oplus {\bf \bar{3}}} \ , \\
 E_{\mu m} & : 2 ({\bf 1},{\bf 2})_{\bf 4}  \ ,  \\
 E_{m \mu } & : 2 ({\bf 2},{\bf 1})_{\bf 4}  \ ,  \\
 E_{mn} & : 4 ({\bf 2},{\bf 2})_{\bf 1}  \ .
\end{aligned}\end{equation}
Projecting out the doublets eliminates the six-dimensional vectors
$E_{\mu m}$ and $E_{m \mu }$, and we are left with $E_{\mu \nu}$,
i.e.\ the metric, the six-dimensional dilaton and the antisymmetric two-tensor, which are part of the
gravity multiplet, and the scalars $E_{mn}$, which reside in vector
or tensor multiplets. Since the latter ones correspond to the
internal metric and $B$-field components, they can be associated with
deformations of the $SU(2) \times SU(2)$ background.

Finally, in the RR-sector we need to decompose the $({\bf 8}^s,{\bf
8}^c)$ representation in type IIA and the $({\bf 8}^s,{\bf 8}^s)$ in type IIB. One finds
\begin{equation}\begin{aligned}\label{RRdec}
  ({\bf 8}^s,{\bf 8}^c) & \to 4 ({\bf 1},{\bf 1})_{\bf 4} \oplus 2 ({\bf 1},{\bf 2})_{\bf \bar{3}} \oplus 2 ({\bf 1},{\bf 2})_{\bf 1} \oplus 2 ({\bf 2},{\bf 1})_{\bf 3} \oplus 2 ({\bf 2},{\bf 1})_{\bf 1} \oplus ({\bf 2},{\bf 2})_{\bf 4} \ , \\
  ({\bf 8}^s,{\bf 8}^s) & \to 4 ({\bf 1},{\bf 1})_{\bf \bar{3}} \oplus 4 ({\bf 1},{\bf 1})_{\bf 1} \oplus 2 ({\bf 1},{\bf 2})_{\bf 4} \oplus 2 ({\bf 2},{\bf 1})_{\bf 4} \oplus \phantom{2}({\bf 2},{\bf 2})_{\bf 3} \oplus ({\bf 2},{\bf 2})_{\bf 1} \ .
\end{aligned}\end{equation}
We see that in type IIA only six-dimensional vectors in the
RR-sector survive the projection. Those which are in the
$({\bf 1},{\bf 1})$ representation form the graviphotons in the
gravity multiplet, those in the $({\bf 2},{\bf 2})$ give the vectors
in the vector multiplets.

Projecting out all $SU(2) \times SU(2)$
doublets leaves a spectrum that  for type IIA combines into
a gravitational multiplets plus a vector multiplet of the non-chiral
$d=6, N =2$ supergravity. For type IIB we obtain instead
a gravitational multiplets and two tensor multiplet of the chiral
$N =2$ supergravity.
To be more precise, in type IIA the gravitational multiplet contains
the graviton, an antisymmetric tensor, two (non-chiral) gravitini, four vector
fields, four Weyl fermions and a real scalar. These degrees of freedom
precisely correspond to the $({\bf 1},{\bf 1})$ representation of the
decompositions given in \eqref{fermiondec}, \eqref{NSdec} and
\eqref{RRdec}.
The vector multiplet contains a vector field, four gaugini and four
real scalars. These arise in the $({\bf 2},{\bf 2})$ representation of
the above decompositions.
In type IIB the gravitational multiplet contains
the graviton, five self-dual antisymmetric tensor and two (chiral)
gravitini. These degree of freedom are found in the $({\bf 1},{\bf 1})$
representation of the above decompositions. In addition there are two
tensor multiplets each containing an anti-self-dual antisymmetric
tensor, two chiral fermions and five scalars. One of them also originates
from the $({\bf 1},{\bf 1})$
representation while the second one comes from the $({\bf 2},{\bf 2})$ representation of
the above decompositions.

Note that  the resulting fields still depend on all
coordinates of the ten-dimensional spacetime, i.e.\ we have not performed
any Kaluza-Klein truncation on the spectrum but really deal with
ten-dimensional backgrounds. This procedure just corresponds to a
rewriting of the ten-dimensional
supergravity in a form where instead of the
ten-dimensional Lorentz group only $SO(1,5)\times SO(4)$ with sixteen
supercharges is manifest.
This ``rewriting'' of the
ten-dimensional theory has been
pioneered in ref.~\cite{deWN} and applied to the case of $SU(3)\times
SU(3)$ structures in refs.~\cite{Grana:2005ny,Koerber:2007xk}.

\subsection{General remarks on $SU(2)$ structures}\label{section:SU2d6}

In this section we review some general facts about the geometry of $SU(2)$ structures
on four-dimensional manifolds $\M_4$. A prominent example of such
manifolds is $K3$ which even has $SU(2)$ holonomy (for review, see
e.g.~\cite{Aspinwall:1996mn}).

The requirement of unbroken supercharges demands that the internal
manifold admits  a globally defined, nowhere vanishing spinor
$\eta$. Four-dimensional manifolds with this property have a structure
group $G$ that is contained in $SU(2)$ and are called
manifolds of $SU(2)$ structure. This is due to the fact that $\eta$
defines an element of the $SO(4)$ spinor bundle  which is a singlet of
the structure group $G \subset SO(4)$.
Since the spin bundle of $SO(4)$ is $SU(2) \times SU(2)$, the choice of
the spinor $\eta$ 
is equivalent to the choice
of a particular unbroken $SU(2)$ subgroup of $SO(4)$, which is
identified with the structure group $G = SU(2)$.

Let us now consider the decomposition of the spinor
representations under $G$.
The
defining spinor $\eta$ and its charge conjugate $\eta^c$ are both
globally defined and nowhere vanishing and therefore they are both
singlets under the structure group $G$. Moreover, they are linearly
independent and have the same chirality.\footnote{Note that $\eta$ and
$\eta^c$ have the same
  chirality in four Euclidean dimensions (see appendix \ref{section:conventions}). The transposed spinors,
i.e.\ $\eta^t$ and
  $\bar{\eta}=\left(\eta^c\right)^t$ are also globally defined and are the singlets of the dual space. }
Together they span the space of Weyl spinors of given chirality, which thus is decomposed into two $SU(2)$ singlets. The Weyl spinor representation of opposite chirality forms a doublet under the $SU(2)$ structure group.
It is just a matter of convention which representation we denote by ${\bf 2}$ and which one by ${\bf \bar{2}}$.
Here and in the following we choose
\begin{equation}
\label{spinor_decomposition_SU2}
 {\bf 2} \to {\bf 2} \ , \qquad \qquad {\bf \bar{2}} \to {\bf 1} \oplus {\bf 1 }
\end{equation}
for the breaking $SO(4) \to SU(2)$.

From the two singlets one can construct three distinct globally
defined
two-forms by
appropriately contracting with $SO(4)$ $\gamma$-matrices
\cite{Gauntlett:2003cy}
\begin{equation}
\label{definition_two-forms}
 \bar{\eta} \gamma_{mn} \eta = - \iu J_{mn} \ , \quad \bar{\eta}^c \gamma_{mn} \eta = \iu \Omega_{mn} \ , \quad \bar{\eta} \gamma_{mn} \eta^c = \iu \bar{\Omega}_{mn} \ , \quad m,n=1,\dots,4 \ ,
\end{equation}
where the normalization $\bar{\eta}\eta = 1$ is chosen.
However, these two-forms are not independent but satisfy
\begin{equation}
\label{relations_forms}
 \Omega \wedge \bar{\Omega} = 2 J \wedge J \ne 0 \ , \qquad \Omega \wedge J = 0 \ , \qquad \Omega \wedge \Omega = 0 \ ,
\end{equation}
which follows from the Fierz identities given
in~\eqref{Fierz_identities_4d}.
Conversely, the Fierz identities also
  show that the choice of a real two-form $J$ and a complex two-form $\Omega$ determines $\eta$
  completely (up to normalization) if they satisfy the above
relations. Therefore, $J$ and $\Omega$
  equivalently define an $SU(2)$ structure on the
  manifold.

Alternatively one can also define an
$SU(2)$-structure in terms of stable forms~\cite{Hitchin:2001rw}. A
stable $p$-form $\omega \in \Lambda^p V^*$ on a vector space $V$ is
defined as a form whose orbit under the action of $Gl(V)$ is open in
$\Lambda^p V^*$.
For a stable two-form $\omega$ on a $2m$-dimensional space this means that
$\omega^m \ne 0$. Thus, a stable two-form on an even-dimensional space
defines a symplectic structure on it.

On a four-dimensional manifold $\M_4$ the stable two-form $J$ satisfies $J \wedge J
\sim \operatorname{vol}_4$  and  locally defines a 
 symplectic structure that reduces the structure group
from $Gl(4)$ to $Sp(4,\mathbb{R})$.\footnote{Note that this symplectic structure may be non-integrable in
  the sense that
  $\diff \omega \ne 0$. Therefore, our notion of a symplectic structure differs from the usual mathematical
  terminology.}
The existence of additional
stable forms can reduce the structure group even further. In this
case one has to ensure that these stable forms do not reduce the
structure group in the same way. For example, one can take two
linearly independent stable
 two-forms $J_i, i =1,2$ that satisfy
\begin{equation}\label{stable_forms}
J_i \wedge J_j = \delta_{ij} \operatorname{vol}_4 \  .
\end{equation}
$J_1$ and $J_2$ then define a holomorphic two-form $\Omega =
J_1 + \iu J_2$, which globally defines a holomorphic subbundle in the
tangent space and therefore breaks the structure group to
$Sl(2,\mathbb{C}) \equiv Sp(2,\mathbb{C})$.

Analogously, in the case of three stable two-forms $J_i, i=1,2,3$,
which satisfy \eqref{stable_forms}
the structure group is reduced even further. Since $J_3$ is orthogonal
to $\Omega = J_1 + \iu J_2$ and  its complex conjugate, it defines a
product between the holomorphic and the anti-holomorphic tangent
bundle. Therefore, the $Sl(2,\mathbb{C})$ is further broken to the
$SU(2)$ subgroup which preserves this product.\footnote{Of course, this breaking is just the well-known
  relation $Sl(n,\mathbb{C}) \cap Sp(2n,\mathbb{R}) = SU(n)$.}
If one defines
\begin{equation}
\label{stable_forms_definition}
 J= J_3 \ , \quad \Omega = J_1+ \iu J_2 \ ,
\end{equation}
it is straightforward to check that~\eqref{stable_forms}
and~\eqref{relations_forms} are indeed equivalent.

In terms of stable forms it is easy to identify the parameter space of $SU(2)$ structures. A triple of stable forms $J_i$ has to fulfill \eqref{stable_forms} in order to define an $SU(2)$ structure on $\M_4$. Thus the $J_i$ span a three-dimensional subspace in the space of two-forms. By choosing some volume form $\operatorname{vol}_4$, i.e.\ some orientation on $\M_4$, we can interpret the wedge product as a scalar product of split signature on the space of two-forms.
With respect to this scalar product, the $J_i$ form an orthonormal basis for a space-like subspace.
The $SO(3,3)$ orbit of such a triple of $J_i$ gives all possible configurations that respect \eqref{stable_forms}. Thus, the configuration space can be written as $SO(3,3)$ divided by the stabilizer of the $J_i$.
The stabilizer consists of $SO(3)$ rotations in the subspace orthogonal to the $J_i$, which leave the $SU(2)$ structure invariant. Therefore, the configuration space for the $J_i$ is $SO(3,3)/SO(3)$.
The $SO(3)$ rotations in the stabilizer correspond to the action of the $SU(2)$ structure group on the space of two-forms. The $J_i$ are singlets under the $SU(2)$ structure group while the space orthogonal to them forms an $SU(2)$ triplet.

One should
  note that there is some redundancy in  the descriptions of
  $SU(2)$ structures on a manifold.
Any
rescaling of the $J_i$ does not change the unbroken $SU(2)$ and
therefore does not correspond to a degree of freedom for the $SU(2)$ structure. Hence we can fix
the normalization by~\eqref{stable_forms}. Furthermore,
there is a rotational $SO(3)$
symmetry between the three forms $J_i$. However, this symmetry is not
obvious from the definition \eqref{definition_two-forms}.  It
corresponds to $SU(2)$
rotations on the Weyl-spinor doublet $(\eta, \eta^c)$ which is a
symmetry because $\eta$ and $\eta^c$ have the same chirality on a
four-dimensional manifold.
One can check that the three two-forms $J_i$ indeed form a triplet under the action of this $SU(2)$.
By modding out this symmetry, we arrive at the parameter space of an $SU(2)$ structure over a point on the manifold $\M_4$, which is
\begin{equation} \label{moduli_space_geom_local}
 \mathcal{M}_{J_i} = \frac{SO(3,3)}{SO(3) \times SO(3)} \ .
\end{equation}

By consideration of the corresponding spin groups of $SO(3,3)$ and $SO(3)$ one expresses this result as
\begin{equation}
 \mathcal{M}_{J_i} = \frac{Sl(4,\mathbb{R})}{SO(4)} \ .
\end{equation}
If one compares this with the parameter space $Gl(4,\mathbb{R})/SO(4)$ of the metric over a point of $\M_4$, we see that the parameter space of $SU(2)$ structures incorporates all metric degrees of freedom except the volume factor. The missing degree of freedom corresponding to the volume factor can be associated with the normalization of the $J_i$ in \eqref{stable_forms}.\footnote{Note that the choice of a triple of normalized $J_i$ is just equivalent to the choice of a Hodge operator $\ast$ on the space of two-forms. This is reflected by the fact that the $J_i$ just span the positive eigenspace of $\ast$ in the space of two-forms.}

At the end of section \ref{section:moduliSU2d6} we derive the global moduli space of an $SU(2)$ structure over $\M_4$ from \eqref{moduli_space_geom_local}. However, before we do so, we first generalize our local discussion to general $SU(2)\times SU(2)$ structures.

\subsection{$SU(2)\times SU(2)$ structures and pure
  spinors}\label{section:SU2SU2d6}
One can generalize the $SU(2)$ structures discussed in the previous
section by assuming that the manifold admits two globally defined,
nowhere vanishing spinors $\eta_1$ and $\eta_2$. Each of them defines
an $SU(2)$ structure on its own and if they are identical everywhere
on the manifold, this reduces to the case discussed in the previous
section.
In the other limiting case where $\eta_1$ and $\eta_2$ are orthogonal
at each point, the two $SU(2)$ structures intersect in some
identity structure, which means that the spinor bundle is trivial and
compactification on this backgrounds preserves all 32 supercharges
(corresponding to  $N =8$ in $d=4$). However, in principle one can
also have the intermediate case of two globally defined,
nowhere vanishing spinors $\eta_1$ and $\eta_2$ that are linearly
independent at most points but become parallel at some points on the
manifold.

Analogously to the last section, there is an equivalent formulation of
$SU(2) \times SU(2)$ structures in terms of globally defined stable
forms. This is elegantly captured   by the notion of pure spinors and
generalized geometry~\cite{Gualtieri:2003dx,Hitchin:2004ut,Grana:2006kf}.
Let us briefly review this concept for a $2n$-dimensional manifold $\M$
and then return to the case of $SU(2)\times SU(2)$ structures
afterwards.

\subsubsection{Generalized geometry and the pure spinor approach}\label{section:SUNSUN}
In
generalized geometry  one considers a generalized tangent bundle
$\mathcal{T}\M$ which locally looks like $T\M \oplus T^*\M$ and
therefore admits a scalar product $\mathcal{I}$ of split signature
that is induced by the canonical pairing between tangent and cotangent
space.
On a $2n$-dimensional manifold $\M$ this bundle thus has a structure
group contained in $SO(2n,2n)$.

Similar to our discussion in the last section, one can introduce
further objects that break the structure group. For example, an almost
complex structure $\mathcal{J}$, fulfilling
\begin{equation}
\mathcal{J}^2 = -1 \ ,
\end{equation}
 can be defined if (and only if)
the complexified generalized tangent bundle globally splits as
\begin{equation}
\label{decomposition_generalized_acs}
\left( \mathcal{T}\M \right)_\mathbb{C} = L_+ \oplus L_- \ ,
\end{equation}
where $L_\pm$ are the eigenspaces of $\mathcal{J}$ with the eigenvalues
$\pm \iu$. If $\mathcal{J}$ is globally defined on $\M$, the structure group
of $\mathcal{T}\M$ is broken from $SO(2n,2n)$ to $U(n,n)$.

When two generalized almost complex structures $\mathcal{J}_{1,2}$ exist, the notion of
compatibility can be defined. More precisely,
$\mathcal{J}_1$ and
$\mathcal{J}_2$ are called compatible if
\begin{enumerate}
 \item $\mathcal{J}_1$ and $\mathcal{J}_2$ commute and
 \item $\mathcal{G} := \mathcal{I} \mathcal{J}_1 \mathcal{J}_2$ is a positive definite metric on $\mathcal{T}\M$, where $\mathcal{I}$ is the canonical scalar product on $\mathcal{T}\M$.
\end{enumerate}
The first condition ensures that the splittings~\eqref{decomposition_generalized_acs} can be done simultaneously, i.e.\ that
\begin{equation}
\label{decomposition_comp_generalized_acs}
\left( \mathcal{T}\M \right)_\mathbb{C} = L_{++} \oplus L_{-+} \oplus L_{+-} \oplus L_{--}\ ,
\end{equation}
where the indices correspond to the eigenvalues of $\mathcal{J}_{1,2}$.
The second condition ensures that each of the four components
in~\eqref{decomposition_comp_generalized_acs} is $n$-dimensional
such that the two compatible generalized almost complex structures
reduce the structure group to $U(n) \times U(n)$, where each $U(n)$
acts on two of the four components.

Let us now briefly review how to reformulate  generalized geometry
in terms of pure spinors $\Phi$ \cite{Gualtieri:2003dx,Hitchin:2004ut}. One first defines the annihilator
space  $L_\Phi$ of a complex $SO(2n,2n)$ Weyl spinor $\Phi$ as the
subspace of complexified gamma-matrices which map
$\Phi$ to zero, i.e.\
\begin{equation}
 L_\Phi \equiv \left\lbrace \Gamma \in \left(\mathcal{T}\M \right)_\mathbb{C} | \Gamma \Phi = 0 \right\rbrace \ .
\end{equation}
Note that  $L_\Phi$ is always isotropic as for each element
$\Gamma$ of $L_\Phi$ we have
\begin{equation}
 0 = \Gamma^2 \Phi = \mathcal{I}(\Gamma,\Gamma)\, \Phi \ ,
\end{equation}
which implies  $\mathcal{I}(\Gamma,\Gamma)=0$ for all $\Gamma \in
L_\Phi$.

A complex Weyl spinor $\Phi$ of $SO(2n,2n)$ is called pure if its annihilator space
has maximal dimension, i.e.\ $\operatorname{dim} L_\Phi = 2n$.
$\Phi$ is called normalizable if
\begin{equation}
\langle \Phi, \bar{\Phi} \rangle > 0 \ ,
\end{equation}
where the brackets denote the usual spinor product.
As a consequence of Chevalley's theorem~\cite{Chevalley:1996}, which states
\begin{equation}
\label{Chevalley_theorem}
 \operatorname{dim} L_\Phi \cap L_\Psi = 0 \quad \Leftrightarrow \quad \langle \Phi , \Psi \rangle \ne 0 \ ,
\end{equation}
normalizable pure spinors define a splitting
\begin{equation}
 \left( \mathcal{T}\M \right)_\mathbb{C} = L_\Phi \oplus L_{\bar{\Phi}} \ .
\end{equation}
By matching the annihilator space $L_\Phi$ with the $+\iu$
eigenspace of some generalized almost complex structure, one can show
that both are equivalent up to the normalization factor of $\Phi$.
Thus, generalized almost complex structures are equivalent to
lines of pure $SO(2n,2n)$ spinors. A pure spinor therefore breaks the
structure group of $\mathcal{T}\M$ further from $U(n,n)$ to $SU(n,n)$
in that fixing its
phase eliminates the $U(1)$ factor.

The compatibility conditions for two generalized almost complex
structures translate into a compatibility condition
on the corresponding pure spinors.
Two normalizable pure $SO(2n,2n)$ spinors $\Phi_{1,2}$ are compatible if and only
if their annihilator spaces intersect in a space of dimension
$n$, i.e.\
\begin{equation}
\label{Compatibility_gacs}
\operatorname{dim}(L_{\Phi_1} \cap L_{\Phi_2}) =n \ .
\end{equation}
Thus the pair $\Phi_{1,2}$
breaks the structure group to $SU(n) \times SU(n)$
(instead of $U(n) \times U(n)$).
Therefore, pure spinors of generalized geometry provide a convenient
framework to deal with $SU(n) \times SU(n)$ structures.
Whenever $\mathcal{T}\M=T\M \oplus T^*\M$ globally,
both $SU(n)$ factors can be projected to the
tangent space $T\M$. In this case the intersection of these
projections defines the
structure group of the tangent bundle.

The compatibility condition of two pure spinors also restricts their
chirality.  Since $SO(n,n)$ transformations do not mix
chiralities, one can always assume $\Phi_1$ and $\Phi_2$ to be
of definite chirality.
Furthermore, two pure spinors $\Phi_1$ and $\Phi_2$ have the same chirality if and only if \cite{Charlton:1996PhD}
\begin{equation}
\label{compatible_pure_spinors_chirality}
 \operatorname{dim}(L_{\Phi_1} \cap L_{\Phi_2}) = 2 k
\end{equation}
for $k \in \mathbb{N}$.
Therefore, two compatible pure spinors are of the same chirality if
$n$ is even and of opposite chirality for
$n$ being odd.

One can construct pure $SO(2n,2n)$ spinors out of the two globally
defined $SO(2n)$ spinors $\eta_1$ and $\eta_2$ discussed at the
beginning of this section. More precisely, one has
\begin{equation} \label{tensor_product_spinors}
\eta_1 \otimes \bar{\eta}_2 = \frac{1}{4}\sum_{k=0}^{2n} \frac{1}{k!} \left(
\bar{\eta}_2 \gamma_{m_1\dots m_k} \eta_1 \right) \gamma^{m_k\dots m_1}   \ ,
\end{equation}
where $\gamma^{m_1\dots m_k}$ is the totally antisymmetric product of
$SO(2n)$ $\gamma$-matrices.
One can act with $SO(2n)$ gamma-matrices from the left or from the
right which in turn defines an $SO(2n,2n)$ action
on the bi-spinor $\eta_1 \otimes
\bar{\eta}_2$. By
extensive use of the Fierz identities given in appendix \ref{section:conventions}, one can show that $\eta_1 \otimes \bar{\eta}_2$ is pure and normalizable. The same holds for
$\eta_1 \otimes \bar{\eta}_2^c$ and these two pure spinors are
moreover compatible. Thus they can be used to discuss $SU(n)\times
SU(n)$ structures.

The map
\begin{equation}\label{mapping_bispinors_forms}
\tau \ : \ \eta_1 \otimes \bar{\eta}_2 \longmapsto \tau(\eta_1 \otimes \bar{\eta}_2) \equiv \frac{1}{4}\sum_{k=0}^{2n} \frac{1}{k!} \left( \bar{\eta}_2 \gamma_{m_1\dots m_k} \eta_1 \right) e^{m_k} \wedge \dots \wedge e^{m_1} \ ,
\end{equation}
with $e^{m_k}$ being a local basis of one-forms, identifies
$SO(2n,2n)$ spinors with formal sums of differential forms. This
isomorphism is canonical up to the choice of a volume form on the
manifold \cite{Grana:2005ny}.
Note that \eqref{mapping_bispinors_forms} maps negative (positive)
chirality spinors to differential forms of odd (even) degree.
Moreover, it is an isometry with respect to the spinor
product and the so-called Mukai pairing. The latter is defined by\footnote{Here, $[ \cdot ]$ is the floor function which rounds down its argument to the next integer number.}
\begin{equation}\label{Mukai_pairing}
 \langle \Psi , \chi \rangle = \sum_p (-)^{[(p-1)/2]} \Psi_p \wedge
 \chi_{2n - p}\ ,
\end{equation}
which maps two formal sums of differential forms to a form of top degree. Like the spinor product, it is symmetric for $n$ even, and anti-symmetric for $n$ being odd.
Using the definition
\begin{equation} \label{lambda_definition}
 \lambda\, \alpha_{p} = (-1)^{[(p-1)/2]} \alpha_{p}
\end{equation}
for a $p$-form $\alpha_{p}$,
we can write the Mukai pairing also in the form
\begin{equation}\label{Mukai_pairing_2}
 \langle \Psi , \chi \rangle = \left[ \Psi \wedge \lambda\, \chi \right]_{\operatorname{deg}=2n} \ .
\end{equation}
In the following we frequently use the isomorphism \eqref{mapping_bispinors_forms}.

Before we go on, let us state two important facts. One can show that a pure spinor $\Phi$ is always of the form~\cite{Gualtieri:2003dx}
\begin{equation} \label{general_pure_spinor_form}
 \Phi = \e^{-B} \wedge \e^{- \iu J} \wedge \Omega \ ,
\end{equation}
where $B$ and $J$ are real two-forms and $\Omega$ is some complex
$k$-form, $k \leq 2n$, that is locally decomposable into complex
one-forms.
Furthermore, one can prove that two compatible
pure spinors are always of the form~\cite{Grana:2006kf}
\begin{equation} \label{general_comp_pure_spinor_form}
 \Phi_1 = \e^{-B} \wedge \tau( \eta_1 \otimes \bar{\eta}_2 )\ ,\qquad
 \Phi_2 = \e^{-B} \wedge \tau( \eta_1 \otimes \bar{\eta}^c_2)\ ,
\end{equation}
where the isomorphism $\tau$ is defined in \eqref{mapping_bispinors_forms}.

Before we close this section, let us define the
generalized Hodge operator~\cite{Jeschek:2004wy,Cassani:2007pq}
\begin{equation}
\label{Hodge_star_generalized}
\ast_B = \e^{-B} \ast \lambda\, \e^B\
\end{equation}
which acts on the space of forms, with $\lambda$ defined by \eqref{lambda_definition}.
Under the isomorphism given in \eqref{mapping_bispinors_forms} the generalized Hodge operator is mapped to charge conjugation on the space of $SO(2n,2n)$ spinors.
Analogously to the conventional Hodge operator, the generalized version can define a positive definite metric $G(\cdot , \cdot ) \equiv \langle \cdot , \ast_B\, \cdot \rangle$ on the space of forms, which is just the composition of $\ast_B$ with the Mukai pairing.
From \eqref{Mukai_pairing_2} it is easy to see that $G$ acts
on the space of forms by
\begin{equation}
G( \e^{-B}\wedge \Psi , \e^{-B} \wedge \chi ) = \langle \e^{-B} \wedge \Psi , \ast_B\,  \e^{-B} \wedge  \chi \rangle =
  \left[ \Psi \wedge \ast\, \chi \right]_{\operatorname{deg}=2n} = \sum_{p=0}^{2n} \Psi_p \wedge \ast\, \chi_p \ ,
\end{equation}
which indeed is positive definite. Therefore, the Mukai pairing and the generalized Hodge operator have the same signature.

Since $\ast_B^2 = 1$ on forms of even degree, we see that the generalized Hodge operator
corresponds to an almost
product structure on $\Lambda^\textrm{even}T^*\M$. For $B=0$, it coincides on two-forms with the conventional Hodge star operator, which is of split signature over each point. In this case, the forms $1\pm \operatorname{vol}$ are eigenvectors of $\ast_B$ to the eigenvalue $\mp 1$ and we see that $\ast_B$ has split signature on $\Lambda^\textrm{even}T^*\M$ over each point of $\M$. Since $B$ can be continously switched on, the signature is independent of $B$, and the eigenspaces of $\ast_B$ (with eigenvalue $\pm 1$)
at a given point on the manifold have the same
dimension.
As the standard Hodge operator, the generalized Hodge operator $\ast_B$ can be globally defined on $\M$ and therefore must be invariant under the $SU(2) \times SU(2)$ structure group. Hence, $\ast_B$ leaves the $SU(2)\times SU(2)$ representations invariant, and its eigenspaces coincide with these representations.

As stated at the beginning of this section the generalized tangent bundle
$\mathcal{T}\M$ locally has the structure $T\M \oplus T^*\M$.
In the following, we first perform a
  local analysis  and consider the
  algebraic structure of the bundle over a point on the manifold.
Therefore, we
  use the notation $T\M \oplus T^*\M$ instead of
  $\mathcal{T}\M$.

\subsubsection{Pure spinors on a manifold of dimension four} \label{section:SU2SU2d6_sub}
Let us now apply the previous discussion to the case of $SU(2)\times
SU(2)$ structures on a four-dimensional manifold $\M_4$. However, let us
start with the simpler case of $SU(2)$ structures or in other words
with the case where a single $SO(4)$ spinor $\eta$ exists on $\M_4$. With
its help we can construct the two pure $SO(4,4)$ spinors $\eta \otimes
\bar{\eta}$ and $\eta \otimes \bar{\eta}^c$.
Using the definitions \eqref{definition_two-forms} and \eqref{mapping_bispinors_forms} we
identify\footnote{This can be seen from the definition
  \eqref{definition_two-forms} together with the normalization
$\bar{\eta} \eta = 1$ and  the fact that $\eta$ and $\eta^c$ are
  orthogonal, i.e.\ $\eta^t \eta = 0$.
Furthermore, using the Fierz identities \eqref{Fierz_identities_4d} one can show $J \wedge J = J \wedge \ast J =\operatorname{vol}_4$ which allows us
  to write the first tensor product as an exponential.}
\begin{equation} \label{pure_SU2_spinors_4d}
\tau(\eta \otimes \bar{\eta}) = \tfrac{1}{4}\, \e^{- \iu J} \ , \qquad
\tau(\eta \otimes \bar{\eta}^c ) = \tfrac{1}{4}\, \iu \Omega \ .
\end{equation}
We can additionally shift these pure $SO(4,4)$ spinors by a $B$-field
leaving all conditions unchanged. Thus, we arrive at
\begin{equation}
\label{pure_spinors_4d}
\Phi_1 =  \tfrac{1}{4}\, \e^{-B - \iu J} \ , \qquad
\Phi_2 = \tfrac{\iu}{4} \, \e^{-B} \wedge \Omega \ .
\end{equation}

Let us now turn to the case  of general $SU(2) \times SU(2)$ structures.
We first analyze the conditions for spinors to
be pure and compatible. For the case at hand this is simplified by the triality
property of $SO(4,4)$ which isomorphically
permutes the spinor representation, its
conjugate and the  vector representation among each other.
In particular, the quadratic form $\langle \cdot
, \cdot \rangle$ on the spinor space is mapped to the usual scalar product on
the vector space. This fact is used in the following.

For a pure spinor $\Phi$ the annihilator space $L_\Phi$
has dimension four. In addition Chevalley's theorem
\eqref{Chevalley_theorem} implies
\begin{equation}
\label{pure_spinor_SO44}
 \langle \Phi , \Phi \rangle = 0 \ .
\end{equation}
As shown in~\cite{Charlton:1996PhD}, this condition is also sufficient for $\Phi$ to be pure. Thus, pure $SO(4,4)$ spinors correspond to complex light-like vectors under the triality map.
Furthermore,  for $\Phi$ to be normalizable we need
\begin{equation}
 \label{normalization_pure_spinor_SO44}
 \langle \Phi , \bar{\Phi} \rangle > 0 \ .
\end{equation}

Now let us consider two pure normalizable spinors $\Phi_i$, $i=1,2$
which by definition satisfy
\begin{equation} \label{conditions_two_pure_spinors}
 \langle \Phi_i,\Phi_i\rangle = 0 \ ,\qquad
 \langle \Phi_i,\bar{\Phi}_i\rangle > 0  \ .
\end{equation}
If they are compatible, they also satisfy
\eqref{Compatibility_gacs}, which on $\M_4$ reads
\begin{equation}\label{dim2}
 \operatorname{dim} (L_{\Phi_1} \cap L_{\Phi_2}) = 2 \ .
\end{equation}
{}From \eqref{compatible_pure_spinors_chirality} we conclude that both $\Phi_i$ have
the same chirality which, together with \eqref{Chevalley_theorem},
implies that \eqref{dim2} is equivalent to
\begin{equation} \label{comp_pure_spinor_conditions}
 \langle \Phi_1, \Phi_2\rangle  = 0  \ , \qquad
 \langle \Phi_1, \bar{\Phi}_2\rangle  = 0 \ .
\end{equation}
Finally, we can choose the normalization
\begin{equation}\label{normalization_pure_spinors}
 \langle \Phi_1 , \bar{\Phi}_1 \rangle = \langle \Phi_2 , \bar{\Phi}_2 \rangle \ne 0 \ .
\end{equation}

Let us close this section by analyzing which possible cases of $SU(2)\times
SU(2)$ structures can occur on four-dimensional manifolds.
We just argued that $\Phi_1$ and $\Phi_2$ have the same chirality
so that the corresponding forms are of odd or even degree.
We start with the case where both spinors $\Phi_1$ and $\Phi_2$ have
negative chirality.
From \eqref{general_pure_spinor_form} we see that both pure spinors
are of the form
\begin{equation} \label{4d_spinors_odd_degree}
 \Phi_i = U_i \wedge \e^{-\iu J_i} \ ,\quad i=1,2
\ ,
\end{equation}
where $U_i$ are two complex one-forms while $J_i$ are two
non-vanishing real two-forms.\footnote{For simplicity we ignore the $B$-field which
however can be easily included.}
In addition, the compatibility condition~\eqref{comp_pure_spinor_conditions} implies
\begin{equation} \label{4d_spinors_odd_degree_comp}
 U_1 \wedge U_2 \wedge (J_1 - J_2) =0 \ , \qquad U_1 \wedge \bar{U}_2 \wedge (J_1 + J_2) =0 \ ,
\end{equation}
while the normalization \eqref{normalization_pure_spinors} translates into
\begin{equation} \label{4d_spinors_odd_degree_norm}
 U_1 \wedge \bar{U}_1 \wedge J_1  = U_2 \wedge \bar{U}_2 \wedge J_2 \ne 0 \ .
\end{equation}
Since $U_2 = a U_1 + b \bar{U}_1$ does not solve \eqref{4d_spinors_odd_degree_comp} and
  \eqref{4d_spinors_odd_degree_norm} we conclude
that $U_2$ is linearly independent of $U_1$ and
$\bar{U}_1$ and therefore $U_1,\bar{U}_1,U_2,\bar{U}_2$ form a basis
of $T^*\M_4$.
Thus, we can find four one-forms that form a basis at every point
of $\M_4$ and hence the manifold $\M$ is parallelizable. This means
that the two factors of the $SU(2) \times SU(2)$ structure
just intersect in the identity. Thus, the structure group of the
manifold is trivial. This in turn implies that $\M_4$ admits four
globally defined
$SO(4)$ spinors corresponding to string backgrounds with $32$
supercharges. This fact can also be seen from
\eqref{general_comp_pure_spinor_form}. Since $\Phi_{1,2}$ are of odd
degree, $\eta_1$ and $\eta_2$ are of opposite chirality. Together with
their charge conjugated spinors they lead to four globally defined spinors.
Since in this paper we focus on backgrounds with $16$ supercharges, we
do not discuss this case any further.

Let us turn to the case where both spinors are of even degree. The
most general form for those two spinors is given in
\eqref{general_comp_pure_spinor_form}, where now $\eta_1$ and $\eta_2$
are of the same chirality to ensure that $\Phi_1$ and $\Phi_2$ are of
even degree. As we explained above Eq.~\eqref{spinor_decomposition_SU2} a spinor $\eta_1$ and
its charge conjugate $\eta^c_1$ span the whole space of Weyl spinors
of a given chirality. Therefore, $\eta_2$ has to be a linear
combination of $\eta_1$ and $\eta_1^c$. However, this  means that
we can rotate $\Phi_1$ and $\Phi_2$ in such a way that they are of the
form
\begin{equation}
 \Phi_1 = \e^{-B} \wedge \tau(\eta_1 \otimes \bar{\eta}_1) \ , \qquad  \Phi_2 = \e^{-B} \wedge \tau(\eta_1 \otimes \bar{\eta}_1^c) \ .
\end{equation}
Therefore, they give a proper $SU(2)$
structure on the manifold, which takes the form
\eqref{pure_spinors_4d}.\footnote{Strictly speaking, we can only call this
a proper $SU(2)$ structure for geometric compactifications since for
non-geometric backgrounds there is globally no projection map
$\mathcal{T} \M \to T \M$ such that we can compare the two $SU(2)$
factors. However, we can do this projection locally, and thus may
compare both $SU(2)$ structures pointwise. In this sense, we can
define proper $SU(2)$ structures even for non-geometric backgrounds.
}

To summarize, due to the fact that the pure spinors have definite chirality
there is no case which interpolates between the trivial structure and
the $SU(2)$ structure case. This can also be understood from the fact
that a pair of nowhere vanishing spinors $\eta, \eta^c$ spans
the space of given chirality. Therefore, all linearly independent spinors
have to be of opposite chirality and thus cannot be parallel to
$\eta$ at any point in $\M_4$.
Thus, generic $SU(2)\times SU(2)$ structures cannot exist but always
have to be $SU(2)$ or trivial structures.
Note that our conclusion crucially depends on the assumption that $\eta_1$ and $\eta_2$ are nowhere vanishing. Therefore the case of a generic warp factor deserves a separate analysis
which, however, we do not go into here.

\subsection{Moduli space of $SU(2)$ structures}\label{section:moduliSU2d6}
The aim of this section is to determine the moduli space of $SU(2)$
 structures. Let us first observe that an eight-dimensional
 Weyl spinor of $SO(4,4)$ decomposes under  $SU(2) \times SU(2)$
as\footnote{In section \ref{section:SU2SU2d6_sub} we showed that for
backgrounds with $16$ supercharges both $SU(2)$ factors must be the
same after projection to the tangent space. However, as long as we
stay in the framework of generalized geometry and consider pure $SO(4,4)$
spinors, these two factors are different. Therefore we do
the decomposition for $SU(2) \times SU(2)$.
}
\begin{equation}
  {\bf 8}^s \to ({\bf 2},{\bf 2}) \oplus 4 ({\bf 1},{\bf 1}) \ , \qquad {\bf 8}^c \to 2 ({\bf 2},{\bf 1}) \oplus 2 ({\bf 1},{\bf 2}) \ .
\end{equation}
Note that, exactly as in~\eqref{spinor_decomposition_SU2},
the two conjugate spinors decompose differently. Eq. \eqref{choice_chirality} gives a canonical choice for the sign of the chirality operator. Hence,
the ${\bf 8}^s$ (${\bf 8}^c$) representation corresponds to forms of even
(odd) degree.
Let us denote the space of forms transforming in the $({\bf r}, {\bf
s})$ representation of $SU(2) \times SU(2)$
 by ${ U_{{\bf r},{\bf s}}}$. They can be arranged in a diamond as
given in Table~\ref{diamond}, where the prime is used to distinguish the several singlets.
\begin{table}[htdp]
\begin{center}
\begin{tabular}{ccccc}
 &&  $U_{ {\bf 1},{\bf 1'}}$  && \\
& $U_{ {\bf 2},{\bf 1'}}$ && $U_{ {\bf 1},{\bf 2}}$ & \\
$U_{ {\bf 1'},{\bf 1'}}$ &&  $U_{ {\bf 2},{\bf 2}}$  && $U_{ {\bf 1},{\bf 1}}$ \\
& $U_{ {\bf 1'},{\bf 2}}$ && $U_{ {\bf 2},{\bf 1}}$ & \\
 &&  $U_{ {\bf 1'},{\bf 1}}$  &&
\end{tabular}
 \caption{\small
\textit{Generalized $SU(2) \times SU(2)$ diamond.}}\label{diamond}
\end{center}
\end{table}

In section~\ref{section:spectrumd6} we showed that
for a background to have 16 supercharges  it is necessary to
remove  all massive gravitino
multiplets which corresponds to projecting out all $SU(2)$
doublets. This
eliminates the entire ${\bf 8}^c$ representation (or equivalently all
odd forms in $U_{ {\bf 2},{\bf 1'}}, U_{ {\bf 1},{\bf 2}},
U_{ {\bf 1'},{\bf 2}}, U_{ {\bf 2},{\bf 1}}$) leaving only  the
${\bf 8}^s$ (i.e.\ the even forms in Table~\ref{diamond}).
This is consistent with the result of the previous section
that backgrounds with 16 supercharges require an $SU(2)$
  structure described by pure spinors of positive chirality.

Now we are able to derive the moduli space. For this, let us first discuss the parameter space of one single normalizable pure $SO(4,4)$ spinor.
The two conditions~\eqref{pure_spinor_SO44}
and~\eqref{normalization_pure_spinor_SO44} have a natural
interpretation in the isomorphic picture where $\Phi$ is a complex
vector. Equation~\eqref{pure_spinor_SO44}
and~\eqref{normalization_pure_spinor_SO44} ensure that the real
and imaginary part of $\Phi$ form a pair of space-like orthogonal
vectors. Therefore, $\Phi$ is left invariant by the group
$SO(2,4)$. From section~\ref{section:SU2SU2d6} we know that a pure normalizable $SO(4,4)$ spinor breaks the structure group to $SU(2,2)$. Both pictures are consistent with each other since $SU(2,2)$ is just the double cover of $SO(2,4)$.
The pure spinor $\Phi$ therefore parameterizes the space
$SO(4,4)/{SO(2,4)}$.
However, the phase of $\Phi$ does not affect the $SU(2,2)$
structure. Hence the actual parameter space of a single pure spinor is
\begin{equation}
{\cal M}_\Phi = \frac{SO(4,4)}{SO(2) \times SO(2,4)} \ .
\end{equation}

The moduli space of $SU(2)$ structures
 is more conveniently
discussed in terms of the real and imaginary
parts of the two spinors $\Phi_i$ or in other words in terms of four
real vectors $\Psi_a, a=1, \dots, 4$ in the space of even
forms. Then~\eqref{comp_pure_spinor_conditions}
and~\eqref{normalization_pure_spinors} just translate into the
conditions
\begin{equation}
\label{conditions_real_spinors}
 \langle \Psi_a, \Psi_b \rangle = c \, \delta_{ab}\, \operatorname{vol}_4 \ ,
\end{equation}
where $c$ parameterizes the scale of the $\Psi_a$.
The four $\Psi_a$ form the singlet corners in Table~\ref{diamond}
since they are globally defined and thus must be singlets of the
structure group.

In order to understand the signature of the $SU(2)\times
SU(2)$ diamond (Table~\ref{diamond}) we use the
generalized Hodge operator $\ast_B$ defined in \eqref{Hodge_star_generalized}.
As we explained in section~\ref{section:SU2SU2d6_sub}, the operator $\ast_B$ leaves the $SU(2)\times SU(2)$ representations invariant and its eigenspaces coincide with the representations. The eigenvalues of $\ast_B$ are $\pm 1$, and the corresponding eigenspaces are of the same dimension.

The eigenspace with eigenvalue $+1$ is spanned by the four
spinors $\Psi_a$, i.e.\ by the $SU(2) \times SU(2)$
singlets, which is consistent
  with~\eqref{conditions_real_spinors}.
This can be calculated using
the form~\eqref{pure_spinors_4d} and the fact that the $J_i$,
$i=1,2,3$, defined in~\eqref{stable_forms_definition}, are self-dual
with respect to the standard Hodge operator.
Therefore, the orthogonal complement $U_{{\bf 2},{\bf 2}}$ is the
eigenspace with eigenvalue $-1$.
This shows that a choice of $\Psi_a$ already determines the eigenspaces of
the generalized Hodge operator and thus the operator itself.
Since the composition of the Mukai pairing with $\ast_B$ is positive definite, the eigenvalue corresponding to some eigenvector of $\ast_B$ gives also its signature under the Mukai
pairing. Therefore we conclude that the Mukai pairing is positive definite on the
$SU(2) \times SU(2)$ singlets and negative definite on $U_{{\bf 2},{\bf 2}}$.

Thus, we see that the $\Psi_a$, which respect the
condition~\eqref{conditions_real_spinors}, define a space-like
four-dimensional subspace in $\Lambda^\textrm{even}T^*\M_4$ or in other words
they parameterize the space
${SO(4,4)}/{SO(4)}$.
However, we also need to divide out the rotational $SO(4)$ symmetry
among the $\Psi_a$ since it does
not change the $SU(2) \times SU(2)$ structure. This leaves as the
physical parameter space
\begin{equation}\label{local_moduli_space_4d}
{\cal M}_{\Psi_a} =  \frac{SO(4,4)}{SO(4)\times SO(4)} \ .
\end{equation}

So far we have analyzed the deformation space of $SU(2)\times SU(2)$
structures.
Let us now discuss its appearance in the low energy effective theory.
There are basically two different effective theories one can write
down. On the one hand it is possible to rewrite the ten-dimensional
supergravity in a form where instead of the
ten-dimensional Lorentz group only $SO(1,5)\times SO(4)$ with $16$
supercharges is manifest. In this formulation no Kaluza-Klein truncation
is performed but instead all fields still carry the full
ten-dimensional coordinate dependence
of the background \eqref{stringbackground}. This is the approach of ref.~\cite{deWN}, which
we already discussed at the end of section \ref{section:spectrumd6}.
In the corresponding effective theory the space
\eqref{local_moduli_space_4d} appears as the target space  of
the Lorentz-scalar deformations which is consistent with the
constraints of the corresponding $d=6$ supergravity.

Alternatively, one can perform a Kaluza-Klein truncation and only keep
the light modes on $\M_4$. This amounts to a truncation of the space of even
forms to a finite subspace $\Lambda^\textrm{even}_\textrm{finite} T^* \M_4 $
with the  assumption
that the Mukai pairing is non-degenerate on it \cite{Grana:2005ny}.\footnote{This just
  corresponds to the assumption that charge conjugation, i.e.\ the
  generalized Hodge star operator, preserves
  $\Lambda^\textrm{even}_\textrm{finite}T^* \M_4 $.}
Concretely this means that we can expand the four real spinors
$\Psi_a$ in a basis of
$\Lambda^\textrm{even}_\textrm{finite}T^* \M_4 $.
The  generalized Hodge
star operator $\ast_B$,
which, via the $\Psi_a$, is globally defined,
splits $\Lambda^\textrm{even}_\textrm{finite}T^* \M_4 $
into the eigenspaces of $\ast_B$, i.e.\
\begin{equation} \label{split_KKmodes}
\Lambda^\textrm{even}_\textrm{finite}T^* \M_4  = \Lambda^\textrm{even}_+ T^* \M_4 \oplus \Lambda^\textrm{even}_- T^* \M_4 \ ,
\end{equation}
where the subscript denotes the eigenvalue with respect to $\ast_B$. Furthermore,
these eigenspaces are orthogonal to each other with respect to the Mukai pairing.
$\Lambda^\textrm{even}_+ T^* \M_4 $  consists of $SU(2)\times SU(2)$ singlets
only, and thus over each point it is spanned by the
$\Psi_a$. Therefore, each element of $\Lambda^\textrm{even}_+ T^* \M_4 $ can be
written as a linear combination of the $\Psi_a$ where the
coefficients may depend on the base point on $\M_4$, i.e.\ are functions
on $\M_4$. However, only constant coefficients survive the Kaluza-Klein
truncation, and thus $\Lambda^\textrm{even}_+ T^* \M_4 $ is spanned by the
$\Psi_a$ only and has dimension four.

Now let us turn to the eigenspace $\Lambda^\textrm{even}_- T^* \M_4 $. It
consists of sections in $U_{ {\bf 2},{\bf 2}}$ and therefore we
cannot make the same argument as for $\Lambda^\textrm{even}_+ T^* \M_4 $. In
contrast to the bundle of $SU(2)\times SU(2)$ singlets, $U_{ {\bf
2},{\bf 2}}$ might be twisted over the manifold and the dimension of
$\Lambda^\textrm{even}_- T^* \M_4 $ may differ from four,
say $n+4$.
Thus, $\Lambda^\textrm{even}_\textrm{finite} T^* \M_4 $ is a vector space
of signature $(4,n+4)$. The four spinors $\Psi_a$
satisfy~\eqref{conditions_real_spinors}, and therefore span a
four-dimensional space-like subspace in
$\Lambda^\textrm{even}_\textrm{finite} T^* \M_4 $. The parameter space
describing these configurations is just $ \mathbb{R}_+ \times {SO(4,n+4)}/{SO(n+4)}$,
where the $\mathbb{R}_+$ factor corresponds to the gauge freedom contained in the choice of the parameter $c$ in~\eqref{conditions_real_spinors}.

In order to find  the moduli space we still have to remove all gauge redundancies. First of all, the constant factor $c$ is just the rescaling
of the pure spinors, and thus is of no physical significance.
Furthermore, the $SO(3)$ symmetry of the geometric
two-forms is promoted to an $SO(4)$-symmetry between the four real
components of the compatible pure spinors.
Modding out both redundancies  we finally arrive at the moduli space
\begin{equation}
\label{moduli_space_4d_NS}
  \mathcal{M}^\textrm{NS}_{d=6}\ =\ \frac{SO(4,n+4)}{SO(4) \times SO(n+4)} \times \mathbb{R}_+  \ ,
\end{equation}
where we now also included the dilaton via the $\mathbb{R}_+$
factor consistently with $N=4$ supergravity.

The first factor is a quaternionic space which is related to
the fact that the heterotic string compactified on the same $\M_4$
includes this factor as a subspace. The
eight supercharges of the heterotic compactification require that this
scalar field space is quaternionic. Correspondingly two different
superconformal cones can be constructed from \eqref{moduli_space_4d_NS}.
In appendix
\ref{section:Hitchind4} we discuss the $N=2$ hyperk\"ahler cone
\cite{Swann:1990,deWit:1999fp}, and show that the corresponding hyperk\"ahler
potential is determined in terms of the pure spinors and the Mukai
pairing. It turns out that imposing purity and compatibility
conditions are part of taking the hyperk\"ahler $SU(2)$ quotient
\cite{Hitchin:1986ea,Anguelova:2002kd} of flat space.
In appendix~\ref{N4cone} we discuss the flat cone which arises in the
corresponding $N=4$ superconformal theory.

The derivation presented so far did not use the absence of $SU(2)$
doublets. We merely confined our attention to deformations of the
two-forms. However, let us note that
$d\Psi_a$ is an $SU(2)\times SU(2)$ doublet and therefore it
cannot correspond to a deformation
parameter after projecting out the doublets. At best it can be related to the
warp factor, which we so far ignored in the paper. Thus,
in the absence of a warp factor and without any doublets we have
$d\Psi_a=0$ and $Y_4$ has to be a $K3$ manifold. This is consistent
with the moduli space \eqref{moduli_space_4d_NS}, which for $n=16$
coincides with the moduli space of K3 manifolds (modulo the
$\mathbb{R}_+$ factor). We will return to this issue in \cite{LST}.

We want to stress that the arguments given below \eqref{split_KKmodes} can be made for any
vector bundle that consists of only singlets under the structure
group. Since the structure group does not act on the vector bundle, it
must be the trivial bundle and we can give a number of nowhere
vanishing sections that form an orthonormal basis at every
point. Since these sections are globally defined and nowhere vanishing,
they can be associated with objects that define the structure group
and therefore they (or locally rescaled versions thereof) survive the
Kaluza-Klein truncation. Furthermore, we can conclude that any section
that survives the
truncation must be just a linear combination of these
sections. Therefore, the space of light modes resulting from a
Kaluza-Klein truncation on this vector bundle has the same dimension
as the bundle itself.

As an example, let us apply this general theorem to the case of geometric deformations of some $SU(2)$ structure. The moduli space over a point was derived in section \ref{section:SU2d6} to be \eqref{moduli_space_geom_local}.
Again we assume that we can find a consistent Kaluza-Klein truncation to a subspace $\Lambda^2_\textrm{finite} T^* \M_4$ of $\Lambda^2 T^* \M_4$ on which the wedge product is non-degenerate. Then the Hodge star operator $\ast$ gives a decomposition
\begin{equation}
\Lambda^2_\textrm{finite} T^* \M_4 = \Lambda^2_+ T^* \M_4 \oplus \Lambda^2_- T^* \M_4 \ ,
\end{equation}
where $\Lambda^2_+ T^* \M_4 $ consists of $SU(2)$ singlets only and therefore is spanned by the $J_i$. Thus, the signature of $\Lambda^2_\textrm{finite} T^* \M_4 $ is $(3, n+3)$ for some $n$, and the $J_i$ parameterize a space-like three-dimensional subspace in this space. Hence, the geometric moduli space is
\begin{equation}
\label{moduli_space_geom}
  \mathcal{M}^\textrm{geom}_{d=6}\ =\ \frac{SO(3,n+3)}{SO(3) \times SO(n+3)} \times \mathbb{R}_+  \ ,
\end{equation}
where we included the $\mathbb{R}_+$ factor for the volume of $\M_4$.

Note that this argument can already be used to identify the
signature of the second cohomology of $K3$. The space of
two-forms on $K3$ can be decomposed into eigenspaces of the Hodge
operator. Then, the elements of the $+1$ eigenspace are spanned by the
K\"ahler form $J$ and the real and imaginary part of the holomorphic
two-form $\Omega$ defined via \eqref{definition_two-forms}. All other
two-forms in the $+1$ eigenspace must be linear combinations of these
three over each point. If such a two-form is closed, the coefficients
must be constant, and it is just a linear combination of these three
forms in cohomology. Therefore, the $+1$ eigenspace of the Hodge
operator in the second cohomology class is exactly three-dimensional and hence the
signature of $H^2(K3,\mathbb{R})$ is $(3,19)$.

\subsection{Inclusion of the Ramond-Ramond sector}\label{section:RRd6}

Generalized geometry is a natural generalization of $G$-structures
since it covers the complete moduli space of the NS-NS sector of
string theory. This is due to the fact that the group acting on
the generalized tangent bundle coincides with the T-duality group of
string theory.
However, it is also possible to include the RR sector of Type II
string theories by extending the T-duality group to the larger
U-duality group which also includes transformations between the NS-NS
and the RR sector~\cite{Hull:1994ys}. To do so, one extends the
generalized tangent bundle $\mathcal{T}\M$ to the exceptional
generalized tangent bundle~\cite{Hull:2007zu,Pacheco:2008ps}. The spin
group over this bundle is then the U-duality group which coincides
with the (non-compact version of the) exceptional group $E_{d+1}$. It seems natural that the formalism of
pure spinors should extend to the case of exceptional generalized
geometry \cite{Grana:2008tbp}.

Let us examine this for the case of $SU(2)$-structures on $\M_4$. In
this case the U-duality group is $E_{5(5)} = SO(5,5)$ with the T-duality subgroup being $SO(4,4)$. Let us first look at the decomposition of the representations of $SO(5,5)$ in terms of its maximal subgroup $SO(4,4) \times \mathbb{R}_+$. The extra $\mathbb{R}_+$-factor corresponds to shifts of the dilaton.
The vector representation of $SO(5,5)$ decomposes as \cite{slansky}
\begin{equation}
 {\bf 10} \to {\bf 8}^v_{0} \oplus {\bf 1}_{+2} \oplus {\bf 1}_{-2}  \ ,
\label{splitting_vector_rep}
\end{equation}
while for the spinor representation we have
\begin{equation}
 {\bf 16} \to {\bf 8}^c_{+1} \oplus {\bf 8}^s_{-1}  \ .
\label{splitting_fund_rep}
\end{equation}
The subscript denotes the charge of the representation under shifts of
the dilaton.
Finally, the adjoint of $SO(5,5)$ decomposes as
\begin{equation} \label{splitting_adjoint_rep}
 {\bf 45} \to {\bf 28}_0 \oplus {\bf 8}^v_{+2} \oplus {\bf 8}^v_{-2} \oplus {\bf 1}_0  \ .
\end{equation}
Note that because of $SO(4,4)$ triality, the three ${\bf 8}$
representations can be interchanged pairwise, which, however,
has to be done in all three decompositions simultaneously.

Let us now determine the geometric realizations of these
representations. For the T-duality group $SO(4,4)$, the vector
representation ${\bf 8}^v$ is (locally) given in geometrical terms by
$T \M_4 \oplus T^* \M_4$ and analogously the spinor representations ${\bf
8}^s$ and ${\bf 8}^c$ by $\Lambda^\textrm{even}T^* \M_4$ and
$\Lambda^\textrm{odd}T^* \M_4$, respectively.
 However, $SO(4,4)$-triality can interchange the three
eight-dimensional bundles $T \M_4 \oplus T^* \M_4$, $\Lambda^\textrm{even}T^*
\M_4$ and $\Lambda^\textrm{odd}T^* \M_4$.
To assign them to the representations in the right way, we note -- as explained in~\cite{Hull:1994ys} -- that the NS-NS and RR charges
together form the ${\bf 16}$ representation of
$E_{5(5)}$.\footnote{With charges we mean those solutions which are
  point-like particles in six dimensions that are charged under the
  NS and RR vectors.
In the NS sector, the charges are formed by the momentum and winding modes of the fundamental string that are charged under $g_{m\mu}$ and $B_{m\mu}$, respectively, while the RR charges descend from ten-dimensional D-brane solutions.} Since the NS
charges come from  Kaluza-Klein and winding modes, they live in the
geometrical bundle $T \M_4 \oplus T^* \M_4$. The RR charges arise from D-branes
wrapped on internal cycles and therefore sit in
$\Lambda^\textrm{even}T^* \M_4$ for type IIA and $\Lambda^\textrm{odd}T^*\M_4$
for type IIB, respectively.
Hence the
${\bf 16}$ representation corresponds to $T \M_4 \oplus T^* \M_4 \oplus \Lambda^\textrm{even}T^* \M_4 $ in type IIA and to $T \M_4 \oplus T^* \M_4 \oplus \Lambda^\textrm{odd}T^* \M_4 $ in type IIB~\cite{Hull:2007zu}. Consequently, the representation ${\bf 8}^v$ is associated
with $\Lambda^\textrm{odd}T^* \M_4$  in type IIA and
$\Lambda^\textrm{even}T^* \M_4$ in type IIB, respectively.
Altogether we thus have
\begin{equation}\begin{aligned}
 \label{representation_IIA_geom}
 {\bf 10} &= (\Lambda^\textrm{odd}T^* \M_4 )_0 \oplus (\mathbb{R})_{+2}  \oplus (\mathbb{R})_{-2}  \ , \\
 {\bf 16} &= (T \M_4 \oplus T^* \M_4)_{+1}  \oplus (\Lambda^\textrm{even}T^* \M_4)_{-1} \ , \\
{\bf 45} &= (so(T \M_4 \oplus T^* \M_4))_0 \oplus (\Lambda^\textrm{odd}T^*
\M_4)_{+2} \oplus (\Lambda^\textrm{odd}T^* \M_4)_{-2} \oplus
(\mathbb{R})_0\ ,
\end{aligned}\end{equation}
for type IIA, while in type IIB we have
\begin{equation}\begin{aligned}
 \label{representation_IIB_geom}
 {\bf 10} &= (\Lambda^\textrm{even}T^* \M_4)_0 \oplus (\mathbb{R})_{+2}  \oplus (\mathbb{R})_{-2}  \ , \\
 {\bf 16} &= (T \M_4 \oplus T^*\M_4)_{+1} \oplus (\Lambda^\textrm{odd}T^*\M_4)_{-1} \ , \\
 {\bf 45} &= (so(T \M_4 \oplus T^* \M_4))_0 \oplus (\Lambda^\textrm{even}T^* \M_4)_{+2} \oplus (\Lambda^\textrm{even}T^* \M_4)_{-2} \oplus (\mathbb{R})_0 \ .
\end{aligned}\end{equation}
Here $(so(T \M_4 \oplus T^* \M_4))$ denotes the Lie-Algebra of $SO(4,4)$
that acts on  $T \M_4 \oplus T^* \M_4$.
The subscripts give the charges under shifts of the dilaton, which do not have a geometric interpretation.
Note that the bundle $\Lambda^\textrm{even}T^* \M_4$ appears in different representations in \eqref{representation_IIA_geom} and in \eqref{representation_IIB_geom}. This shows that the embedding of the pure $SO(4,4)$ spinors
$\Phi_1, \Phi_2 \in \Lambda^\textrm{even}T^* \M_4$ has to be different for type IIA and type IIB.

In type IIA backgrounds with 16 supercharges the situation is
straightforward. We already argued that in this case we have to
project out all $SU(2)\times SU(2)$ doublets or correspondingly
$\Lambda^\textrm{odd}T^* \M_4$ together with $T \M_4 \oplus T^* \M_4$.
Eq.~\eqref{representation_IIA_geom} then implies
that $SO(5,5)$ is broken to $SO(4,4) \times \mathbb{R}_+$ by the
projection.
This in turn says that all scalar degrees of freedom
coming from the RR-sector are projected out together with
the massive gravitinos. Of course this conclusion is also reached by
direct inspection of the massless type IIA spectrum
discussed in section~\ref{section:spectrumd6}.
This observation also immediately says that the local moduli space is unchanged
and given by \eqref{moduli_space_4d_NS},
i.e.
\begin{equation}
\mathcal{M}^\textrm{IIA}_{d=6}\ =\ \mathcal{M}^\textrm{NS}_{d=6}\ .
\end{equation}

The analogous discussion in  type IIB is slightly more involved. From
\eqref{representation_IIB_geom} we see that neither the additional
generators of $SO(5,5)$ are projected out nor can we embed the pure
spinors into the spinor representation of $SO(5,5)$. However,
from~\eqref{representation_IIB_geom} we see that we can embed the
$SO(4,4)$ spinors into the vector representation of
$SO(5,5)$.
More precisely,
we can either embed the complex  pure spinors $\Phi_1$ and
$\Phi_2$ into complex $SO(5,5)$ vectors or, alternatively, use their
real and imaginary parts denoted by  $\Psi_a$ in the previous section
and embed them into real $SO(5,5)$ vector representations.
We use \eqref{representation_IIB_geom} to decompose the $SO(5,5)$ vector into its components
\begin{equation} \label{decompose_SO55_vector}
 \zeta=(\zeta^+, \zeta^s, \tilde{\zeta}^s ) \ ,
\end{equation}
where $\zeta^+$ lives in $\Lambda^\textrm{even}T^* \M_4$ while $\zeta^s,
\tilde{\zeta}^s$ are the two singlets.
Then the embedding of the four $\Psi_a$ into $SO(5,5)$ vectors $\zeta_a$ is
given by
\begin{equation} \label{SO55_embedding_SO44_spinors}
 \zeta_a = (\Psi_a , 0 , 0) \ , \qquad a=1,\dots ,4 \ .
\end{equation}
This
results in a set of four orthonormal space-like $SO(5,5)$ vectors
$\zeta_a$ which -- after modding out the rotational symmetry between
them -- parameterize the space
\begin{equation}\label{IIBint}
\mathcal{M}_{\zeta_a} = \frac{SO(5,5)}{SO(4) \times SO(1,5)} \ .
\end{equation}

However, this cannot be the correct parameter space yet. As one can
read off from~\eqref{representation_IIB_geom}, the four vectors
$\zeta_a$ are not charged under the dilaton shift.
Thus, the dilaton is not  yet included in the parameter space \eqref{IIBint}.
Reconsidering~\eqref{splitting_vector_rep} shows that the
two singlets are charged under
dilaton shifts, and together form a real $SO(1,1)$ vector. If we
impose a normalization condition on this vector, it parameterizes
$SO(1,1)$ and therefore the dilaton degree of freedom $\phi$.
We can embed this $SO(1,1)$ vector into an $SO(5,5)$ vector
$\zeta_5$ using \eqref{decompose_SO55_vector}, i.e.\
\begin{equation}
\label{SO55_embedding_SO11_vector}
 \zeta_5 = \tfrac{1}{\sqrt{2}} (0 , \e^{\phi} , \e^{-\phi} ) \ .
\end{equation}
We see that the $\zeta_I$, $I=1, \dots, 5$, are all space-like and
satisfy
\begin{equation}
\label{relations_SO55_vectors}
 \langle \zeta_I , \zeta_J \rangle_5 = \delta_{IJ} \ ,
\end{equation}
where we gauge-fixed the parameter $c$ in
\eqref{conditions_real_spinors} to be $1$.

The stabilizer of this set of vectors is naturally given by $SO(5)
\subset SO(5,5)$ which are the rotations in the space perpendicular to
all $\zeta_I$. The generators of this $SO(5)$ are the generators of
the $SU(2) \times SU(2)$ structure which lie in the T-duality
subgroup $SO(4,4)$ together with the generators of the following
transformation
\begin{equation}\label{Btrans}
 \delta_{\ST} \  : \ \zeta \longmapsto \tfrac{1}{\sqrt{2}} \left( (\e^{-\phi} \zeta^s - \e^{\phi} \tilde{\zeta}^s) \ST  ,\ \e^{\phi} \langle \zeta^+ , \ST \rangle_4 ,\ - \e^{-\phi} \langle \zeta^+ , \ST \rangle_4  \right)
\end{equation}
where $\ST \in U_{ {\bf 2},{\bf 2}}$. By a
straightforward calculation one can check that these transformations
are indeed generators of $SO(5,5)$ which stabilize the $\zeta_I$. Since
$\ST$ also transforms as an $SO(4)\equiv
SU(2)\times SU(2)$ vector, the transformations \eqref{Btrans} together
with the generators of $SO(4)$ form the adjoint of $SO(5)$. Therefore, the $\zeta_I$, $I=1, \dots, 5$, obeying \eqref{relations_SO55_vectors}, span the space
${SO(5,5)}/{SO(5)}$.

The embedding of the NS sector into $SO(5,5)$ given in
\eqref{SO55_embedding_SO44_spinors} and
\eqref{SO55_embedding_SO11_vector} is not yet generic. Consider
the following $SO(5,5)$
transformations that are not part of $SO(4,4) \times SO(1,1)$ and that
are not of the form \eqref{Btrans}:\footnote{
Note that the specific parameterization chosen depends on the
dilaton value and the decomposition of $\Lambda^\textrm{even}T^* \M_4$
into the $({\bf 1},{\bf 1})$ and $({\bf 2},{\bf 2})$ representations.}
\begin{equation}\label{C_transforms}\begin{aligned}
 \delta_\A \  : \ \zeta \longmapsto & \tfrac{1}{\sqrt{2}} \left( -(\e^{-\phi} \zeta^s + \e^{\phi} \tilde{\zeta}^s) \A ,\ \e^{\phi} \langle \zeta^+ , \A \rangle_4 ,\ \e^{-\phi} \langle \zeta^+ , \A \rangle_4  \right) \ , \\
 \delta_\G \  : \ \zeta \longmapsto & \tfrac{1}{\sqrt{2}} \left( -(\e^{-\phi} \zeta^s + \e^{\phi} \tilde{\zeta}^s) \G ,\ \e^{\phi} \langle \zeta^+ , \G \rangle_4 ,\ \e^{-\phi} \langle \zeta^+ , \G \rangle_4  \right) \ , \\
 \delta_\B \  : \ \zeta \longmapsto & \tfrac{1}{\sqrt{2}} \left( (\e^{-\phi} \zeta^s - \e^{\phi} \tilde{\zeta}^s) \B ,\ \e^{\phi} \langle \zeta^+ , \B \rangle_4 ,\ -\e^{-\phi} \langle \zeta^+ , \B \rangle_4  \right) \ ,
\end{aligned}\end{equation}
where the superscript of the deformation parameter denotes in which
$SU(2) \times SU(2)$ representation the spinor parameter lies.
These transformations modify the embedding and therefore introduce additional
degrees of freedom. As we will see shortly they correspond to the
RR-scalars. However, let us first observe that
$\delta_\G$ rotates $\zeta_5$ and the $\zeta_a$, $a=1,\dots,4$, into each
other. Therefore, $\G$ parameterizes gauge
degrees of freedom which enhance the $SO(4)$ symmetry between the
$\zeta_a$ to an $SO(5)$ symmetry between the $\zeta_I$.
The remaining parameters $\A$ and $\B$ genuinely modify the embedding and therefore correspond
to additional physical scalar degrees of freedom. They are precisely
the RR-scalars  of type IIB. To see this, we split
the formal sum of RR-fields
\begin{equation}
 C = C_0 + C_2 + C_4 \ \in \Lambda^\textrm{even} T^* \M_4  \ ,
\end{equation}
into $SU(2)\times SU(2)$ representations and associate them with the parameters, i.e.\
\begin{equation}
 C = C^{({\bf 2},{\bf 2})} + C^{({\bf 1},{\bf 1})} \equiv \A + \B \ .
\end{equation}
Note that we only considered the infinitesimal transformations in
\eqref{C_transforms}. Therefore a finite shift in the $C$-fields
corresponds to the exponentiation of \eqref{C_transforms}.

The previous discussion shows that in type IIB apart from the pure
spinors $\Psi_a$ also the dilaton and the RR scalars $C$ are part of
the moduli space and therefore the space given in \eqref{IIBint} has to be modified.
We just argued that the basic objects are five $SO(5,5)$ vectors $\zeta_I$
that satisfy \eqref{relations_SO55_vectors} and which are stabilized
by $SO(5)$.
In addition there is an $SO(5)$ gauge symmetry rotating
the vectors into each other.
Therefore the physical parameter space is
\begin{equation}
{\cal M}_{\zeta_I}\ =\  \frac{SO(5,5)}{SO(5)\times SO(5)}  \ .
\end{equation}

Finally, let us derive the moduli space of type IIB $SU(2)$-structure
backgrounds in six dimensions. As in section~\ref{section:moduliSU2d6}
we assume that a consistent Kaluza-Klein truncation which leaves
the Mukai-pairing non-degenerate exists. By repeating the argument given at
the end of section~\ref{section:moduliSU2d6}
we  conclude that the subspace of positive signature is globally
spanned by the $\zeta_I$ and thus has dimension five.
Therefore, the  $\zeta_I$ span a space-like five-dimensional subspace
in a vector space of signature $(5,n+5)$.
Hence, the moduli space is
\begin{equation}
\label{moduli_space_IIB_6d}
 \mathcal{M}^{\textrm{IIB}}_{d=6}\ =\ \frac{SO(5,n+5)}{SO(5)\times SO(n+5)}  \ .
\end{equation}
This result is consistent with the corresponding chiral supergravity in $d=6$ \cite{Romans:1986er}.

Both $ \mathcal{M}^{\textrm{IIA}}_{d=6}$ and $ \mathcal{M}^{\textrm{IIB}}_{d=6}$ are the
base of a flat cone corresponding to the associated superconformal
supergravity. This is further discussed in appendix~\ref{N4cone}.

This concludes our analysis of type II compactified on an $SU(2)\times SU(2)$ structure manifold $\M_4$ of dimension four.
Let us now turn to compactifications on six-dimensional manifolds $\M_6$.

\section{Compactifications on $SU(2) \times SU(2)$ structure manifolds
  in four space-time dimensions}
\label{section:compd4}
So far we discussed backgrounds with six space-time dimensions and 16
supercharges. Let us now  study backgrounds of the form $M_{3,1} \times
\M_{6}$ but with the same number of supercharges. Thus, in this section
we focus on
six-dimensional manifolds $\M_{6}$ which have  $SU(2) \times SU(2)$
structure.

To be more precise, we consider backgrounds whose Lorentz group is $SO(1,3)\times SO(6)$. The ten-dimensional spinor representation
decomposes accordingly as
\begin{equation}
{\bf 16} \to ({\bf 2}, {\bf 4}) \oplus ({\bf \bar 2},{\bf \bar 4})\ ,
\end{equation}
where the ${\bf 2}$ is a Weyl spinor of $SO(1,3)$ while the ${\bf 4}$ denotes the spinor representation of $SO(6)$.

Backgrounds allowing for sixteen supercharges must admit two or four globally defined spinors, which
corresponds to manifolds $\M_6$ with a reduced structure group $SU(2)$
or $SU(2)\times SU(2)$, respectively. Similar to section~\ref{section:compd6}, we start with a general analysis of the spectrum of type II supergravities in such backgrounds, and then determine their moduli space explicitly by use of the pure spinor formalism of generalized geometry.

\subsection{Field decomposition for $d=4$}\label{section:spectrumd4}
We start by analyzing  the massless type II supergravity fields in
terms of their representations under the four-dimensional Lorentz
group and the structure group and show analogously to
section~\ref{section:spectrumd6} how they assemble in $N =4$
multiplets, in the spirit of~\cite{Grana:2005ny}.

Again, we use the light-cone gauge and only consider
representations of $SO(2)$ instead of the whole $SO(1,3)$ Lorentz
group. Since we treat the case of an $SU(2) \times SU(2)$ structure group, we examine the decomposition of massless type II supergravity fields under the group $SO(2) \times SU(2) \times SU(2)$. For this, let us recall the decomposition of the two Majorana-Weyl representations ${\bf 8}^s$ and ${\bf 8}^c$ and the vector representation ${\bf 8}^v$ under the breaking $SO(8) \to SO(2) \times SO(6) \to SO(2) \times SU(2)$. We get
\begin{equation}\begin{aligned} \label{decomposeSO8toSU2d4}
   {\bf 8}^s & \to {\bf 4}_{\bf + \frac{1}{2}} \oplus {\bf
\bar{4}}_{\bf - \frac{1}{2}} \to 2 \, {\bf 1}_{\bf \pm \frac{1}{2}}
\oplus 2 \, {\bf 2}_{\bf \pm \frac{1}{2}} \ , \\
   {\bf 8}^c & \to {\bf 4}_{\bf - \frac{1}{2}} \oplus {\bf
\bar{4}}_{\bf + \frac{1}{2}} \to 2 \, {\bf 1}_{\bf \pm \frac{1}{2}}
\oplus 2 \, {\bf 2}_{\bf \pm \frac{1}{2}}  \ , \\
   {\bf 8}^v & \to {\bf 1}_{\bf \pm 1} \oplus {\bf 6}_{\bf 0} \to {\bf
1}_{\bf \pm 1}  \oplus 2 \, {\bf 1}_{\bf 0} \oplus  {\bf 2}_{\bf 0} \oplus {\bf \bar{2}}_{\bf 0} \ ,
\end{aligned}\end{equation}
where the subscript denotes the $SO(2)$ charge, i.e.\ the helicity. We
note that both Majorana-Weyl representations decompose in the same way
under the $SU(2)$ structure group. Therefore, we expect type
IIA and type IIB compactifications to give the same theory in $d=4$.

In type IIA the massless fermionic degrees of freedom originate from the $({\bf 8}^s,{\bf 8}^v)$ and $({\bf 8}^v,{\bf 8}^c)$ representation of $SO(8)_L \times SO(8)_R$, while in type IIB they form the $({\bf 8}^s,{\bf 8}^v)$ and $({\bf 8}^v,{\bf 8}^s)$ representation.
Under the decomposition $SO(8)_L \times SO(8)_R \to SO(2) \times SU(2)_L \times SU(2)_R$ they split as
\begin{equation}\begin{aligned}
  ({\bf 8}^s,{\bf 8}^v) & \to 2 ({\bf 1},{\bf 1})_{\bf \pm \frac{3}{2} , \pm \frac{1}{2}} \oplus 4 ({\bf 1},{\bf 1})_{\bf \pm \frac{1}{2}} \oplus 4 ({\bf 1},{\bf 2})_{\bf \pm \frac{1}{2}} \oplus ({\bf 2},{\bf 1})_{\bf \pm \frac{3}{2}, \pm \frac{1}{2}} \oplus 2 ({\bf 2},{\bf 1})_{\bf \pm \frac{1}{2}} \oplus 2 ({\bf 2},{\bf 2})_{\bf \pm \frac{1}{2}} \ , \\
  ({\bf 8}^v,{\bf 8}^c) & \to 2 ({\bf 1},{\bf 1})_{\bf \pm \frac{3}{2}, \pm \frac{1}{2}} \oplus 4 ({\bf 1},{\bf 1})_{\bf \pm \frac{1}{2}} \oplus ({\bf 1},{\bf 2})_{\bf \pm \frac{3}{2}, \pm \frac{1}{2}} \oplus 2 ({\bf 1},{\bf 2})_{\bf \pm \frac{1}{2}} \oplus 4 ({\bf 2},{\bf 1})_{\bf \pm \frac{1}{2}} \oplus 2 ({\bf 2},{\bf 2})_{\bf \pm \frac{1}{2}} \ ,
\end{aligned}\end{equation}
with $({\bf 8}^v,{\bf 8}^s)$ decomposing like the $({\bf 8}^v,{\bf 8}^c)$ due to \eqref{decomposeSO8toSU2d4}.
We see that half of the gravitinos, come in the $({\bf 1},{\bf 1})$ representation while the other half is in the doublet representations $({\bf 1},{\bf 2})$ and $({\bf 2},{\bf 1})$ of $SU(2)_L \times SU(2)_R$. The latter ones again correspond to massive gravitino multiplets that must be projected out to end up with standard $N =4$ supergravity.
After this projection, the fermionic components in the $({\bf 1},{\bf
1})$ become part of the gravity multiplet, while the $({\bf 2},{\bf
2})$ components correspond to the fermionic degrees of freedom in the
$N =4$ vector multiplets.

The  bosonic fields can be treated
in the same way. In the NS-sector we decompose  the $({\bf 8}^v,{\bf
8}^v)$ which corresponds to $E_{MN} = g_{MN} + B_{MN} + \phi \eta_{MN}$ as
\begin{equation}\begin{aligned}
 E_{\mu \nu} & : ({\bf 1},{\bf 1})_{\bf \pm 2}  \oplus ({\bf 1},{\bf 1})_{\bf T} \oplus ({\bf 1},{\bf 1})_{\bf 0} \ , \\
 E_{\mu m} & : 2 ({\bf 1},{\bf 1})_{\bf \pm 1} \oplus 2 ({\bf 1},{\bf 2})_{\bf \pm 1}  \ ,  \\
 E_{m \mu } & : 2 ({\bf 1},{\bf 1})_{\bf \pm 1} \oplus 2 ({\bf 2},{\bf 1})_{\bf \pm 1}  \ ,  \\
 E_{mn} & : 4 ({\bf 1},{\bf 1})_{\bf 0} \oplus 4 ({\bf 1},{\bf 2})_{\bf 0} \oplus 4 ({\bf 2},{\bf 1})_{\bf 0} \oplus 4 ({\bf 2},{\bf 2})_{\bf 0} \ ,
\end{aligned}\end{equation}
where ${\bf T}$ denotes the antisymmetric tensor, and the singlet $({\bf 1},{\bf 1})_{\bf 0}$ corresponds to the four-dimensional dilaton $\phi$.
After the projection we are left with the four-dimensional
metric, the antisymmetric two-tensor, four vectors and five scalars $E_{mn}^{({\bf
1},{\bf 1})}$ and $\phi$ all in the $({\bf 1},{\bf 1})$ representation. In
addition we also have four scalars $E_{mn}^{({\bf 2},{\bf 2})}$ in the
$({\bf 2},{\bf 2})$ representation.
As in $d=6$, they can be associated with deformations of the $SU(2) \times SU(2)$ structure background.

In  the RR-sector we need to consider the decomposition of $({\bf
8}^s,{\bf 8}^c)$ for type IIA and of $({\bf 8}^s,{\bf 8}^s)$ for type
IIB. They both decompose similarly as
\begin{equation} \begin{aligned}
  ({\bf 8}^s,{\bf 8}^c) \to & 4 ({\bf 1},{\bf 1})_{\bf \pm 1} \oplus 8 ({\bf 1},{\bf 1})_{\bf 0} \oplus 2 ({\bf 1},{\bf 2})_{\bf \pm 1} \oplus 4 ({\bf 1},{\bf 2})_{\bf 0} \\
& \oplus 2 ({\bf 2},{\bf 1})_{\bf \pm 1} \oplus 4 ({\bf 2},{\bf 1})_{\bf 0} \oplus ({\bf 2},{\bf 2})_{\bf \pm 1} \oplus 2 ({\bf 2},{\bf 2})_{\bf 0} \ .
\end{aligned}\end{equation}
Projecting out all $SU(2) \times SU(2)$ doublets leaves us with four
vectors and eight scalars in the ${({\bf 1},{\bf 1})}$ representation
and one vector and two scalars in the ${({\bf 2},{\bf 2})}$ representation.

Together these fields can be arranged into an $N=4$ gravity multiplet
plus three $N=4$ vector multiplets. The gravity multiplet contains the
graviton, four gravitini, six vector fields, four Weyl fermions and
two scalars all
in the  ${({\bf 1},{\bf 1})}$ representation.
The vector multiplets each contain one vector, four gaugini and six scalars.
Two of them are also in the  ${({\bf 1},{\bf 1})}$ representation
while the third vector multiplet is in the ${({\bf 2},{\bf 2})}$
representation. We see that, in contrast to $d=6$, not all fields in
the ${({\bf 1},{\bf 1})}$
representation are part of the gravity multiplet but they also form
two vector multiplets. This corresponds to the fact that the
six-dimensional gravity multiplet reduces to a four-dimensional
gravity multiplet plus two vector multiplets.

As we already discussed in section \ref{section:spectrumd6}, these multiplets still consist of ten-dimensional fields that are reordered in such a way that they form $N =4$ multiplets. In the corresponding rewriting of the action only $SO(1,3)\times SO(6)$ symmetry and $N =4$ supersymmetry are manifest. Then we projected out the $SU(2)\times SU(2)$ doublets to achieve a theory that actually allows only for $N =4$ supersymmetry.

\subsection{$\M_6$ with $SU(2)$ structure }\label{section:SU2d4}
A compact six-dimensional manifold $\M_6$  has a structure group $G$ that is a subgroup of $SO(6)$, which acts on spinors via its double cover $SU(4)$.
One globally defined and nowhere vanishing spinor $\eta$ on $\M_6$
defines an $SU(3)$ structure \cite{CS,Gauntlett:2003cy}. This is due to the fact that the
structure group $G$ must admit a singlet  and this requires
$SU(4)$ to be broken to a subgroup contained in $SU(3)$. Therefore,
the existence of $\eta$ implies  $G \subset SU(3)$.

If we assume that there are two such spinors $\eta_{1}$ and $\eta_{2}$
that are orthogonal at each point,  the structure group is broken
further. Now there must be two singlets of the structure group and therefore, $G$ is
contained in $SU(2)$. The spinors $\eta_{1}$ and $\eta_{2}$
each define an $SU(3)$ structure which intersect in an $SU(2)$
structure.
The two distinct $SU(3)$-structures are defined by~\cite{Gauntlett:2003cy,Bovy:2005qq}
\begin{equation}\begin{aligned}
\label{SU3_structure_forms_definition}
 J^{(1)}_{mn} &:= - \iu \bar{\eta}_{1} \gamma_{mn} \eta_{1} \ , \qquad \Omega^{(1)}_{mnp} :=  \iu \bar{\eta}^c_{1} \gamma_{mnp} \eta_{1} \ , \qquad m,n,p=1,\dots,6 \ ,\\
 J^{(2)}_{mn} &:= - \iu \bar{\eta}_{2} \gamma_{mn} \eta_{2} \ , \qquad \Omega^{(2)}_{mnp} :=  \iu \bar{\eta}^c_{2} \gamma_{mnp} \eta_{2} \ ,
\end{aligned}\end{equation}
where our spinor convention are summarized in appendix~\ref{section:conventions}.
With the use of the Fierz identities \eqref{Fierz_identities_6d} one can express them in
terms of an $SU(2)$ structure:
\begin{equation}\begin{aligned}
\label{SU(3)_structure_forms_SU2_splitting}
 J^{(1)} &= J + \tfrac{\iu}{2} K \wedge \bar{K} \ , \quad \Omega^{(1)} = \Omega \wedge K \ ,\\
 J^{(2)} &= J - \tfrac{\iu}{2} K \wedge \bar{K} \ , \quad \Omega^{(2)} = \Omega \wedge \bar{K} \ .
\end{aligned}\end{equation}
The $SU(2)$ structure is defined by the complex one-form
\cite{Gauntlett:2003cy,Bovy:2005qq}
\begin{equation} \label{definition_one-form_K}
 K_m := \bar{\eta}^c_{2} \gamma_m \eta_{1} \ ,
\end{equation}
and the two-forms $J$ and $\Omega$ given by
\begin{equation} \label{definition_two-forms_6d}
 J_{mn} = - \tfrac{1}{2} \iu \left(\bar{\eta}_{1} \gamma_{mn} \eta_{1}
  + \bar{\eta}_{2} \gamma_{mn} \eta_{2}\right) \ ,\qquad
 \Omega_{mn} = \iu \bar{\eta}_{2} \gamma_{mn} \eta_{1} \ .
\end{equation}
$J$ and $\Omega$ fulfill~\eqref{relations_forms}, while $K$ satisfies
\begin{equation}\label{K_compatible}
   K_m K^m =0 \ , \qquad \bar{K}_m K^m = 2 \ ,\qquad
 \iota_K J =0 \ , \qquad \iota_K \Omega = \iota_{\bar{K}} \Omega =0 \ .
\end{equation}
$K$ also specifies an almost product structure
\begin{equation}
\label{almost_product_structure}
 P_m^{\phantom{m}n} := K_m\bar{K}^n + \bar{K}_m K^n - \delta_m^{\phantom{m}n} \ ,
\end{equation}
in that
\begin{equation}
 P_m^{\phantom{m}n} P_n^{\phantom{n}p} = \delta_m^{\phantom{m}p} \ .
\end{equation}
As can be seen from \eqref{SU(3)_structure_forms_SU2_splitting}, this almost product structure is related to the almost complex structures $J^{(i)}$ of the two $SU(3)$ structures by
\begin{equation} \label{almost_product_structure_relation}
 P_m^{\phantom{m}n} = - J^{(1) \phantom{m}p}_{\phantom{(1)}m} \ J^{(2) \phantom{p}n}_{\phantom{(2)}p} \ .
\end{equation}
From \eqref{K_compatible} we can see that $K_m$ and $\bar{K}_m$ are
both eigenvectors of $P_m^{\phantom{m}n}$ with eigenvalue $+1$. The
vectors orthogonal to $K_m, \bar{K}_m$ have eigenvalue $-1$ as can be seen from \eqref{almost_product_structure}. Therefore, $K_m$
and $\bar{K}_m$ even span the $+1$ eigenspace.

In terms of stable forms, an $SU(2)$-structure on $\M_6$ can be defined
by a global complex one-form\footnote{Note that every one-form is stable by definition.} $K$ which breaks the structure group
$SO(6)$ to $SO(4)$  and -- as on $\M_4$ -- by three
stable two-forms $J_i$ that reduce this group further to
$SU(2)$. To assure this breaking of the structure group, all of
these forms have to be compatible with each other in that
they satisfy~\eqref{stable_forms}
and~\eqref{K_compatible}.

Actually, an almost product structure
$P_m^{\phantom{m}n}$ that has a positive eigenspace of dimension two
and a globally defined, nowhere vanishing spinor
$\eta$ are enough to define an $SU(2)$ structure on a manifold of
dimension six. The reason is that $P_m^{\phantom{m}n}$ reduces the structure
group to $SO(2) \times SO(4)$ and therefore also reduces the $SU(3)$ structure
defined by $\eta$ to an $SU(2)$ structure.
This fits nicely with the
fact that the existence of $P_m^{\phantom{m}n}$ is already enough to
assure that the quantities defined
in~\eqref{SU3_structure_forms_definition} are of the
form~\eqref{SU(3)_structure_forms_SU2_splitting} and thereby indeed
define an $SU(2)$ structure on the manifold.
Correspondingly, the two globally defined spinors that reduce the structure group to $SU(2)$ are $\eta$ and $(v_{m} \gamma^{m} \eta^c)$ with $v_m$ is any (real) $+1$-eigenvector of $P$.

Now let us derive the parameter space of $SU(2)$ structures. As before, we have to ensure that we compactify to $N=4$, and therefore project out
all $SU(2)$ doublets, as explained in section \ref{section:spectrumd4}.
As we show in appendix \ref{section:productstructured6} this projection forces the almost product structure $P$ to be rigid. Therefore, the parameter space splits into a part for the two-dimensional identity structure and one for deformations of the $SU(2)$ structure in the four-dimensional subspace. The former is parameterized by $K$, the latter one by $J$ and $\Omega$.
The local parameter space of the $SU(2)$ structure part was already derived in section \ref{section:SU2d6} and is given by \eqref{moduli_space_geom_local}. The identity structure is parameterized by the complex one-form $K$ in a two-dimensional space. Its length corresponds to $K \wedge \bar{K}$ and parameterizes the volume of the two-dimensional space. The group $SU(1,1) \equiv Sl(2,\mathbb{R})$ leaves $K \wedge \bar{K}$ invariant, while it acts freely on $K$. Therefore, its action parameterizes the remaining freedom in choosing $K$.
Since the phase of $K$ is of no relevance, we have to mod out this degree of freedom, and end up with the parameter space ${Sl(2,\mathbb{R})}/{SO(2)}$. Therefore, after including the degree of freedom that correspond to the volume of the four-dimensional subspace, we end up with the parameter space
\begin{equation}
\mathcal{M}_{K, J_i} = \frac{SO(3,3)}{SO(3) \times SO(3)} \times \mathbb{R}_+ \times \frac{Sl(2,\mathbb{R})}{SO(2)} \times \mathbb{R}_+ \ .
\end{equation}
By the argument we presented in section \ref{section:moduliSU2d6}, all subspaces which consist of $SU(2)$ singlets remain unchanged when we perform a Kaluza-Klein truncation. Hence, as in section \ref{section:moduliSU2d6} only the $SU(2)$ structure parameter space can change and the moduli space is
\begin{equation}
\mathcal{M}^\textrm{geom}_{d=4} = \frac{SO(3,3+n)}{SO(3) \times SO(3+n)} \times \mathbb{R}_+ \times \frac{Sl(2,\mathbb{R})}{SO(2)} \times \mathbb{R}_+ \ .
\end{equation}

\subsection{Generalized geometry and $SU(2)\times SU(2)$ structures on $\M_6$}
\label{section:SU2SU2d4}

As in section~\ref{section:SU2SU2d6} we now generalize the previous discussion
to the case of $SU(2)\times SU(2)$ structures using pure $SO(6,6)$ spinors.
In six dimensions, the condition for the existence of pure spinors has been
analyzed in great detail by Hitchin~\cite{Hitchin:2004ut}. The result
is that a normalizable pure $SO(6,6)$ spinor $\Phi$ is in one-to-one
correspondence to
a real stable $SO(6,6)$ spinor and hence looses half of its degrees of freedom.

As we already noted in \eqref{Compatibility_gacs} two normalizable pure $SO(6,6)$ spinors $\Phi^+$ and $\Phi^-$ are
compatible if and only if \cite{Gualtieri:2003dx}
\begin{equation}\label{compatibilty6}
  \operatorname{dim} L_{\Phi^+} \cap L_{\Phi^-} = 3 \ .
\end{equation}
Eq.~\eqref{compatible_pure_spinors_chirality} then implies that $\Phi^\pm$
must be of opposite chirality.
From \eqref{compatible_pure_spinors_chirality} we know that for two spinors $\Phi^\pm$ of opposite chirality the annihilator spaces $L_{\Phi^\pm}$ intersect in an odd-dimensional space.
Thus, Eq.~\eqref{compatibilty6}
can be understood as telling us that $L_{\Phi^+} \cap L_{\Phi^-}$ must be neither one- nor
five-dimensional. From~\eqref{Chevalley_theorem} one can deduce \cite{Grana:2005sn,Grana:2005ny}
\begin{equation}
\label{compatibility_conditions}
 \langle \Phi^+, \Gamma^M \Phi^- \rangle = \langle \bar{\Phi}^+, \Gamma^M \Phi^- \rangle = 0   \ ,
\end{equation}
where $\Gamma^M, M=1,\dots,12,$ are gamma-matrices of $SO(6,6)$. This is a more convenient form for the compatibility condition.
In addition, we can impose the normalization condition
\begin{equation}
\label{compatibility_conditions_normalization}
 \langle \Phi^+, \bar{\Phi}^+ \rangle = \langle \Phi^-, \bar{\Phi}^- \rangle  \ .
\end{equation}

So far, \eqref{compatibility_conditions} and
\eqref{compatibility_conditions_normalization} only define
an $SU(3) \times SU(3)$
structure on $\M_6$. In order to construct an $SU(2) \times
SU(2)$ structure, one has to introduce further objects that are
globally defined and compatible with the spinors
introduced so far. One way to proceed is by mimicking the $SU(2)$
structure construction  and define
two $SU(3) \times SU(3)$ structures with compatibility conditions
imposed such that they intersect in an $SU(2) \times SU(2)$
structure. Each $SU(3) \times SU(3)$ structure already defines a
generalized metric on $T \M_6 \oplus T^*\M_6$, and these two generalized
metrics must coincide in a well-defined string background.
We could express each $SU(3) \times SU(3)$ structure in terms of a
pair of compatible pure spinors $\Phi^\pm_{(i)}$, $i=1,2,$ and formulate
conditions on these pure spinors to ensure that we end up with an
$SU(2) \times SU(2)$ structure. However, it is more elegant to start
with four generalized almost complex structures
$\mathcal{J}_{\Phi^\pm_{(i)}}$
(cf.~section~\ref{section:SU2SU2d6}) and switch to the notion of pure
spinors later.

First of all, to ensure that we are able to diagonalize all four
$\mathcal{J}_{\Phi^\pm_{(i)}}$ simultaneously, we demand that they commute
with each other. This implies that we can decompose $(T\M_6 \oplus
T^*\M_6)_\mathbb{C}$ in the way of
\eqref{decomposition_comp_generalized_acs} for each pair
$\mathcal{J}_{\Phi^\pm_{(i)}}$, $i=1,2$, simultaneously. Each of the four
components in \eqref{decomposition_comp_generalized_acs} is of complex dimension $3$, and one of the $SU(3)$ factors of the
$SU(3) \times SU(3)$ structure group acts on $L_{++}$ and $L_{--}$ while the other one acts
on  $L_{+ -}$ and $L_{-+}$.
We want to define both pairs in such a way that together they break
the structure group to $SU(2) \times SU(2)$. For this, the complex
three-dimensional $SU(3)$ representation $L_{++}$ must split into a two-
and a one-dimensional part. The same must hold for the other components
in~\eqref{decomposition_comp_generalized_acs}.
This corresponds to
\begin{equation}\begin{aligned}
\label{intersection_annihilator_spaces}
 \operatorname{dim}_\mathbb{C} \left( L_{++}^{(1)} \cap L_{--}^{(2)}
 \right) &= 2 =
 \operatorname{dim}_\mathbb{C} \left( L_{--}^{(1)} \cap L_{++}^{(2)}
 \right) \ ,\\
 \operatorname{dim}_\mathbb{C} \left( L_{+-}^{(1)} \cap L_{-+}^{(2)} \right) &=
 2= \operatorname{dim}_\mathbb{C} \left( L_{-+}^{(1)} \cap L_{+-}^{(2)} \right) \ .
\end{aligned}\end{equation}
Note that the first equation implies the second and the third one
implies the fourth.

Analogously to~\eqref{almost_product_structure_relation}, we can also define two
generalized almost product structures by
\begin{equation}
\label{generalized_almost_product_structure}
 \mathcal{P}_\pm = - \mathcal{J}_{\Phi^\pm_{(1)}} \mathcal{J}_{\Phi^\pm_{(2)}} \ ,
\end{equation}
which satisfy
\begin{equation}
 \mathcal{P}_\pm^2 = 1 \ .
\end{equation}
In order to determine the dimension of the eigenspaces let us observe
that \eqref{intersection_annihilator_spaces} implies
\begin{equation} \begin{aligned}
 \operatorname{dim}_\mathbb{C} L_{\Phi^\pm_{(1)}} \cap L_{\bar{\Phi}^\pm_{(2)}} =
 \operatorname{dim}_\mathbb{C} L_{\bar{\Phi}^\pm_{(1)}} \cap L_{\Phi^\pm_{(2)}} = 4 \ ,
\end{aligned} \end{equation}
where we used the relations
\begin{equation}\begin{aligned}
L_{\Phi^+_{(i)}} = L_{++}^{(i)} \oplus L_{+-}^{(i)} \ , \qquad
L_{\Phi^-_{(i)}} = L_{++}^{(i)} \oplus L_{-+}^{(i)}  \ ,\qquad i=1,2 \ .
\end{aligned} \end{equation}
Therefore, both $\mathcal{P}_+$ and $\mathcal{P}_-$ have an
eigenspace  of dimension eight for the eigenvalue $-1$, and
correspondingly, an eigenspace of dimension four for the eigenvalue
$+1$.

Let us now show that the ${\cal P}_\pm$ are identical. To see this we recall
that both  pairs of pure spinor must define the same generalized metric
$\mathcal{G}$ on $T\M_6 \oplus T^*\M_6$, i.e.\
\begin{equation}
 \mathcal{G} = \mathcal{I} \mathcal{J}_{\Phi^+_{(1)}} \mathcal{J}_{\Phi^-_{(1)}} = \mathcal{I} \mathcal{J}_{\Phi^+_{(2)}} \mathcal{J}_{\Phi^-_{(2)}}  \ ,
\end{equation}
where $\mathcal{I}$ is the bilinear form of split signature that is induced by the canonical pairing of tangent and cotangent space.
Therefore, $\mathcal{J}_{\Phi^+_{(2)}}$ is already determined by the other pure spinors, via the equation
\begin{equation}
\label{map_fourth_spinor}
 \mathcal{J}_{\Phi^+_{(2)}} =  - \mathcal{J}_{\Phi^+_{(1)}} \mathcal{J}_{\Phi^-_{(1)}} \mathcal{J}_{\Phi^-_{(2)}} \ .
\end{equation}
This implies
\begin{equation}
 \mathcal{P}^+ = - \mathcal{J}_{\Phi^+_{(1)}} \mathcal{J}_{\Phi^+_{(2)}} = \mathcal{J}_{\Phi^+_{(1)}} \mathcal{J}_{\Phi^+_{(1)}} \mathcal{J}_{\Phi^-_{(1)}} \mathcal{J}_{\Phi^-_{(2)}} = - \mathcal{J}_{\Phi^-_{(1)}} \mathcal{J}_{\Phi^-_{(2)}} = \mathcal{P}^- \ .
\end{equation}
Therefore we see that an $SU(2) \times SU(2)$ structure can alternatively be
defined by a pair of compatible pure spinors $\Phi_{(1)}^+$, $\Phi_{(1)}^-$
and a generalized almost product structure $\mathcal{P}$ which has
the following properties:
\begin{enumerate}
 \item $\mathcal{P}^2 = 1$ .
 \item $\mathcal{P}$ is symmetric with respect to $\mathcal{I}$.
 \item $\mathcal{P}$ commutes with the generalized almost complex structures $\mathcal{J}_{\Phi^\pm_{(1)}}$.
 \item The eigenspaces of $\mathcal{P}$ to the eigenvalues $-1$ and $+1$ are of dimension $8$ and $4$, respectively.
\end{enumerate}
These conditions replace \eqref{intersection_annihilator_spaces}. Note that the
second and third conditions ensure that $\mathcal{P}$ is also
symmetric with respect to the metric defined by
$\mathcal{J}_{\Phi^+_{(1)}}$ and $\mathcal{J}_{\Phi^-_{(1)}}$.

Since the generalized almost complex structures
$\mathcal{J}_{\Phi^-_{(i)}}$ are skew-symmetric with respect to the
canonical pairing and commute with each other, $\mathcal{P}$ is indeed
symmetric with respect to the canonical pairing by construction. This
implies that the canonical pairing is block-diagonal with respect to
the splitting of the bundle induced by $\mathcal{P}$. Therefore,
$\mathcal{P}$ reduces the structure group to $SO(4,4) \times
SO(2,2)$. Since it commutes with $\mathcal{J}_{\Phi^+_{(1)}}$ and
$\mathcal{J}_{\Phi^-_{(1)}}$, both generalized almost complex structures
are similarly block-diagonal with respect to this splitting.

Thus, we conclude that reducing an $SU(3) \times SU(3)$ structure to
some $SU(2) \times SU(2)$ structure corresponds to the fact that one
is able to globally split  $T\M_6 \oplus T^*\M_6$ into
\begin{equation}
\label{splitting_generalized_tangent_bundle}
 T\M_6 \oplus T^*\M_6 = (T_2 \M_6 \oplus T_2^* \M_6) \oplus (T_4 \M_6 \oplus
 T_4^* \M_6) \ ,
\end{equation}
where $T_4 \M_6 \oplus T_4^* \M_6$ is the eight-dimensional vector bundle that is the $-1$ eigenspace of $\mathcal{P}$ at every point, and $T_2 \M_6 \oplus T_2^* \M_6$ is correspondingly the four-dimensional vector bundle that forms the $+1$ eigenspace of $\mathcal{P}$ at every point.\footnote{Properly written, \eqref{splitting_generalized_tangent_bundle}~reads
$\mathcal{T}\M_6 = \mathcal{T}_2 \M_6 \oplus \mathcal{T}_4 \M_6.$}
The pure spinor pair $\Phi^\pm_{(1)}$, corresponding to $\mathcal{J}_{\Phi^\pm_{(1)}}$, defines some $SU(2)\times SU(2)$ structure on $T_4 \M_6 \oplus T_4^* \M_6$ and some trivial structure on $T_2 \M_6 \oplus T_2^* \M_6$, i.e.\ $T_2 \M_6 \oplus T_2^* \M_6$ is the trivial bundle.
On $T_4 \M_6 \oplus T_4^* \M_6$, we can redo the analysis of
section~\ref{section:compd6} since the dimension of $T_4 \M_6 \oplus T_4^* \M_6$ is eight.

Let us make this structure slightly more explicit by considering
the pure spinors $\Phi^\pm$ that correspond to $\mathcal{J}_{\Phi^\pm}$.\footnote{Here and in the following, we drop the subscripts of $\Phi^+_{(1)}$ and $\Phi^-_{(1)}$.}
First, let us fix the generalized almost product structure
$\mathcal{P}$ and investigate the structure of $\Phi^+$ and
$\Phi^-$. Eq.~\eqref{splitting_generalized_tangent_bundle} induces a
splitting of
the $SO(6,6)$ spinor space $\Lambda^\bullet T^* \M_6$, i.e.\
\begin{equation}
 \Lambda^\bullet T^* \M_6 = \Lambda^\bullet T^*_2 \M_6 \wedge \Lambda^\bullet T^*_4 \M_6 \ ,
\end{equation}
where $\Lambda^\bullet T^*_2 \M_6$ and $\Lambda^\bullet T^*_4 \M_6$ are the $SO(2,2)$ and the $SO(4,4)$ spinor bundle over $\M_6$, respectively.
This decomposition carries over to the chiral subbundles
\begin{equation}\begin{aligned}
\label{spinor_bundle_decomposition}
 \Lambda^{\textrm{even}} T^* \M_6\ &=\  \Lambda^{\textrm{even}} T^*_2 \M_6 \wedge  \Lambda^{\textrm{even}} T^*_4 \M_6\ \oplus\  \Lambda^{\textrm{odd}} T^*_2 \M_6 \wedge \Lambda^{\textrm{odd}} T^*_4 \M_6 \ , \\
 \Lambda^{\textrm{odd}} T^* \M_6\ &=\ \Lambda^{\textrm{even}} T^*_2 \M_6 \wedge  \Lambda^{\textrm{odd}} T^*_4 \M_6\ \oplus \ \Lambda^{\textrm{odd}} T^*_2 \M_6 \wedge \Lambda^{\textrm{even}} T^*_4 \M_6  \ .
\end{aligned}
\end{equation}
The direct sum on the right-hand side holds globally,
since, by use of $\mathcal{P}$, we can define chirality operators for
$\Lambda^\bullet T^*_2 \M_6$ and $\Lambda^\bullet T^*_4 \M_6$ independently. In other words, the structure group does not mix the spinor bundles $\Lambda^{\textrm{even}} T^*_4 \M_6$ and $\Lambda^{\textrm{odd}} T^*_4 \M_6$ and the spinor bundles $\Lambda^{\textrm{even}} T^*_2 \M_6 $ and $\Lambda^{\textrm{odd}} T^*_2 \M_6$.

Moreover, since the generalized almost complex structures commute with
$\mathcal{P}$, they split
under~\eqref{splitting_generalized_tangent_bundle} into a generalized
almost complex structure on each component. Correspondingly,
using~\eqref{spinor_bundle_decomposition}, the pure spinors $\Phi^+$
and $\Phi^-$ globally decompose into pure spinors on the spinor
subbundles.
As already argued above, the spinor bundles on the right-hand side of Eq.~\eqref{spinor_bundle_decomposition} do not mix under the action of the structure group, and therefore, the components of $\Phi^+$ and $\Phi^-$ on the subbundles can be analyzed separately. Their components on $\Lambda^\bullet T^*_4 \M_6$ must define an $SU(2) \times SU(2)$ structure.
However, we already discussed the case of an $SU(2) \times SU(2)$ structure group on some vector bundle of dimension eight in section~\ref{section:compd6}.
We know from section~\ref{section:SU2SU2d6} that an $SU(2) \times SU(2)$ structure group on $T_4 \M_6 \oplus T_4^* \M_6$ is already defined by two pure spinors that must have the same chirality. Any additional nowhere vanishing, globally defined pure spinor would break the structure group further. Thus, we can distinguish two cases: Either both spinor components on $\Lambda^\bullet T^*_4 \M_6$ lie in $\Lambda^\textrm{odd} T^*_4 \M_6$ or in $\Lambda^\textrm{even} T^*_4 \M_6$.
Note that in both cases we are left with two pure spinors of opposite chirality in $\Lambda^{\bullet} T^*_2 \M_6$ which define a trivial structure on $T_2 \M_6 \oplus T_2^* \M_6$.

In the first case, both pure spinors
on $T_4 \M_6 \oplus T_4^* \M_6$ are of negative chirality. As we showed in
section~\ref{section:SU2SU2d6}, these two pure spinors define an
$SU(2)\times SU(2)$ structure where the two $SU(2)$ factors have
trivial intersection.
Thus $\M_6$ admits a trivial structure, i.e.\ is parallelizable, and the background has 32 supercharges. As in section~\ref{section:SU2SU2d6}, we do not discuss this case any further.

The second possibility is that both spinor components are of positive
chirality and define -- analogously to section~\ref{section:SU2SU2d6} -- a proper $SU(2)$ structure on the
manifold. Thus, also on $\M_6$ the possibility of an intermediate $SU(2)\times SU(2)$
structures does not exist. Instead one can only have an $SU(2)$
structure or a trivial structure, as we already concluded in our analysis for $\M_4$ in
section~\ref{section:SU2SU2d6}.
In the $SU(2)$ structure case we can write
\begin{equation}\label{SUtwoY6}
 \Phi^+ = \Theta_+ \wedge \Phi_1 \ , \qquad
 \Phi^- = \Theta_- \wedge \Phi_2 \ ,
\end{equation}
where $\Theta_\pm$ are $SO(2,2)$ spinors of opposite chirality and
therefore define a trivial structure on $T_2 \M_6 \oplus T_2^* \M_6$. The
$SO(4,4)$ spinors $\Phi_1$ and $\Phi_2$ are pure and of even
chirality and define the $SU(2)$ structure on $T_4 \M_6
\oplus T_4^* \M_6$. This is precisely the situation we already discussed
in section~\ref{section:SU2d4}. There the $SU(2)$ structure was defined in terms of the two spinors $\eta_i$. The relation between the $\eta_i$ and the $\Phi_\pm$ is analogously to \eqref{general_comp_pure_spinor_form} described by
\begin{equation}\label{SUtwoY6_eta}
 \Phi^+ = \e^{-B} \wedge \tau(\eta_1 \otimes \bar{\eta}_2) \ , \qquad
 \Phi^- = \e^{-B} \wedge \tau(\eta_1 \otimes \bar{\eta}_2^c) \ ,
\end{equation}
where $B$ is the NS two-form, which is not determined by the $\eta_i$. We can insert \eqref{mapping_bispinors_forms} and relate the components in \eqref{SUtwoY6} to the quantities $K, J, B$ and $\Omega$ via
\eqref{definition_one-form_K} and \eqref{definition_two-forms_6d}.
We end up with
\begin{equation}\label{thetaplus}
\Theta_+ = \e^{- B_{(2)} +\tfrac{1}{2} K \wedge \bar{K}} \ , \quad \Theta_- = K \ ,  \quad \Phi_1 = \tfrac{\iu}{4}\, \e^{-B_{(4)}} \wedge \Omega \ , \quad \Phi_2 = \tfrac{1}{4}\, \e^{-B_{(4)} - \iu J } \ ,
\end{equation}
and therefore
\begin{equation}
\label{pure_spinors_6d}
\Phi^+=
\tfrac{\iu}{4}\, \e^{-B_{(2)} + \tfrac{1}{2} K\wedge \bar{K}} \wedge \e^{-B_{(4)}} \wedge \Omega \ , \qquad
\Phi^-=
\tfrac{1}{4}\, K\wedge \e^{-B_{(4)} - \iu J} \ ,
\end{equation}
where we denoted the components of the $B$ field on $\Lambda^2 T^*_2 \M_6$ by $B_{(2)}$ and on $\Lambda^2 T^*_4 \M_6$ by $B_{(4)}$, respectively.
As mentioned earlier, there is some gauge freedom in choosing $\eta_1$ and $\eta_2$ out of the space of $SU(2)$ singlets, which translates into a rotational gauge freedom between $\Phi_1$ and $\Phi_2$. Therefore, it is more convenient in the following not to specify the $\Phi_i$ in terms of $J$ and $\Omega$.

\subsection{Moduli space of $\M_6$} \label{section:moduliSU2d4}
Now we are in the position to discuss the moduli space of $SU(2)$
structures on $\M_6$.
By the splitting we described above, we can do this independently for
the pure spinors on $T_2 \M_6 \oplus T_2^* \M_6$ and the ones on $T_4 \M_6
\oplus T_4^* \M_6$. On the eight-dimensional subspace $T_4 \M_6 \oplus
T_4^* \M_6$ the arguments are the same as on $\M_4$ and
thus $\Phi_1$ and $\Phi_2$ form the moduli
space~\eqref{local_moduli_space_4d}.

Additionally, the $SO(2,2)$ spinors
$\Theta_+$ and $\Theta_-$ each parameterize a moduli space on their own. The Lie algebra splits according to
\begin{equation} \label{SO22_splitting}
so(2,2) = sl(2,\mathbb{R})_T \oplus sl(2,\mathbb{R})_U \ ,
\end{equation}
where the generators of the two sub-algebras in \eqref{SO22_splitting} are given explicitly in \eqref{Sl2R_e} and \eqref{Sl2R_o}.
The first sub-algebra $sl(2,\mathbb{R})_T$ just acts on $\Theta_+$, while
the second $sl(2,\mathbb{R})_U$ acts on $\Theta_-$.
The degrees of freedom in
$\Theta_+$ correspond to a two-form acting on the
negative eigenspace of $\mathcal{P}$ and a form of
degree zero. Together they form an $Sl(2,\mathbb{R})_T$ doublet which is naturally normalized. Furthermore, we have to mod out
the gauge degree of freedom corresponding to the phase of $\Theta_+$.
From \eqref{general_pure_spinor_form} we learn that
the remaining complex
degree of freedom of $\Theta_+$ is given by the volume and the
$B$-field. It spans the parameter space
$Sl(2,\mathbb{R})_T/SO(2)$.
Similarly, $\Theta_-$ can be expanded in the basis of
one-forms on the negative eigenspace of $\mathcal{P}$, which is
two-dimensional, and therefore defines an $Sl(2,\mathbb{R})_U$ doublet analogously to $\Theta_+$, exhibiting the same normalization and gauge degree of freedom. Hence, $\Theta_-$ spans the moduli space ${Sl(2,\mathbb{R})_U}/{SO(2)}$.\footnote{Note that
for $\M_6= K3\times T^2, \Theta_\pm$ parameterize the K\"ahler and
complex structure deformations of the $T^2$, respectively.}

Additionally to the parameter space of the pure spinors, we have the dilaton field $\phi$ in the NS-NS sector, which is complexified by the dualized $B$ field in four dimensions, and forms the moduli space
${Sl(2,\mathbb{R})_S}/{SO(2)}$.
So altogether we have the (local) moduli space
\begin{equation}
\label{local_moduli_space_6d}
   \mathcal{M}_{\Theta_\pm,\Phi_i} = \frac{SO(4,4)}{SO(4)\times SO(4)} \times \frac{Sl(2,\mathbb{R})_S}{SO(2)} \times \frac{Sl(2,\mathbb{R})_T}{SO(2)}  \times \frac{Sl(2,\mathbb{R})_U}{SO(2)}  \ .
\end{equation}

To derive the moduli space, we need to perform a Kaluza-Klein
truncation as we already discussed in detail in
section~\ref{section:moduliSU2d6}. The first factor describes the
deformations of the $SU(2)\times SU(2)$ structure on $T_4 \M_6 \oplus
T_4^* \M_6$, analogously
to~\eqref{local_moduli_space_4d}. Therefore the first factor leads to the same expression for the global moduli space of the truncated theory as in~\eqref{moduli_space_4d_NS}.
To derive the global moduli space for the other three factors, we note that $T_2 \M_6 \oplus T_2^* \M_6$ and $\Lambda^\bullet
T^*_2 \M_6$ consist of $SU(2) \times SU(2)$ singlets only and thus are trivial
bundles. By the general argument on spaces that consist only of structure group singlets which we presented in section~\ref{section:moduliSU2d6}, we conclude that the moduli
spaces of $\Theta_\pm$ are unchanged. Since the dilaton and the
$B$-field transform as scalars under the internal Lorentz group of $\M_6$, their moduli space also
stays the same after Kaluza-Klein truncation.
Therefore, the moduli space of the Kaluza-Klein truncated theory is
\begin{equation}
\label{moduli_space_6d_NS}
   \mathcal{M}^\textrm{NS}_{d=4} = \frac{SO(4,n+4)}{SO(4)\times SO(n+4)} \times \frac{Sl(2,\mathbb{R})_S}{SO(2)} \times \frac{Sl(2,\mathbb{R})_T}{SO(2)}  \times \frac{Sl(2,\mathbb{R})_U}{SO(2)}  \ .
\end{equation}

So far we left $\mathcal{P}$ fixed for the whole discussion. Indeed,
in appendix~\ref{section:productstructured6} we show that projecting out all massive gravitinos
also eliminates all degrees of freedom that correspond to
deformations of $\mathcal{P}$. Therefore, \eqref{moduli_space_6d_NS}
indeed describes the NS moduli space of $SU(2)$ structure
compactifications in $d=4$.

As we already noted at the end of section~\ref{section:moduliSU2d6},
the components of
the exterior derivatives $\diff \Phi_{1,2}$ which are in $\Lambda^\textrm{odd} T_4^* \M_6$, i.e.\ with all legs
in the directions of the
four-dimensional part of the tangent space, are $SU(2)$ doublets and
therefore in the absence of a warp factor are projected out.
This additionally constrains $\M_6$ in that its
four-dimensional component has to be a $K3$ manifold. Or in other words
$\M_6$ has to be a $K3$ fibered over a (twisted) torus.\footnote{We thank
  Danny Mart\'inez-Pedrera for discussions on this issue.}
This situation was already analyzed in
ref.~\cite{ReidEdwards:2008rd} and we will return to it in \cite{LST}.

\subsection{R-R scalars and exceptional generalized
  geometry} \label{section:RRd4}

Now we want to include the RR-fields by extending the
pure spinor formalism again to exceptional generalized geometry.
 We will mainly use the Ansatz proposed
in~\cite{Grana:2008tbp} but applied to the situation of some
$SU(2)$-structure. The main difference will be the existence of a generalized almost
product structure $\mathcal{P}$ which has already been introduced above and the projection
procedure to $N=4$.

The U-duality group in $d=4$ is $E_{7(7)}$ with the
T-duality subgroup being $SO(6,6)$. Let us first recall the
decomposition of the representations of $E_{7(7)}$ in terms of the
maximal subgroup $Sl(2,\mathbb{R})_S \times SO(6,6)$. The factor
$Sl(2,\mathbb{R})_S$ is the S-duality subgroup acting on the
four-dimensional dilaton $\phi$ complexified by the dualized $B$-field.
The fundamental representation of $E_{7(7)}$ decomposes as~\cite{Cremmer:1978ds}
\begin{equation}
\label{E7_decomposition_fundamental}
 {\bf 56} \to ({\bf 2},{\bf 12}) + ({\bf 1},{\bf 32})  \ ,
\end{equation}
while the adjoint of $E_{7(7)}$ decomposes as
\begin{equation}
\label{E7_decomposition_adjoint}
 {\bf 133} \to ({\bf 3},{\bf 1}) + ({\bf 1},{\bf 66}) + ({\bf 2}, {\bf \bar{32}}) \ .
\end{equation}

As in section \ref{section:RRd6}, we can derive the geometrical
realization of these representations by considering the charges in
$d=4$.
It was shown in ref.~\cite{Hull:1994ys} that the electric and magnetic
charges form the $ {\bf 56}$ representation of $E_{7(7)}$.
The $({\bf 2},{\bf 12})$ part in \eqref{E7_decomposition_fundamental}
represents the NS-NS charges, i.e.\ winding and momentum modes as well as NS$5$-branes and KK-monopoles, and thus corresponds to a doublet in
$T \M_6 \oplus T^*\M_6$.\footnote{In contrast to~\cite{Hull:2007zu}, we do not distinguish the bundles $T \M_6 \oplus T^* \M_6$
  and $\Lambda^5 T \M_6 \oplus \Lambda^5 T^* \M_6$ because they are related by a volume form on $\M_6$. Such a volume form we
  already chose to identify the $SO(6,6)$ spinor bundles with $\Lambda^{\textrm{even}}T^* \M_6$ and $\Lambda^{\textrm{odd}}T^*
  \M_6$. Thus, we can identify the bundles $T \M_6 \oplus T^* \M_6$ and $\Lambda^5 T \M_6 \oplus \Lambda^5 T^* \M_6$, and write them as
  a doublet under the S-duality group.
}
The $({\bf 1},{\bf 32})$ represents
the RR charges, which
correspond to ten-dimensional D-brane solutions.
In type IIA, they are elements of  $\Lambda^{\textrm{even}}T^* \M_6$, while in type IIB they live in the bundle $\Lambda^{\textrm{odd}}T^* \M_6$~\cite{Hull:1994ys}. Therefore, \eqref{E7_decomposition_fundamental} is realized geometrically by~\cite{Hull:2007zu}
\begin{equation}
\label{E7_decomposition_fundamental_geometrically_IIA}
{\bf 56}^{\textrm{IIA}} = (T \M_6 \oplus T^* \M_6)_{\bf 2} \oplus (\Lambda^{\textrm{even}}T^* \M_6)_{\bf 1}
\end{equation}
for type IIA, and by
\begin{equation}
\label{E7_decomposition_fundamental_geometrically_IIB}
{\bf 56}^{\textrm{IIB}} = (T \M_6 \oplus T^* \M_6)_{\bf 2} \oplus (\Lambda^{\textrm{odd}}T^* \M_6)_{\bf 1}
\end{equation}
in type IIB. The subscript denotes the representation under the S-duality group $Sl(2,\mathbb{R})_S$, which has no geometric realization.
Correspondingly, the decomposition of the adjoint of the U-duality group is realized geometrically by
\begin{equation}
\label{E7_decomposition_geometrically_IIA}
 {\bf 133}^{\textrm{IIA}} = ( \mathbb{R} )_{\bf 3} \oplus (so(T\M_6 \oplus T^*\M_6))_{\bf 1} \oplus (\Lambda^{\textrm{odd}}T^* \M_6 )_{\bf 2}
\end{equation}
for type IIA, and
\begin{equation}
\label{E7_decomposition_geometrically_IIB}
 {\bf 133}^{\textrm{IIB}} = ( \mathbb{R} )_{\bf 3} \oplus (so(T\M_6 \oplus T^*\M_6))_{\bf 1} \oplus (\Lambda^{\textrm{even}}T^* \M_6 )_{\bf 2}
\end{equation}
for type IIB.
As on four-dimensional manifolds $\M_4$, we see that the spinor
representations of $SO(6,6)$ are related to the RR-fields $C$. In type IIA, the $C$-fields define an $SO(6,6)$ spinor of odd chirality via\footnote{Note that we use the ``democratic'' formulation for the RR-fields, and that we only consider scalar degrees of freedom. Therefore, all legs of the forms in \eqref{C_fields_IIA} and \eqref{C_fields_IIB} are internal.}
\begin{equation}
\label{C_fields_IIA}
 C^{IIA} = C_1 + C_3 + C_5 \ \in \Lambda^\textrm{odd}T^*\M_6 \ ,
\end{equation}
while in type IIB, the spinor is of even chirality and defined by
\begin{equation}
\label{C_fields_IIB}
 C^{IIB} = C_0 + C_2 + C_4 + C_6 \ \in \Lambda^\textrm{even}T^*\M_6 \ .
\end{equation}
This fits nicely with the $SO(6,6)$ spinors appearing
in~\eqref{E7_decomposition_geometrically_IIA}
and~\eqref{E7_decomposition_geometrically_IIB}. However, in both
\eqref{E7_decomposition_geometrically_IIA}
and \eqref{E7_decomposition_geometrically_IIB} there appears a
doublet of
$SO(6,6)$ spinors in the adjoint of $E_{7(7)}$. As we will see below, one linear combination
of these spinors is in the stabilizer of the $SU(2) \times SU(2)$
structure, while the remaining linearly independent linear combination
corresponds to the RR scalar fields.

We have argued in section~\ref{section:spectrumd4} that in
order to compactify on a non-trivial $SU(2)\times SU(2)$ structure,
we have to project out all $SU(2)\times SU(2)$
doublets. We already know from section~\ref{section:SU2SU2d4}
that  only the $SO(2,2)
\times SO(4,4)$ subgroup of $SO(6,6)$ survives this projection. Therefore, the $({\bf 1},{\bf 66})$ component in \eqref{E7_decomposition_adjoint} is projected to the direct sum of the adjoints of $SO(2,2)$ and $SO(4,4)$.
Furthermore, the first component
in \eqref{E7_decomposition_adjoint} consists of $SO(6,6)$
singlets. Therefore, it is also a singlet under $SU(2) \times SU(2)$ and thus
invariant under the projection. Hence, we are left with the last component
in~\eqref{E7_decomposition_adjoint}, which is a doublet of $SO(6,6)$ spinors.

Due to the existence of the generalized almost product structure
$\mathcal{P}$, we can decompose the $SO(6,6)$ spinor bundles as done
in~\eqref{spinor_bundle_decomposition}. Furthermore, as shown in
section~\ref{section:moduliSU2d6} we know that the projection to $N
=4$ removes $\Lambda^\textrm{odd}_4 T^* \M_6$ and we are left  with only
half of the degrees of freedom
\begin{equation}
 \begin{aligned}
 \Lambda^{\textrm{even}} T^* \M_6 &\longrightarrow \Lambda^{\textrm{even}} T^*_2 \M_6 \wedge  \Lambda^{\textrm{even}} T^*_4 \M_6 \ , \\
 \Lambda^{\textrm{odd}} T^* \M_6 &\longrightarrow \Lambda^{\textrm{odd}} T^*_2 \M_6 \wedge \Lambda^{\textrm{even}} T^*_4 \M_6  \ .
\end{aligned}
\end{equation}
Therefore, only part of the U-duality group $E_{7(7)}$ survives this
projection. In type IIA, we end up with the subgroup
$G^{\textrm{IIA}}$ whose adjoint is the subspace
of~\eqref{E7_decomposition_geometrically_IIA} given by
\begin{equation}
\label{SO66_subgroup_IIA}
 g^{\textrm{IIA}} =  ( \mathbb{R} )_{\bf 3} \oplus so(T_2 \M_6 \oplus T^*_2 \M_6)_{\bf 1} \oplus so(T_4 \M_6 \oplus T^*_4 \M_6 )_{\bf 1}
  \oplus  (\Lambda^{\textrm{odd}}T_2^* \M_6 \wedge \Lambda^{\textrm{even}}T_4^* \M_6 )_{\bf 2}  \ .
\end{equation}
In type IIB, we find the subgroup $G^{\textrm{IIB}}$ whose adjoint is
the subspace of~\eqref{E7_decomposition_geometrically_IIB} given by
\begin{equation}
\label{SO66_subgroup_IIB}
 g^{\textrm{IIB}} = ( \mathbb{R} )_{\bf 3} \oplus so(T_2 \M_6 \oplus T^*_2 \M_6)_{\bf 1} \oplus so(T_4 \M_6 \oplus T^*_4 \M_6)_{\bf 1}
  \oplus  (\Lambda^{\textrm{even}}T_2^* \M_6 \wedge \Lambda^{\textrm{even}}T_4^* \M_6)_{\bf 2} \ .
\end{equation}
In appendix~\ref{section:E7_group} we show that both $G^{\textrm{IIA}}$ and $G^{\textrm{IIB}}$ define $SO(6,6) \times Sl(2,\mathbb{R})_{T/U}$ subgroups of $E_{7(7)}$. The $Sl(2,\mathbb{R})_{T/U}$ factor is generated by one of the two sub-algebras in \eqref{SO22_splitting}, depending on whether one considers type IIA or type IIB. $Sl(2,\mathbb{R})_{T}$ acts on the K\"ahler part of the identity structure on $T_2 \M_6 \oplus T^*_2 \M_6$ and forms the extra factor in type IIA, while $Sl(2,\mathbb{R})_{U}$ acts on its complex structure part and forms the extra factor in type IIB.

\subsection{Including the R-R sector}
\label{section:moduliRRd4}

Finally, in this section we determine the moduli space of $SU(2)
\times SU(2)$ structure compactifications of type II theories in
$d=4$. We will mainly consider type IIA but the type
IIB results are easily obtained by changing some chiralities and an
exchange of $sl(2,\mathbb{R})_T$ with $sl(2,\mathbb{R})_U$ in the Lie
algebra $so(2,2)$. (For more details, see appendix \ref{section:E7_group}.)

We start by embedding the pure $SO(6,6)$ spinors into some $E_{7(7)}$ representations. The spinor $\Phi^+$ of positive chirality is embedded into the fundamental representation via
\begin{equation}
\label{E7_embedding_spinor_fund}
\lambda =  (\lambda_i^A, \lambda^+) = ( 0 , \Re ( \Phi^+ )) \
, \quad i=1,2 \ ,
\end{equation}
where we used the decomposition~\eqref{E7_decomposition_fundamental}
and \eqref{decomp56}.\footnote{Note that pure $SO(6,6)$ spinors are completely determined by their real or imaginary part (or any linear combination)~\cite{Hitchin:2004ut}.}
The stabilizer of $\lambda$ is
determined in appendix~\ref{section:E7_stabilizer_fundamental} to be
$SO(4,6) \times SO(2) \subset SO(6,6) \times Sl(2,\mathbb{R})_T$.
Furthermore the phase of $\Phi^+$ is just some gauge degree of freedom. In the $E_{7(7)}$ covariant formalism this gauge freedom manifests itself in the fact that $\lambda$ and
\begin{equation}
\tilde{\lambda} = ( 0 , \Im (\Phi^+ ) )
\end{equation}
describe the same $SU(2) \times SU(2)$ structure. They are related by the generalized almost complex structure $\mathcal{J}_{\Phi^+}$ which embeds into the adjoint of $E_{7(7)}$.
Therefore, after modding out the transformations generated by $\mathcal{J}_{\Phi^+}$, the parameter space for $\lambda$ is
\begin{equation}
 \mathcal{M}_\lambda = \frac{SO(6,6)}{SO(2) \times SO(4,6)} \times \frac{Sl(2,\mathbb{R})_T}{SO(2)} \ .
\end{equation}
In appendix~\ref{section:E7_stabilizer_fundamental} we determine the action of the $E_{7(7)}$ transformations on the embedding \eqref{E7_embedding_spinor_fund}.

The pure $SO(6,6)$ spinor of negative chirality cannot be embedded
into the fundamental of $E_{7(7)}$ but only into its adjoint. However, we see from \eqref{E7_decomposition_adjoint} that
we must embed it as an $Sl(2,\mathbb{R})_S$ doublet. Therefore, we
introduce some complex vector $u^i, {i=1,2},$ which is stable and normalized, i.e.\
\begin{equation} \label{E7_normalization_u}
|u|^2 = u^i \epsilon_{ij} \bar{u}^j = 1 \ .
\end{equation}
$u^i$ describes the complexified dilaton degree of freedom.
Then we embed $\Phi^-$ via
\begin{equation}
\label{E7_embedding_spinor_adjoint}
 \U_1 = (\admui^i_{\phantom{i}j} , \admuA^A_{\phantom{A}B} , \admuspin^{i-}) = ( 0 , 0 ,  \Re ( u^i \Phi^- ) ) \ .
\end{equation}
Note that the overall phase of $u^i$ is just a choice of gauge.
The calculation of the moduli space however is a bit more involved than expected. Naively one would think that analogously to the gauge freedom of $\lambda$ the gauge freedom in $\U_1$ is some phase rotation which relates $\U_1$ to
\begin{equation} \label{SU2embeddingU2}
\U_2 = ( 0 , 0 ,  \Im ( u^i \Phi^- ) ) \ .
\end{equation}
However, these two elements of the adjoint do not commute, and therefore determine a third one which reads
\begin{equation} \label{SU2embeddingU3}
 \U_3 = \tfrac{\iu}{4k} \langle \bar{\Phi}^- , \Phi^- \rangle ( u^i \bar{u}_j + \bar{u}^i u_j ,  \iu |u|^2  (\mathcal{J}_{\Phi^-})^A_{\phantom{A}B} , 0) \ ,
\end{equation}
where $\mathcal{J}_{\Phi^-}$ is the generalized almost complex structure corresponding to $\Phi^-$ and defined by \cite{Grana:2005ny}
\begin{equation}
(\mathcal{J}_{\Phi^-})_{AB} = \iu \frac{\langle \bar{\Phi}^- , \Gamma_{AB} \Phi^- \rangle}{\langle \bar{\Phi}^- , \Phi^- \rangle} \ .
\end{equation}
The normalization $k$ is defined as
\begin{equation}
k = \sqrt{\tfrac{1}{2} |u|^2 \langle \bar{\Phi}^- , \Phi^- \rangle} \ .
\end{equation}
As explained in section \ref{section:SUNSUN}, $\mathcal{J}_{\Phi^-}$ determines $\Phi^-$ up to a phase. As a consequence, $\U_3$ determines $\U_1$ and $\U_2$ up to a rotation between those two. Hence, each $\U_a$ determines the other two. It turns out that the $\U_a$ define a highest weight $SU(2)$ embedding of $u^i$ and the pure spinor $\Phi^-$ in $E_{7(7)}$ \cite{Swann:1990,Grana:2008tbp}. Indeed, the $\U_a$ fulfill the $su(2)$ algebra
\begin{equation}
[\U_a , \U_b] = 2 k \epsilon_{abc} \U_c \ .
\end{equation}
Purity of $\Phi^-$ together with \eqref{E7_normalization_u} is equivalent to the fact that the $\U_a$ indeed form an $su(2)$ algebra. Furthermore, the $\U_a$ share the same stabilizer and make the $SU(2)$ gauge freedom manifest.
By calculating the stabilizer and modding out all gauge degrees of freedom, we end up with the parameter space
\begin{equation}
 \mathcal{M}_\admu = \frac{SO(6,6)}{SO(4) \times SO(2,6)} \ .
\end{equation}
The calculations are a bit lengthy and are presented in
appendix~\ref{section:E7_stabilizer_adjoint}. There, we also give
the action of the additional $E_{7(7)}$ transformations on the $\U_a$.

Finally, we consider the common parameter space of both objects. By imposing \cite{Grana:2008tbp}
\begin{equation}\label{E7_compatibility_conditions}
 \U_a \cdot \lambda = 0 \ ,
\end{equation}
we generalize the usual compatibility conditions \eqref{compatibility_conditions}.
We compute in appendix~\ref{section:E7_stabilizer_both} that the (local) moduli space is
\begin{equation} \label{E7_local_moduli_space}
 \mathcal{M}_{\lambda ,\admu} = \frac{SO(6,6)}{SO(6) \times SO(6)} \times \frac{Sl(2,\mathbb{R})_T}{SO(2)} \ ,
\end{equation}
where we also modded out the additional gauge transformations that arise as symmetries
between the $\Lambda^\textrm{even}_4 T^*\M$ components of $\Phi^+$ and
$\Phi^-$, as we already did in the case of compactification to six
dimensions.
The embeddings \eqref{E7_embedding_spinor_fund} and \eqref{E7_embedding_spinor_adjoint} can be deformed by some $E_{7(7)}$ transformation. This corresponds to the $SO(6,6)$ deformations which we discussed already in the last sections and to additional degrees of freedom that can be identified with the RR scalars.

It is now easy to understand the moduli space. For this we note that the subspace of positive signature in the first factor in \eqref{E7_local_moduli_space} is spanned by $SU(2) \times SU(2)$ singlets. By applying the general argument presented in section~\ref{section:moduliSU2d6}, we know that after the Kaluza-Klein truncation this subspace is still of dimension six.
However, the space of negative signature is spanned by $({\bf 2}, {\bf
2})$ representations of $SU(2) \times SU(2)$. Therefore, its dimension
can be different globally, say $n+6$.
For the second factor, we can argue in the same way as in section~\ref{section:moduliSU2d4} to show that it gives the same moduli space globally.
Therefore, the moduli space is
\begin{equation} \label{E7_moduli_space_IIA}
 \mathcal{M}^{IIA}_{d=4}= \frac{SO(6,6+n)}{SO(6) \times SO(6+n)}  \times \frac{Sl(2,\mathbb{R})_T}{SO(2)} \ .
\end{equation}
In appendix~\ref{section:E7_group} we give a prescription how to obtain the corresponding type IIB moduli space
\begin{equation} \label{E7_moduli_space_IIB}
 \mathcal{M}^{IIB}_{d=4}= \frac{SO(6,6+n)}{SO(6) \times SO(6+n)}  \times \frac{Sl(2,\mathbb{R})_U}{SO(2)} \ .
\end{equation}
Similar to the interpretation of the pure spinor formalism as hyperk\"ahler $SU(2)$ quotient in the case of $d=6$, which we discussed in appendix \ref{section:Hitchind4}, we can interpret the purity and compatibility conditions as leading from the superconformal cone to the moduli space of supergravity. This is discussed in more detail in appendix \ref{N4cone}.

\section{Conclusions}\label{section:conclusions}

In this paper we showed that $SU(2) \times SU(2)$ structures always reduce to either $SU(2)$ or identity structures and we derived the general form of the moduli space for all $SU(2)$ backgrounds in four, five and six space-time dimensions.

For compactifications to six dimensions, we derived the geometric subspace of scalar degrees of freedom to be
\begin{equation}
  \mathcal{M}^\textrm{geom}_{d=6}\ =\ \frac{SO(3,n+3)}{SO(3) \times SO(n+3)} \times \mathbb{R}_+  \ ,
\end{equation}
where $n$ is some integer number and the $\mathbb{R}_+ $ corresponds to the volume factor.
We included the $B$ field into this moduli space by switching to the
framework of generalized geometry. The dilaton gives an extra degree
of freedom that does not mix with the rest of the NS-sector and we found
that the NS moduli space reads
\begin{equation}\label{conclusions_NSd6}
  \mathcal{M}^\textrm{NS}_{d=6}\ =\ \frac{SO(4,n+4)}{SO(4) \times SO(n+4)} \times \mathbb{R}_+  \ .
\end{equation}
We also argued in appendix \ref{section:Hitchind4} that the pure spinor formalism describes the analogue
of a hyperk\"ahler $SU(2)$ quotient in $d=6$ and that the corresponding hyperk\"ahler potential can be naturally expressed in terms of pure spinors.

We further showed that by embedding both pure spinors into representations of
the U-duality group $SO(5,5)$ we are able to incorporate the
RR-scalars into the moduli space.
We showed that in the IIA case no RR-scalars arise and the moduli
space is still given by \eqref{conclusions_NSd6},
i.e.~$\mathcal{M}^\textrm{IIA}_{d=6} = \mathcal{M}^\textrm{NS}_{d=6}$.
For type IIB additional RR-scalars exist and 
they enlarge the moduli space to
\begin{equation}
 \mathcal{M}_{d=6}^{\textrm{IIB}}\ =\ \frac{SO(5,n+5)}{SO(5) \times SO(n+5)}  \ .
\end{equation}

We used the same strategy to determine the moduli spaces for $SU(2)$
structure compactifications to $d=4$. Additionally, we had to
introduce an almost product structure and generalizations thereof to
force the structure group to be $SU(2)$.
We then derived the geometric moduli space of these compactifications to be
\begin{equation}
\mathcal{M}^\textrm{geom}_{d=4} = \frac{SO(3,3+n)}{SO(3) \times SO(3+n)} \times \mathbb{R}_+ \times \frac{Sl(2,\mathbb{R})}{SO(2)} \times \mathbb{R}_+ \ .
\end{equation}
Incorporating the $B$ fields of the NS-sector enlarges the moduli space to
\begin{equation}
   \mathcal{M}^\textrm{NS}_{d=4} = \frac{SO(4,n+4)}{SO(4)\times SO(n+4)} \times \frac{Sl(2,\mathbb{R})_S}{SO(2)} \times \frac{Sl(2,\mathbb{R})_T}{SO(2)}  \times \frac{Sl(2,\mathbb{R})_U}{SO(2)}  \ .
\end{equation}

Within the $E_{7(7)}$ covariant formalism of exceptional generalized
geometry we derived the moduli space to be of the form
\begin{equation}
 \mathcal{M}_{d=4}^\textrm{IIA/IIB}=\frac{SO(6,n+6)}{SO(6) \times SO(n+6)} \times \frac{Sl(2,\mathbb{R})_{T/U}}{SO(2)}  \ ,
\end{equation}
where the extra factor is either $\frac{Sl(2,\mathbb{R})_T}{SO(2)} $
or $\frac{Sl(2,\mathbb{R})_U}{SO(2)}$  depending on whether we
consider type IIA or type IIB.
We also showed that we can interpret the flat superconformal cone over this space in terms of pure spinors and their embeddings into $E_{7(7)}$ representations.
Furthermore, we  explicitly identified how shifts in the fields
correspond to shifts in the moduli space and determined how the
RR-scalars are fibered over the rest of the space. We also briefly
commented on the appearance of string/string/string triality
\cite{Duff:1995sm} in these
compactifications.

In the appendix we performed the analogous analysis for compactifications to $d=5$. We argued that the moduli space is the same for type IIA and IIB due to the absence of chirality in five dimensions and -- after incorporating all scalar degrees of freedom -- reads
\begin{equation}
   \mathcal{M}_{d=5} = \frac{SO(5,n+5)}{SO(5) \times SO(n+5)} \times \mathbb{R}_+ \ .
\end{equation}

Note that this work covers all compactifications of type II theories
that lead to $16$ supercharges in the low energy effective action. The
moduli spaces which we derived here could already have been predicted from the general form of supergravity theories with $16$ supercharges. However, here we showed explicitly how these moduli spaces arise in the compactification procedure. More precisely, we gave an example how the U-duality covariant formalism can be used to determine the moduli space for backgrounds that break part of the supersymmetry.

In the derivation we mainly used algebraic properties of the pure
spinors but did not impose explicitly any differential constraint.
The reason being that the metric on scalar field space
is determined by the algebraic properties while differential
constraints affect the potential of the effective action.
However, by analyzing the light spectrum of the effective supergravity
we argued that we have to project out all $SU(2)$ doublet degrees of
freedom in order to remove the massive gravitino multiplets. Their
presence would alter the standard supergravity with $16$ supercharges
and in particular change the scalar geometry. Since the exterior
derivative of the pure spinors $d\Phi$ is an $SU(2)$ doublet this
effectively also constrains the class of compactification manifolds.
In the absence of a warp factor it implies that $K3$ is the
four-dimensional compactification manifold $Y_4$, while for
the higher-dimensional $Y_{5,6}$ a component of the almost
product structure appears locally as $K3$.

For all spaces the number of light modes is determined by the integer
$n$ with $n=16$ for $K3$. Generically, this number is related to the global twisting of the bundle of forms that are in the $({\bf 2},{\bf 2})$ representation of $SU(2) \times SU(2)$.
All other details of the dimensional reduction are encoded in the
possible gauging of the supergravity action and in the warp factor.
It would be interesting to analyze the possible differential
conditions induced by consistent flux configurations and to classify
the gauged supergravities that can arise as low energy effective
actions. A first step in this direction will be presented in \cite{LST}.
It would also be interesting to study possible non-perturbative
dualities with the heterotic string as was done, for example,
in ref.~\cite{ReidEdwards:2008rd}.
\vfill
\newpage

\section*{Acknowledgments}
This work was partly supported by the German Science Foundation (DFG)
under the Collaborative Research Center (SFB) 676 ``Particles, Strings and the Early Universe'' and by the European Union
6th Framework Program MRTN-CT-503369 ``Quest for unification".

We have greatly benefited from conversations with Vicente Cort\'es, Danny Mart\'inez-Pedrera,
Paul-Andy Nagy, Paul Smyth, Antoine Van Proeyen and  Daniel Waldram.

\vskip 2cm

\appendix
\noindent
{\bf\Large Appendix}

\section{Spinor conventions and Fierz identities}
\label{section:conventions}
In this appendix we collect our spinor conventions and Fierz identities used throughout the
paper.
For $SO(N)$ the gamma-matrices $\gamma_m$ satisfy
\begin{equation} \label{appendix_Clifford_algebra}
 \{ \gamma_m , \gamma_n \} = 2 g_{mn}\ , \quad  m,n=1,\dots,N\ ,
\end{equation}
where $g_{mn}$ is the $SO(N)$ metric, which can be used to raise and
lower the index of the gamma-matrices.
For $N$ even the chirality operator is given by $\gamma_0 =
\iu^{N/2} \frac{1}{N!} \epsilon^{m_1 \dots m_N} \gamma_{m_1 \dots
m_N}$, where $\epsilon$ specifies the orientation of the
manifold.
For antisymmetric products of gamma-matrices we abbreviate
\begin{equation}
 \gamma_{m_1 \dots m_k} = \gamma_{[m_1} \dots \gamma_{m_k]} \ .
\end{equation}
The antisymmetric products of two gamma-matrices $\gamma_{mn}$ fulfill the $SO(N)$ commutation relations and generate the action of $SO(N)$ on spinors $\eta$.

As explained e.g.\ in \cite{VanProeyen:1999ni}, for any $N$ one can define the charge conjugation matrix, which maps a spinor $\eta$ to its charge conjugate $\eta^c$.
For $N = 4k, k\in \mathbb{N}_0$, the charge conjugation matrix commutes with the chirality operator and therefore charge conjugation preserves the chirality of a spinor. For $N= 4k+ 2$, charge conjugation anti-commutes with the chirality operator and thus exchanges the chirality of spinors.

With~\eqref{definition_two-forms} and \eqref{tensor_product_spinors} we can
compute the $SO(4)$ Fierz identities to be
\begin{equation} \begin{aligned}\label{Fierz_identities_4d}
 \eta_\alpha \bar{\eta}^\beta & = \tfrac{1}{2}(P_-)_{\alpha}^{\phantom{\alpha} \beta} - \tfrac{1}{8} \iu J_{mn}\left( \gamma^{mn} P_- \right)_{\alpha}^{\phantom{\alpha} \beta} \ , \qquad \textrm{for} \ \alpha , \beta =1, \dots , 4 \ , \\
 (\eta^c)_\alpha \bar{\eta}^\beta & = \tfrac{1}{8} \iu \bar{\Omega}_{mn} \left( \gamma^{mn} P_- \right)_{\alpha}^{\phantom{\alpha} \beta} \ , \\
  \eta_\alpha (\bar{\eta}^c)^\beta & = \tfrac{1}{8} \iu \Omega_{mn} \left( \gamma^{mn} P_- \right)_{\alpha}^{\phantom{\alpha} \beta}   \ ,
\end{aligned} \end{equation}
where $P_\pm = \tfrac{1}{2} \left( 1 \pm \gamma_0 \right)$ are the
chiral
projection operators.

Analogously, the $SO(6)$ Fierz identities for two spinors $\eta_1$ and
$\eta_2$ can be derived  by using \eqref{tensor_product_spinors} together with the definitions \eqref{SU3_structure_forms_definition}, \eqref{definition_one-form_K} and \eqref{definition_two-forms_6d} to be
\begin{equation} \begin{aligned}\label{Fierz_identities_6d}
(\eta_i)_\alpha (\bar{\eta}_i)^\beta & = \frac{1}{2}(P_-)_{\alpha}^{\phantom{\alpha} \beta} - \frac{1}{4} \iu J^{(i)}_{mn}\left( \gamma^{mn} P_- \right)_{\alpha}^{\phantom{\alpha} \beta} \qquad \textrm{for} \ i=1,2, \ \textrm{and} \ \alpha , \beta =1, \dots , 8 \ ,  \\
(\eta_i^c)_\alpha (\bar{\eta}_i)^\beta & = \frac{1}{24} \iu \bar{\Omega}^{(i)}_{mnp} \left( \gamma^{mnp} P_- \right)_{\alpha}^{\phantom{\alpha} \beta} \ ,  \\
(\eta_i)_\alpha (\bar{\eta}^c_i)^\beta & = \frac{1}{24} \iu \Omega^{(i)}_{mnp} \left( \gamma^{mnp} P_+ \right)_{\alpha}^{\phantom{\alpha} \beta} \ ,  \\
(\eta_1)_\alpha (\bar{\eta}_2^c)^\beta & =  \frac{1}{2} K_m \left( \gamma^{m} P_+ \right)_{\alpha}^{\phantom{\alpha} \beta}
- \frac{1}{8} \iu K_m J_{np} \left( \gamma^{mnp} P_+ \right)_{\alpha}^{\phantom{\alpha} \beta} \ ,  \\
(\eta_2^c)_\alpha (\bar{\eta}_1)^\beta & =  \frac{1}{2} \bar{K}_m \left( \gamma^{m} P_- \right)_{\alpha}^{\phantom{\alpha} \beta} - \frac{1}{8} \iu \bar{K}_m J_{np} \left( \gamma^{mnp} P_- \right)_{\alpha}^{\phantom{\alpha} \beta} \ ,  \\
(\eta_2)_\alpha (\bar{\eta}_1)^\beta & =  \frac{1}{4} \iu \bar{\Omega}_{mn}\left( \gamma^{mn} P_- \right)_{\alpha}^{\phantom{\alpha} \beta} \ , \\
(\eta_1)_\alpha (\bar{\eta}_2)^\beta & =  \frac{1}{4} \iu \Omega_{mn}\left( \gamma^{mn} P_- \right)_{\alpha}^{\phantom{\alpha} \beta} \ .
\end{aligned} \end{equation}
With the help of
\begin{equation}
(\eta_1)_\alpha (\bar{\eta}_1)^\beta = (\eta_1)_\alpha (\bar{\eta}_2^c)^\delta (\eta_2^c)_\delta (\bar{\eta}_1)^\beta \ , \qquad
(\eta_1)_\alpha (\bar{\eta}_1^c)^\beta = (\eta_1)_\alpha (\bar{\eta}_2)^\delta (\eta_2)_\delta (\bar{\eta}^c_1)^\beta \ ,
\end{equation}
etc., we can derive the relations \eqref{K_compatible} and \eqref{SU(3)_structure_forms_SU2_splitting} for the forms involved.

For $SO(N,N)$ spinors, the gamma-matrices $\Gamma_A$ are defined by
\begin{equation}
  \{ \Gamma_A , \Gamma_B \} = 2 \mathcal{I}_{AB}\ , \quad  A,B=1,\dots,2N \ ,
\end{equation}
where $\mathcal{I}$ is the $SO(N,N)$ metric.
We can  
also write the gamma-matrices in terms of raising and
lowering operators $\Gamma_{m^+}$ and $\Gamma_{m^-}$ such that
\begin{equation} \label{appendix_Clifford_split}
\begin{aligned}
\{ \Gamma_{m^+} , \Gamma_{n^+} \} & = 0 \ , \\
\{ \Gamma_{m^-} , \Gamma_{n^-} \} & = 0 \ , \\
\{ \Gamma_{m^+} , \Gamma_{n^-} \} & = 2 g_{mn} \quad \textrm{for all}
\ m,n=1,\dots, N \ ,
\end{aligned} \end{equation}
where $g_{mn}$ is the $SO(N)$ metric.
As for $SO(N)$ gamma-matrices, we abbreviate the antisymmetric product of $SO(N,N)$ gamma-matrices by
\begin{equation}
 \Gamma_{A_1 \dots A_k} = \Gamma_{[A_1} \dots \Gamma_{A_k]} \ .
\end{equation}
The antisymmetric products of two gamma-matrices $\Gamma_{AB}$ fulfill the $SO(N,N)$ commutation relations, and generate the action of $SO(N,N)$ on spinors $\Phi$.
The chirality operator is given by $\Gamma_0 =
\frac{1}{(2N)!} \epsilon^{A_1 \dots A_d} \Gamma_{A_1 \dots
A_d}$, where $\epsilon$ is naturally normalized by
\begin{equation} \label{choice_chirality}
\epsilon^{m_1+m_1- \dots m_N+ m_N-} = 1 \ ,
\end{equation}
if $N$ is even. In this case it defines a canonical choice of positive chirality.

Over a point on a $k$-dimensional manifold $\M_k$ we can define $SO(k,k)$ gamma-matrices via the operators
\begin{equation} \label{Gamma_matrices_geometrical}
\Gamma_{m^+} \equiv \diff x^m \wedge \ , \quad \Gamma_{m^-} \equiv \iota_{x^m} \ ,
\end{equation}
which act on forms and where $\iota_{x_m}$ denotes the insertion of the tangent vector $x_m$. They naturally
fulfill the Clifford algebra \eqref{appendix_Clifford_split} since
\begin{equation} \label{appendix_geometric_clifford_algebra}
[\diff x^m \wedge \ , \ \iota_{x^n}] \omega_p = \delta^m_{\phantom{m}n} \omega_p \ ,
\end{equation}
for any $p$-form $\omega_p$.
Therefore, we can canonically define an $SO(k,k)$ action on the space of forms $\Lambda^\bullet T^* \M_k$. The chirality operator $\Gamma_0$ acts on a $p$-form $\omega_p$ by
\begin{equation}
 \Gamma_0\ \omega_p = (-1)^{p}\ \omega_p \ ,
\end{equation}
hence the Weyl spinor bundle of positive (negative) chirality is given by the bundle of even (odd) forms.
The generators of this $SO(k,k)$ action naturally split into three types according to the number of raising and lowering operators. Transformations of the type $\Gamma_{m+n-}$ preserve the degree of a form and span the algebra of the geometrical group $Gl(k, \mathbb{R})$. The generators $\Gamma_{m+n+}$ and $\Gamma_{m-n-}$ correspond to two-forms and bi-vectors. Hence we conclude
\begin{equation} \label{appendix_decomposition_so_algebra}
 so(k,k) = gl(k, \mathbb{R}) \oplus \Lambda^2 T^* \M \oplus \Lambda^2 T \M \ .
\end{equation}

\section{Superconformal Cones}\label{SCFcone}

In this appendix we briefly discuss the existence of various
higher-dimensional manifolds that arise as cones over the moduli
spaces which we discussed in the main text. These cones appear in the
associated superconformal supergravity where the extra degrees of
freedom correspond to a superconformal compensating multiplet.

In theories with eight supercharges ($N=2$ in $d=4$) the scalars of
the hypermultiplets span a quaternionic manifold which is the base of
a hyperk\"ahler cone (the Swann bundle) \cite{Swann:1990}. The construction of the cone
is well understood in $N=2$ superconformal supergravity \cite{deWit:1999fp}.

The NS-sector of type II compactification is shared by a
compactification of the heterotic string on the same manifold, which
therefore only has half the amount of supersymmetry.
As a consequence the moduli spaces  $\mathcal{M}_{d=6,5,4}^\textrm{NS}$
given in \eqref{moduli_space_4d_NS}, \eqref{moduli_space_5d_NS},
\eqref{moduli_space_6d_NS} all contain the same manifold
$SO(4,n+4)/SO(4)\times SO(4+n)$. The associated hyperk\"ahler cone of
the $N=2$ superconformal supergravity is briefly reviewed in
appendix~\ref{section:Hitchind4}.

The superconformal supergravity for sixteen supercharges ($N=4$ in
$d=4$) is less developed
\cite{de Roo:1984gd}. However, on general grounds one expects the existence of a
flat cone over the $N=4$ scalar manifold. This flat cone is briefly
discussed in appendix~\ref{N4cone}.

\subsection{The hyperk\"ahler cone} \label{section:Hitchind4}

Let us consider the moduli space of $SU(2)$ structures on four-dimensional manifolds $\M_4$, which we argued in section \ref{section:moduliSU2d6} to look like \eqref{moduli_space_4d_NS}.
Here, we want to discuss how specific geometrical constructions over this moduli space arise naturally in the formalism of pure spinors.

Let us start with four real unconstrained but linearly independent spinors $\Psi_a$, $a=1,\dots,4$, living in the space $\Lambda^\textrm{even}_\textrm{finite}T^*\M_4$ of signature $(4,n+4)$. These $\Psi_a$ can be associated with the spinors we defined above \eqref{conditions_real_spinors} but without imposing any purity and compatibility conditions on them.
They parameterize an open subset in the space $\bigoplus_{i=1}^4 \mathbb{R}^{4,n+4}$, which is just a flat cone over \eqref{moduli_space_4d_NS}.
The purity and compatibility conditions~\eqref{conditions_real_spinors} then lead to the deformation space
\begin{equation}
\label{SO_bundle_over_moduli_space}
 \frac{SO(4,n+4)}{SO(n+4)} \times \mathbb{R}_+ \ ,
\end{equation}
where the $\mathbb{R}_+$ factor corresponds to the gauge freedom to choose the parameter $c$.

In order to construct the physical moduli space, we mod out an additional $SO(4)$ between the spinors and the overall $\mathbb{R}_+$ factor. The generators $\left( \mathcal{J}_{ij} \right)^A_{\phantom{A}B}$ for this $SO(4)$ are given by
\begin{equation}
 \mathcal{J}_{ij}^{AB} = \langle \Psi_i , \Gamma^{AB} \Psi_j \rangle \ , \quad \left( \mathcal{J}_{ij} \right)^A_{\phantom{A}B} = \mathcal{J}_{ij}^{AC} \mathcal{I}_{CB} \ .
\end{equation}
They span the Lie algebra $su_1(2) \oplus su_2(2)$, where
\begin{align}
 su_1(2) = & \operatorname{span}(\mathcal{J}_{ij} + \epsilon_{ijkl} \mathcal{J}_{kl}) \ , \\
 su_2(2) = & \operatorname{span}(\mathcal{J}_{ij} - \epsilon_{ijkl} \mathcal{J}_{kl}) \ .
\end{align}
If we mod out only the $SU(2)$ group generated by $su_2(2)$, we end up with the non-Abelian hyperk\"ahler $SU(2)$ quotient of flat space~\cite{Hitchin:1986ea}. This is the hyperk\"ahler cone
\begin{equation}
\label{HKC}
 \frac{SO(4,n+4)}{SU(2) \times SO(n+4)} \times \mathbb{R}_+
\end{equation}
of the quaternionic manifold given by~\eqref{moduli_space_4d_NS}~\cite{Swann:1990,Hitchin:1986ea,Anguelova:2002kd}. By comparing our calculations with those in~\cite{Anguelova:2002kd}, one easily sees that \eqref{comp_pure_spinor_conditions}, \eqref{conditions_two_pure_spinors} and \eqref{normalization_pure_spinors} are just the conditions to set all nine moment maps of the $SU(2)$ generated by $su_2(2)$ to zero.
The three generators of $su_1(2)$ define the hyperk\"ahler structure of~\eqref{HKC}. The hyperk\"ahler potential of the space $\bigoplus_{i=1}^4 \mathbb{R}^{4,n+4}$ we started with is simply
\begin{equation}
 H = \langle \bar{\Phi}_1 , \Phi_1 \rangle  +  \langle \bar{\Phi}_2 , \Phi_2 \rangle = J \wedge J + \Omega \wedge \bar{\Omega} \ .
\end{equation}
By performing the hyperk\"ahler $SU(2)$ quotient, one derives the hyperk\"ahler potential of \eqref{HKC}, which is~\cite{Anguelova:2002kd}
\begin{equation} \label{HKC_Kahlerpotental}
 H = 2 \sqrt{\langle \bar{\Phi}_1 , \Phi_1 \rangle  \langle \bar{\Phi}_2 , \Phi_2 \rangle  - \langle \bar{\Phi}_1 , \Phi_2 \rangle  \langle \bar{\Phi}_2 , \Phi_1 \rangle } \ .
\end{equation}
Here, the second term in the square root is included to make the expression $SO(4)$ invariant.
This hyperk\"ahler potential also determines the metric on the quaternionic manifold \eqref{moduli_space_4d_NS}.

\subsection{Flat cone of $N=4$}\label{N4cone}

For (non-chiral) theories with 16 supercharges in $d$ space-time dimensions the
moduli space is given by $\mathcal{M} = \frac{SO(10-d, n_V)}{SO(10-d)\times
SO(n_V)}\times \mathbb{R}^+$, where $n_V$ counts the number of
vector multiplets \cite{de Roo:1984gd}.  The gravitational multiplet
contains $(10-d)$ vectors while each vector multiplet contains one vector
and $(10-d)$ scalars.  $\mathcal{M}$ is thus spanned by the
$n_V \cdot(10-d)$ scalars of the vector multiplets.

In the superconformal framework one needs to couple $(10-d)$
additional compensating vector multiplets with $(10-d)^2$ additional
scalars. Thus, the scalar manifold in the superconformal supergravity
has dimension $(n_V+ (10-d))\cdot(10-d)$. Since theories with 16
supercharges only have the gauge couplings as free parameters, one
expects the superconformal theory to have a flat scalar manifold. Or
in other words, one expects the existence of a flat
$(n_V+(10-d))\cdot(10-d)$-dimensional cone over $\mathcal{M}$.\footnote{We
  thank Vicente Cort\'es, Paul Smyth and Antoine Van Proeyen  for useful
  discussions on these issues.}
Indeed $\mathcal{M}$ can be viewed as the choice of a space-like
$(10-d)$-dimensional subspace inside a space of signature $(10-d, n_V)$.
If the positions of the vectors which span this subspace are
unconstrained they define a $Gl(10-d)$-bundle over $\mathcal{M}$,
which is an open subset in $\bigoplus_{i=1}^{(10-d)} \mathbb{R}^{(10-d,n_v)} $.

In the construction of the moduli spaces \eqref{moduli_space_4d_NS}, \eqref{moduli_space_5d} and \eqref{E7_moduli_space_IIA} we deal each time with a set of $(10-d)$ space-like vectors in a $(10-d, n_V)$-dimensional space as soon as we project to a theory with sixteen supercharges. These space-like vectors span in each case the superconformal cone over the moduli space. By imposing purity and compatibility conditions for these vectors and modding out the rotational symmetry between them, we project this cone to the underlying moduli space of the corresponding supergravity theory.
Let us discuss this now for each case in more detail.

In appendix \ref{section:Hitchind4} we already explained how the four real vectors $\Psi_a$ defined above \eqref{conditions_real_spinors} parameterize the flat cone over the first factor of \eqref{moduli_space_4d_NS} and how the purity and compatibility conditions together with the removal of gauge freedom project this cone to the moduli space given in \eqref{moduli_space_4d_NS}.
Below \eqref{SO66_local_parameter_space} we explain the analogous procedure for $N=4$ compactifications in the case of $d=4$.
The corresponding discussion for $d=5$ can be found at the end of appendix \ref{section:RRd5}, below \eqref{moduli_space_5d}.

For a chiral theory with 16 supercharges in $6$ space-time dimensions the
moduli space is given by $\mathcal{M} = \frac{SO(5, n_T)}{SO(5)\times
SO(n_T)}$, where $n_T$ counts the number of
tensor multiplets \cite{Romans:1986er}.  The gravitational multiplet
contains $5$ anti-self-dual antisymmetric tensors while each tensor multiplet contains a self-dual antisymmetric tensor and $5$ scalars.  $\mathcal{M}$ is thus spanned by the $ 5 n_T$ scalars of the tensor multiplets.

In the superconformal framework one needs to couple $5$
additional compensating tensor multiplets with $25$ additional
scalars.
As in the non-chiral theories, the corresponding scalar manifold should be a flat cone over $\mathcal{M}$.
Indeed, the $\zeta_I$ introduced in \eqref{SO55_embedding_SO44_spinors} and \eqref{SO55_embedding_SO11_vector} define a $Gl(5)$ bundle over the moduli space \eqref{moduli_space_IIB_6d}.
This bundle is a $5 \cdot (n_T + 5)$ dimensional flat cone over \eqref{moduli_space_IIB_6d}. The condition \eqref{relations_SO55_vectors} and modding out the rotational symmetry between the $\zeta_I$ together remove these compensating multiplets and project the moduli space to \eqref{moduli_space_IIB_6d}.

\section{Generalized almost product structures
} \label{section:productstructured6}
In section~\ref{section:SU2SU2d4} we introduce a generalized almost product structure $\mathcal{P}$ on six-dimensional manifolds. In general, a generalized almost product structure may vary. However, in the following we want to show that for $SU(2) \times SU(2)$ structures, deformations of $\mathcal{P}$ correspond to $SU(2)\times SU(2)$ doublets, which are projected out together with the massive gravitinos to achieve a consistent $N =4$ theory (see section \ref{section:spectrumd4}). In particular, we show the same result also for a conventional almost product structure $P$, which plays a similar role in section \ref{section:SU2d4}.

Let us start with some conventional almost product structure $P$ and afterwards generalize the result to a generalized almost product structure $\mathcal{P}$.
An almost product structure $P$ whose eigenspace with eigenvalue $+1$ has dimension $p$ can be described by some $p$-form $\alpha$ which is locally decomposable into one-forms.
In our case we have to deal with some nowhere-vanishing two-form $\alpha$ which is locally decomposable, i.e.\ whose square is zero
\begin{equation}
 \alpha \wedge \alpha = 0 \ .
\end{equation}
The splitting of the tangent space which defines $P$ is then obtained by
\begin{equation}
 T \M = T_2 \M \oplus T_4 \M \equiv A(T\M ) \oplus \operatorname{kernel}(A) \ ,
\end{equation}
where $A = \frac{1}{2} (1+ P) $ is understood as the linear map that is related to $\alpha$ by some arbitrary but fixed metric.\footnote{We can always use the metric that corresponds to the actual point in moduli space. Therefore, the use of a metric in this argument about allowed deformations does not introduce any further assumption.}

We can now try to understand the moduli space of $\alpha$. Since $\alpha$ must stay locally decomposable under all valid deformations, $\alpha$ can only vary by some two-form that has one leg on each eigenspace of $P$, i.e.\ by some two-form in $T^*_2 \M \wedge T^*_4 \M$. For $SU(2)$ structure compactifications $T^*_4 \M$ consists of $SU(2)$ doublets only, while $T^*_2 \M$ consists of $SU(2)$ singlets. Hence, these two-forms are doublets under the $SU(2)$ structure group.
Since all of these doublets are projected out together with the massive gravitino multiplets, we see that there are no valid global deformations of the almost product structure. This shows that for any $SU(2)$ structure $P$ is fixed.

Now let us consider some generalized almost product structure $\mathcal{P}$.
We consider the standard form for $\mathcal{P}$ that is\footnote{Note that we can always diagonalize $\mathcal{P}$ such that it reduces to some conventional almost product structure. Since the number of deformations is independent of the basis, this argument generalizes to arbitrary $\mathcal{P}$.}
\begin{equation}
 \mathcal{P}_0 = \left(
 \begin{array}{cc}
  P & 0 \\
  0 & P^t
\end{array} \right) \ ,
\end{equation}
where $P$ is some almost product structure on $T\M$ whose eigenspace with eigenvalue $-1$ has dimension four. We want to analyze its orbit under $SO(6,6)$. For this we use the decomposition of the algebra of $SO(6,6)$ given in \eqref{appendix_decomposition_so_algebra}. By $Gl(6)$ transformations we only transform $P$, which is already captured by the fact that we chose $P$ to be a general almost product structure. We showed above that $P$ can only be deformed by $SU(2)$ doublets.
The remaining $SO(6,6)$ transformation of $\mathcal{P}$ are two-form and bi-vector shifts.
Under a two-form shift, $\mathcal{P}$ transforms as
\begin{equation}
\left(
 \begin{array}{cc}
  P & 0 \\
  0 & P^t
\end{array} \right) \to
\left(
 \begin{array}{cc}
  1 & 0 \\
  B & 1
\end{array} \right)
\cdot \left(
 \begin{array}{cc}
  P & 0 \\
  0 & P^t
\end{array} \right)
\cdot \left(
 \begin{array}{cc}
  1 & 0 \\
  -B & 1
\end{array} \right) =
\left(
 \begin{array}{cc}
  P & 0 \\
  (B \cdot P - P^t \cdot B) & P^t
\end{array} \right) \ .
\end{equation}
If $B$ has both legs on $T^*_4 \M$ or both on $T^*_2 \M$, this transformation leaves $\mathcal{P}$ just invariant. The same result holds for bi-vector shifts.
Therefore, the two-form and bi-vector shifts that change $\mathcal{P}$ have one leg on each eigenspace of $P$. However, these transformations are $SU(2)$ doublets. They are projected out together with the massive gravitino multiplets.

Finally, let us remark that the transformations of $SO(6,6)$ that deform $\mathcal{P}$ are those elements that are not in $SO(2,2) \times SO(4,4)$. Thus, projecting out the massive gravitinos corresponds to the projection of $SO(6,6)$ to $SO(2,2) \times SO(4,4)$.

\section{ Calculations in the $E_{7(7)}$ covariant formalism}
\label{section:E7}
The U-duality group is the symmetry group of type II supergravity, which is broken to a discrete subgroup by non-perturbative effects~\cite{Hull:1994ys}. It naturally incorporates the T-duality group and the S-duality group which form the symmetry group of the NS sector. The $E_{7(7)}$ covariant formalism introduced in~\cite{Hull:2007zu} is therefore a natural language to discuss the moduli space of type II theories in $d=4$, and incorporates the pure spinor approach of generalized geometry~\cite{Pacheco:2008ps,Grana:2008tbp}.

In this appendix we present the technical computations which we omitted in section \ref{section:RRd4} and \ref{section:moduliRRd4}.

\subsection{Facts about $E_{7(7)}$} \label{section:E7_group}
 The goal of this appendix is to show that both
\eqref{SO66_subgroup_IIA} and \eqref{SO66_subgroup_IIB} describe
$SO(6,6) \times Sl(2,\mathbb{R})$ subgroups of $E_{7(7)}$ and to
determine the decomposition of its representations in terms of the
projected geometrical group $G^{\textrm{NS}}=Sl(2,\mathbb{R})_S \times
Sl(2,\mathbb{R})_T \times Sl(2,\mathbb{R})_U \times SO(4,4)$, i.e.\
the S-duality group times the projected T-duality group.

Let us first
assemble some  properties of $E_{7(7)}$ which we need later on.
A more complete list of facts can be found in~\cite{Cremmer:1978ds,Grana:2008tbp}.
The fundamental representation of $E_{7(7)}$ decomposes under
$G^{\textrm{TS}}=Sl(2,\mathbb{R})_S \times SO(6,6)$ into the $({\bf 2},{\bf 12})$ and the $({\bf 1},{\bf 32})$ representation (see Eq.~\eqref{E7_decomposition_fundamental}). Thus its elements can be written as
\begin{equation}\label{decomp56}
\lambda = (\lambda_i^A, \lambda^\pm) \ ,\quad  i=1,2, \quad
A=1,\ldots,12\ ,
\end{equation}
where $\lambda_i^A$ is an $Sl(2,\mathbb{R})_S$ doublet of $SO(6,6)$ vectors.
$\lambda^\pm$ denotes a chiral spinor of $SO(6,6)$ where
here and in the following, the upper (lower) chirality sign is
valid for type IIA (IIB).
The adjoint of $E_{7(7)}$ decomposes under $Sl(2,\mathbb{R})_S \times
SO(6,6)$ into the adjoint representations of $Sl(2,\mathbb{R})_S$ and $SO(6,6)$ and into the $({\bf 2}, {\bf \bar{32}})$ representation (cf.~\eqref{E7_decomposition_adjoint}). Therefore
its elements are of the form
\begin{equation}
\label{E7_adjoint_decomposition}
 \adM = (  \admi^i_{\phantom{i} j} , \admA^A_{\phantom{A} B} , \admspin^{i \mp} ) \ ,
\end{equation}
where $\admi^i_{\phantom{i} j}$ is in the adjoint of
$Sl(2,\mathbb{R})_S$, $\admA^A_{\phantom{A} B}$ is in the adjoint of
$SO(6,6)$ and $\admspin^{i \mp}$ is a doublet of $SO(6,6)$ spinors.
The transformation law of $\lambda$ reads
\begin{equation}
\label{variation_fund_E7}
 \delta_\adM \lambda = \big( \admi^i_{\phantom{i} j} \lambda^{jA} + \admA^A_{\phantom{A} B} \lambda^{iB} + \langle \admspin^{i\mp}, \Gamma^A \lambda^{\pm} \rangle \ ,
 \tfrac{1}{4} \admA_{AB} \Gamma^{AB}\lambda^\pm  + \epsilon_{ij} \lambda^{iA} \mathcal{I}_{AB}
\Gamma^B \admspin^{j\mp} \big) \ .
\end{equation}
The transformation law of $M$ is given by the commutator
\begin{equation} \begin{aligned} \label{E7_adjoint_action}
\delta_\adM \adN = [\adM,\adN] = \bigg( & [\admi , \adni]^i_{\phantom{i} j} + \epsilon_{jk}
\left( \langle \admspin^{i\mp} , \adnspin^{k\mp} \rangle - \langle \adnspin^{i\mp} ,
\admspin^{k\mp} \rangle \right) \  ,  \\
 & [\admA,\adnA]^A_{\phantom{A} B} + \epsilon_{ij} \langle \admspin^{i\mp} ,
\Gamma^A_{\phantom{A} B} \adnspin^{j\mp} \rangle \ , \\
 & \admi^i_{\phantom{i} j} \adnspin^{j\mp} -  \adni^i_{\phantom{i} j} \admspin^{j\mp} + \tfrac{1}{4}\ \admA_{AB} \Gamma^{AB} \adnspin^{i\mp} - \tfrac{1}{4}\ \adnA_{AB} \Gamma^{AB} \admspin^{i\mp} \bigg) \ .
\end{aligned} \end{equation}

In the following we will restrict ourself to type IIA. For the
analogous analysis in type IIB one has only to exchange some
chiralities. Specifically one has to change
\begin{equation}\begin{aligned}
 \Phi^+ & \longleftrightarrow \Phi^- \ , \\
 C \in \Lambda^\textrm{odd} T^* \M_6 & \longrightarrow C \in \Lambda^\textrm{even} T^* \M_6 \ , \\
 Sl(2,\mathbb{R})_U & \longleftrightarrow Sl(2,\mathbb{R})_T \ , \\
 \lambda^+ & \longrightarrow \lambda^- \ , \\
 \admspin^{i-}  & \longrightarrow \admspin^{i+}\ , \\
 \admuspin^{i-}  & \longrightarrow \admuspin^{i+} \ .
\end{aligned}\end{equation}

Now consider the decomposition of the adjoint representation of
$SO(6,6)$ under the breaking $SO(6,6) \to SO(4,4) \times SO(2,2)$. In
terms of the Lie algebras,
we can further decompose $so(2,2)$ into $sl(2,\mathbb{R})_T$ and $sl(2,\mathbb{R})_U$, according to \eqref{SO22_splitting}.
The two sub-algebras in terms of their generators are given by
\begin{equation}
 \label{Sl2R_e}
sl(2,\mathbb{R})_T = \operatorname{span} \left( \Gamma^{1+2+}, \Gamma^{1-2-}, \left( \Gamma^{1+1-} + \Gamma^{2+2-} \right) \right)
\end{equation}
and
\begin{equation}
\label{Sl2R_o}
 sl(2,\mathbb{R})_U = \operatorname{span} \left( \Gamma^{1+2-}, \Gamma^{2+1-}, \left( \Gamma^{1+1-} - \Gamma^{2+2-} \right) \right) \ .
\end{equation}
Here, $\Gamma^{i\pm j\pm}$ denotes the anti-symmetric product of two gamma-matrices that fulfill the Clifford algebra \eqref{appendix_Clifford_split} and which can be associated with one-forms and vectors as written in \eqref{Gamma_matrices_geometrical}. The above decomposition can be further understood by consideration of \eqref{appendix_decomposition_so_algebra}. The $sl(2,\mathbb{R})$ sub-algebra of $gl(2,\mathbb{R})$ can be associated with $sl(2,\mathbb{R})_U$, while the generator of the global factor together with the two-form and bi-vector shifts form the $sl(2,\mathbb{R})_T$.
Due to the splitting \eqref{SO22_splitting} we can consider the decomposition of the representations under the breaking $SO(6,6) \to SO(4,4) \times Sl(2,\mathbb{R})_T \times Sl(2,\mathbb{R})_U$.
The vector representation splits as
\begin{equation} \label{Breaking_SO66_S044_SO22_vector}
 {\bf 12} \to ({\bf 1},{\bf 2},{\bf 2}) \oplus ({\bf 8}^v,{\bf 1},{\bf 1}) \ ,
\end{equation}
while the spinor representation decomposes as
\begin{equation} \label{Breaking_SO66_S044_SO22_spinor}
 {\bf 32} \to ({\bf 8}^s,{\bf 2},{\bf 1}) \oplus ({\bf 8}^c,{\bf 1},{\bf 2}) \ .
\end{equation}
The adjoint decomposes as
\begin{equation} \label{Breaking_SO66_S044_SO22}
  {\bf 66} \to ({\bf 28},{\bf 1},{\bf 1}) \oplus ({\bf 1},{\bf 3},{\bf 1}) \oplus ({\bf 1},{\bf 1},{\bf 3}) \oplus ({\bf 8}^v,{\bf 2},{\bf 2}) \ .
\end{equation}

Now we have everything to show that~\eqref{SO66_subgroup_IIA} and~\eqref{SO66_subgroup_IIB} both describe $SO(6,6) \times Sl(2,\mathbb{R})$ subgroups of $E_{7(7)}$. Let us start with the type IIA case, and consider the sub-algebra~\eqref{SO66_subgroup_IIA} of the adjoint representation~\eqref{E7_decomposition_geometrically_IIA} of $E_{7(7)}$. Analogously to~\eqref{SO66_subgroup_IIA}, we write an element in $g^{\textrm{IIA}}$ as
\begin{equation}
\label{decomposition_SO66_subgroup_of_E7}
 \adM = \left( \admi^i_{\phantom{i}j}, \admA^p_{\phantom{p}q}, \admA^a_{\phantom{a}b}, \admspin^{i-}  \right) \ ,
\end{equation}
where we have decomposed  the $\admA^A_{\phantom{A}B}$ of
Eq.~\eqref{E7_adjoint_decomposition} into the $SO(2,2)$
generators $\admA^p_{\phantom{p}q}, p,q=1,\ldots,4$ and  the $SO(4,4)$
generators $\admA^a_{\phantom{a}b}, a,b=1,\ldots,8$. Finally, we note that $\admspin^{i-} \in \Lambda^\textrm{odd} T_2^* \M_6 \wedge \Lambda^\textrm{even} T_4^* \M_6$.

We can determine the commutator of two elements in $g^{\textrm{IIA}}$ by restricting~\eqref{E7_adjoint_action} to elements in $g^{\textrm{IIA}}$. We obtain
\begin{align}
 [\adM,\adN] = \Big( & [\admi , \adni]^i_{\phantom{i} j} +
\epsilon_{jk} \left( \langle \admspin^{i-} , \adnspin^{k-} \rangle + \langle
\admspin^{k-} , \adnspin^{i-} \rangle \right)\ , \
  [\admA,\adnA]^p_{\phantom{p} q} + \epsilon_{ij} \langle \admspin^{i-} , \Gamma^p_{\phantom{p} q} \adnspin^{j-} \rangle\ ,  \nonumber  \\ &
[\admA,\adnA]^a_{\phantom{a} b} + \epsilon_{ij} \langle \admspin^{i-} , \Gamma^a_{\phantom{a} b} \adnspin^{j-} \rangle\ , \
 \admi^i_{\phantom{i} j} \adnspin^{j-} -  \adni^i_{\phantom{i} j} \admspin^{j-} \label{commutator_SO66_subgroup} \\
   & + \tfrac{1}{4}\ \admA_{ab} \Gamma^{ab} \adnspin^{i-} - \tfrac{1}{4}\ \adnA_{ab} \Gamma^{ab} \admspin^{i-}
+ \tfrac{1}{4}\ \admA_{pq} \Gamma^{pq} \adnspin^{i-} - \tfrac{1}{4}\ \adnA_{pq} \Gamma^{pq} \admspin^{i-} \Big)  \, . \nonumber
\end{align}
As explained above, we can further decompose the above expression due to \eqref{SO22_splitting}.
However, because $\admspin^{i-}$ and $\adnspin^{i-}$ live in
$\Lambda^\textrm{odd} T_2^* \M_6 \wedge \Lambda^\textrm{even} T_4^* \M_6$, their
components on $T_2 \M_6 \oplus T^*_2 \M_6$ are of odd chirality. Hence,
they are singlets under $Sl(2,\mathbb{R})_T$ and doublets under
$Sl(2,\mathbb{R})_U$, as can be seen from \eqref{Sl2R_e} and \eqref{Sl2R_o}.
From~\eqref{commutator_SO66_subgroup} we then obtain that elements in
$Sl(2,\mathbb{R})_T$ commute with all other generators of
$G^{\textrm{IIA}}$, i.e.\ the group $G^{\textrm{IIA}}$ is a direct
product with one factor being $Sl(2,\mathbb{R})_T$. Furthermore,
from~\eqref{commutator_SO66_subgroup} we see that $\admspin^{i-}$ is in the
$({\bf 8}, {\bf 2}, {\bf 2})$ of $SO(4,4) \times Sl(2,\mathbb{R})_S
\times Sl(2,\mathbb{R})_U$, and  from~\eqref{Breaking_SO66_S044_SO22}
we see that together with the remaining generators of
$G^{\textrm{IIA}}$ they indeed span the algebra of $SO(6,6)$. Thus we
conclude that $G^{\textrm{IIA}} \equiv SO(6,6) \times
Sl(2,\mathbb{R})_T$.

For type IIB, the arguments are very similar, apart from the fact that
in~\eqref{decomposition_SO66_subgroup_of_E7} the negative chirality
spinor $\admspin^{i-}$ is replaced by  the positive chirality spinor
$\admspin^{i+}$ which lives in $\Lambda^\textrm{even} T_2^* \M_6 \wedge
\Lambda^\textrm{even} T_4^* \M_6$. Therefore, instead of $Sl(2,\mathbb{R})_T$ the factor
$Sl(2,\mathbb{R})_U$ drops out and  we obtain
$G^{\textrm{IIB}} \equiv SO(6,6) \times Sl(2,\mathbb{R})_U$.

Thus, we identified three subgroups of $E_{7(7)}$. First
there is $G^{\textrm{TS}} =Sl(2,\mathbb{R})_S \times SO(6,6)$ which arises
by turning off all RR-degrees of
freedom.
The other two
subgroups arise in $N=4$ compactifications after
projecting out all $SU(2) \times SU(2)$
doublets or equivalently all massive gravitino multiplets. They are
$G^{\textrm{IIA}} = Sl(2,\mathbb{R})_T \times SO(6,6)$  and
$G^{\textrm{IIB}} =Sl(2,\mathbb{R})_U \times SO(6,6)$.
The common subgroup of all three groups is $G^{\textrm{NS}} =
 Sl(2,\mathbb{R})_S \times
Sl(2,\mathbb{R})_T \times Sl(2,\mathbb{R})_U \times SO(4,4)$.

It is instructive to decompose the $E_{7(7)}$ representations with
respect to the breaking $E_{7(7)} \to G^{\textrm{TS}} \to
G^{\textrm{NS}}$. Using
\eqref{E7_decomposition_fundamental} together with
\eqref{Breaking_SO66_S044_SO22_vector} and
\eqref{Breaking_SO66_S044_SO22_spinor},  the fundamental
representation decomposes into
\begin{equation}\label{STU_decomposition_E7_fundamental}
\begin{aligned}
{\bf 56} \to & ({\bf 2},{\bf 12}) + ({\bf 1},{\bf 32})
\\ \to & ({\bf 2},{\bf 2},{\bf 2},{\bf 1}) \oplus ({\bf 2},{\bf 1},{\bf 1},{\bf 8}^v) \oplus ({\bf 1},{\bf 2},{\bf 1},{\bf 8}^s) \oplus ({\bf 1},{\bf 1},{\bf 2},{\bf 8}^c) \ ,
\end{aligned}\end{equation}
and similarly using \eqref{Breaking_SO66_S044_SO22} and
\eqref{Breaking_SO66_S044_SO22_spinor} the adjoint
\eqref{E7_decomposition_adjoint} breaks into
\begin{equation}\label{STU_decomposition_E7_adjoint}
\begin{aligned}
{\bf 133} \to & ({\bf 3},{\bf 1}) + ({\bf 1},{\bf 66}) + ({\bf 2}, {\bf \bar{32}})
\\ \to & ({\bf 3},{\bf 1},{\bf 1},{\bf 1}) \oplus ({\bf 1},{\bf 3},{\bf 1},{\bf 1}) \oplus ({\bf 1},{\bf 1},{\bf 3},{\bf 1}) \oplus ({\bf 1},{\bf 1},{\bf 1},{\bf 28})
\\ & \oplus ({\bf 1},{\bf 2},{\bf 2},{\bf 8}^v) \oplus ({\bf 2},{\bf 1},{\bf 2},{\bf 8}^s) \oplus ({\bf 2},{\bf 2},{\bf 1},{\bf 8}^c) \ .
\end{aligned}\end{equation}
This shows that both \eqref{STU_decomposition_E7_fundamental} and \eqref{STU_decomposition_E7_adjoint} are invariant under the joint action of $SO(4,4)$ triality together with the corresponding $STU$ triality action interchanging the three $Sl(2,\mathbb{R})$ factors.

The part of the triality action interchanging $T$ and $U$ just
corresponds to a generalization of T-duality between type IIA and type
IIB. However, the interchange of $S$ with $T$ or $U$ seems to lead to
some theory where the RR-sector is absent but no
$SU(2)\times SU(2)$ doublets have been projected out. For
type II compactifications on $K3 \times T^2$, this corresponds to the
non-perturbative duality with the heterotic string compactified  on
$T^6$. In this case, the  triality action is known as
string/string/string triality~\cite{Duff:1995sm}.
The heterotic duals of some specific $SU(2)$ structure compactifications have been analyzed in \cite{ReidEdwards:2008rd}.

\subsection{Calculating stabilizers and orbits} \label{section:E7_stabilizers}
Let us now compute the stabilizer of the pure $SO(6,6)$ spinors
embedded in the fundamental and adjoint representation of
$E_{7(7)}$. We consider type IIA compactifications in some detail and
only give the results for type IIB which can easily be obtained by
some chirality changes, as explained in appendix
\ref{section:E7_group}.
Furthermore we
use the reduction $E_{7(7)}\to G^{\textrm{IIA}} =
Sl(2,\mathbb{R})_T \times SO(6,6)$ and
compute the intersection of the stabilizer with this subgroup. Note
that we only have to consider the real or the imaginary part of some
pure $SO(6,6)$ spinor because they are related via the Hitchin
functional~\cite{Grana:2005ny}.

\subsubsection{Embedding into the fundamental}\label{section:E7_stabilizer_fundamental}
Let us start with the pure spinor $\Phi^+$ of positive chirality. According to~\eqref{E7_decomposition_fundamental}, we can embed its real part $\Re (\Phi^+ )$ into the fundamental representation as done in~\eqref{E7_embedding_spinor_fund}.
We can compute the stabilizer using~\eqref{variation_fund_E7}  and
the splitting~\eqref{SO66_subgroup_IIA}. We know that $\Re \Phi^+$ is
a singlet under $Sl(2,\mathbb{R})_S$ and, since $\Re \Phi^+ \in
\Lambda^\textrm{even} T^*_2 \M_6 \wedge \Lambda^\textrm{even} T^*_4 \M_6$, also a
singlet under $Sl(2,\mathbb{R})_U$. Therefore both groups are in the
stabilizer. Due to the same reason, $\Re \Phi^+$ transforms under
$Sl(2,\mathbb{R})_T$ non-trivially such that no element in this group
stabilizes $\Re \Phi^+$. Furthermore, $\Phi_1$ defined in
\eqref{SUtwoY6} is pure and is therefore stabilized by
$SU(2,2)_{\Phi_1} \equiv SO(2,4) \subset SO(4,4)$. Therefore, $\Re
\Phi^+$ is also stabilized by this subgroup.

It remains to understand which elements $\admspin^{i-}\in
\Lambda^\textrm{odd} T^*_2 \M_6 \wedge \Lambda^\textrm{even} T^*_4 \M_6$ stabilize
$\lambda$. We use the parameterization \eqref{SUtwoY6} and
\eqref{thetaplus}. Thus, $\Re \Phi^+$ can be expressed as\footnote{For
simplicity we again switched off the $B$-field which, however, does not change the computation.}
\begin{align}
\label{parametrisation_Re_Phi_plus}
 \Re \Phi^+ = 1 \wedge \Re \Phi_1 - \Re K \wedge \Im K \wedge \Im \Phi_1 \ ,
\end{align}
where $K$ is the holomorphic one-form on $\Lambda^\textrm{odd} T^*_2 \M_6$ defined in \eqref{definition_one-form_K} and related to the pure spinors in \eqref{thetaplus}.
The transformation parameters can be analogously written as
\begin{equation}
\label{parametrisation_C_transforms}
 \admspin^{i-} = \Re K \wedge \alpha^{i1} + \Im K \wedge \alpha^{i2}, \quad i=1,2 \ ,
\end{equation}
where  $\alpha^{ij}$ are spinors in $\Lambda^\textrm{even} T^*_4 \M_6$

From~\eqref{variation_fund_E7} we now see that those $\admspin^{i-}$ are in the stabilizer which fulfill
\begin{equation}
 \langle \admspin^{i-} , \Gamma^A \Re \Phi^+ \rangle = 0 \qquad \textrm{for all} \quad A=1 , \dots , 12 \ .
\end{equation}
These equations are only non-trivial for the gamma-matrices
that act on $\Lambda^\textrm{odd} T^*_2 \M_6$. Thus, we arrive at
\begin{equation}
\label{stabilizer_Phi_plus}
 \langle \alpha^{ij} , \Re \Phi_1 \rangle = \langle \alpha^{ij} , \Im \Phi_1 \rangle = 0 \ .
\end{equation}
This just eliminates all $SU(2,2)_{\Phi_1}$
singlets. Therefore, the elements in $\Lambda^\textrm{odd} T^*_2 \M_6 \wedge
\Lambda^\textrm{even} T^*_4 \M_6$ which are in the stabilizer just form the
representation $({\bf 2}, {\bf 2}, {\bf 6})$ of the group
$Sl(2,\mathbb{R})_S \times Sl(2,\mathbb{R})_U \times
SU(2,2)_{\Phi_1}$. This combines with the adjoint of
$Sl(2,\mathbb{R})_S \times Sl(2,\mathbb{R})_U \times SU(2,2)_{\Phi_1}$
into the adjoint of $SO(4,6)$. There is one further $SO(2)$ factor in
the stabilizer which corresponds to a simultaneous rotation of the
doublets $(\Re \Phi_1, \Im \Phi_1)$ and $(1 , \Re K\wedge \Im K)$ such
that the overall phase factor of $\Phi^+$ stays constant. Therefore,
the whole stabilizer is $SO(4,6) \times SO(2)$.
This can also be understood from the fact that $\Re \Phi^+$
transforms non-trivially under $Sl(2,\mathbb{R})_T$ and therefore
$\lambda$ must be in the $({\bf 2},{\bf 12 }) $ of
$G^{\textrm{IIA}}$. This just gives a doublet of space-like $SO(6,6)$
spinors, which is indeed stabilized by $SO(4,6) \times SO(2)$.

In order to obtain the physical parameter space we need to mod out by
one further gauge degree of freedom which corresponds to the phase of $\Phi^+$, as explained in section \ref{section:moduliRRd4}.
We thus finally get
\begin{equation}
\mathcal{M}_\lambda = \frac{SO(6,6)}{SO(2) \times SO(4,6)} \times \frac{Sl(2,\mathbb{R})_T}{SO(2)} \ .
\end{equation}
The first factor just describes a space-like two-dimensional subspace
in a space of signature $(6,6)$.

Before we move on to the embedding of the other spinor, we want to
know what is the most general form for the embedding of $\lambda$ given in \eqref{E7_embedding_spinor_fund} if it is deformed
by some general $G^{\textrm{IIA}}$ deformation. The
only deformations that are not already caught by the Ansatz
in~\eqref{E7_embedding_spinor_fund} are the elements in the adjoint of
$G^{\textrm{IIA}}$ that are represented by
$\Lambda^\textrm{odd} T^*_2 \M_6 \wedge \Lambda^\textrm{even} T^*_4 \M_6$.
We just showed that the part of $\Lambda^\textrm{odd} T^*_2 \M_6 \wedge
\Lambda^\textrm{even} T^*_4 \M_6$ which is in the $({\bf 2}, {\bf 2}, {\bf
6})$ representation of the group $Sl(2,\mathbb{R})_S\times
Sl(2,\mathbb{R})_U\times SU(2,2)_{\Phi_1}$ stabilizes $\lambda$, and
therefore cannot deform it. Thus, only  the deformations in the $({\bf 2}, {\bf 2}, {\bf 1})$ representation, i.e.\ $SU(2,2)_{\Phi_1}$ singlets, may deform $\lambda$. These are spinors of the form
\begin{equation}
 m^{i-} =  (c^{i2+} \Re K - c^{i1+} \Im K)\wedge \Re \Phi_1 - (c^{i1-}
\Re K +c^{i2-} \Im K )
\wedge \Im \Phi_1  \ ,
\label{RR_deformations_fundamental_singlet}
\end{equation}
where the $c^{i a}, a=1+,1-,2+,2-,$ are real. The action of such
elements $\admspin^{i-}$ in the adjoint on $\lambda$ reads
\begin{equation}
\label{variation_RR_deformations_fundamental_singlet}
 \delta_{\admspin^{i-}} \lambda = \left( c^{i a} , 0 \right) \ ,
\end{equation}
where we
used~\eqref{E7_embedding_spinor_fund},~\eqref{parametrisation_Re_Phi_plus}
and~\eqref{variation_fund_E7}. Note that $c^{i a}$ is an $SO(2,2)$
vector which naturally embeds into an $SO(6,6)$ vector.

By exponentiation of the $m^{i-}$ in
\eqref{RR_deformations_fundamental_singlet} the most general form of
the embedding \eqref{E7_embedding_spinor_fund} can be obtained. These
additional degrees of freedom correspond to the RR-fields and to gauge
degrees of freedom. We discuss their physical significance in appendix
\ref{section:E7_stabilizer_both} when we consider both the embeddings
of $\Phi^+$ and $\Phi^-$.

\subsubsection{Embedding into the adjoint}\label{section:E7_stabilizer_adjoint}

Let us now come to the embedding of the other pure spinor
$\Phi^-$. Due to its negative chirality $\Phi^-$
cannot be embedded into the fundamental representation of $E_{7(7)}$
but only into its adjoint via
$\U_1 =( 0 , 0 ,  \Re( u^i \Phi^- ) )$
as we argued in \eqref{E7_embedding_spinor_adjoint}.\footnote{Note that we distinguish in our notation formally between $\adM$ and $\U_1$ which are both elements in the adjoint of $E_{7(7)}$. The former one corresponds to an unspecified generator of the group action. The latter one parameterizes the $SU(2) \times SU(2)$ structure and is defined by the embedding of the pure spinor $\Phi^-$.}
Now let us analyze the intersection of its stabilizer with
$G^{\textrm{IIA}}$. Since $E_{7(7)}$ acts on $\U_1$ via the Lie
bracket, we can use~\eqref{commutator_SO66_subgroup} to determine its
stabilizer. With help of \eqref{SUtwoY6} and
\eqref{thetaplus} we can write it as $\U_1 = (0,0, \Re( u^i K \wedge \Phi_2) )$. One can easily see that $Sl(2,\mathbb{R})_S$ acts freely
on $\U_1$ since it acts freely on $u^i$. Since $\Re \Phi^- ,\ \Im \Phi^-
\ \in \Lambda^\textrm{odd}_2 T^* \M_6 \wedge \Lambda^\textrm{even}_4 T^* \M_6$, we
also see that  $Sl(2,\mathbb{R})_U$ acts freely on $\U_1$. However,
there are two additional phase rotations inside $Sl(2,\mathbb{R})_S \times
Sl(2,\mathbb{R})_U \times SO(4,4)$ that leave $\U_1$ invariant. The first rotates $u^i$ and
the spinor component $K \in \Lambda^\textrm{odd}_2 T^* \M_6$ of $\Phi^-$
with opposite phases. We call the generator of this transformation
$R_{(+1,-1,0)}$. The second generator, $R_{(+1,+1,-2)}$,
rotates
 $u^i$ and $K$ by the same phase and $\Phi_2$ oppositely.

$Sl(2,\mathbb{R})_T$ acts trivially and therefore is part of the
stabilizer of $\U_1$. Since $\Phi_2$ is pure, we know that it is also
stabilized by an $SU(2,2)_{\Phi_2} \subset SO(4,4)$ subgroup, and so is
$\Phi^-$.
It remains to determine the elements $\admspin^{i-}$ of
$\Lambda^\textrm{odd} T^*_2 \M_6 \wedge \Lambda^\textrm{even} T^*_4 \M_6$ which
leave $\U_1$ invariant.
From~\eqref{commutator_SO66_subgroup} we see that this leads to the following equations
\begin{equation}\begin{aligned}
\label{stabilizer_adjoint_equ1}
 \langle \admspin^{i-} , \Re \left( u^j \Phi^- \right) \rangle + \langle \admspin^{j-} , \Re \left( u^i \Phi^- \right) \rangle &= 0  \ , \\
 \epsilon_{ij} \langle \admspin^{i-} , \Gamma^p_{\phantom{p}q}  \Re \left( u^j \Phi^- \right) \rangle &= 0 \ , \\
 \epsilon_{ij} \langle \admspin^{i-} , \Gamma^a_{\phantom{a}b}  \Re \left( u^j \Phi^- \right) \rangle &= 0 \ .
\end{aligned}\end{equation}
For convenience, we choose $u^1=1$ and $u^2=-\iu$ and insert
\eqref{parametrisation_C_transforms} into the above equations.
Using $\Re \Phi^- = \Re K \wedge \Re \Phi_2 - \Im K \wedge \Im \Phi_2$ and $
 \Im \Phi^- = \Re K \wedge \Im \Phi_2 + \Im K \wedge \Re \Phi_2$,
the first equation in \eqref{stabilizer_adjoint_equ1} implies
\begin{equation}\begin{aligned}\label{parameter}
 \langle \alpha^{11} , \Im \Phi_2 \rangle &= - \langle \alpha^{12} , \Re \Phi_2 \rangle \ , \\
 \langle \alpha^{22} , \Im \Phi_2 \rangle &=   \langle \alpha^{21} , \Re \Phi_2 \rangle \ , \\
 \langle \alpha^{11} , \Re \Phi_2 \rangle - \langle \alpha^{12} , \Im \Phi_2 \rangle  &= \langle \alpha^{22} , \Re \Phi_2 \rangle  +  \langle \alpha^{21} , \Im \Phi_2 \rangle \ .
\end{aligned}\end{equation}
Similarly, the second equation in \eqref{stabilizer_adjoint_equ1} leads to
\begin{equation}\begin{aligned}
 \langle \alpha^{11} , \Im \Phi_2 \rangle &=   \langle \alpha^{21} , \Re \Phi_2 \rangle \ , \\
 \langle \alpha^{22} , \Im \Phi_2 \rangle &= - \langle \alpha^{12} , \Re \Phi_2 \rangle \ , \\
 \langle \alpha^{11} , \Re \Phi_2 \rangle + \langle \alpha^{12} , \Im \Phi_2 \rangle  &=   \langle \alpha^{22} , \Re \Phi_2 \rangle - \langle \alpha^{21} , \Im \Phi_2 \rangle \ .
\end{aligned}\end{equation}
Together, they give five relations between the eight $SU(2,2)_{\Phi_2}$ singlets, which read
\begin{equation}\begin{aligned}
 \langle \alpha^{11} , \Im \Phi_2 \rangle = - \langle \alpha^{12} , \Re \Phi_2 \rangle  &=   \langle \alpha^{21} , \Re \Phi_2 \rangle = \langle \alpha^{22} , \Im \Phi_2 \rangle      \ , \\
 \langle \alpha^{11} , \Re \Phi_2 \rangle &=   \langle \alpha^{22} , \Re \Phi_2 \rangle  \ , \\
 \langle \alpha^{12} , \Im \Phi_2 \rangle &= - \langle \alpha^{21} , \Im \Phi_2 \rangle  \ .
\end{aligned}\end{equation}
This means that only three of the $SU(2,2)_{\Phi_2}$ singlets are
elements in the stabilizer. It remains to analyze the third equation
in \eqref{stabilizer_adjoint_equ1}. By the same method, we can write it as
\begin{equation}
\label{stabilizer_constraint_6}
  \langle \alpha^{11} , \Gamma^a_{\phantom{a}b} \Re \Phi_2 \rangle - \langle \alpha^{12} , \Gamma^a_{\phantom{a}b} \Im \Phi_2 \rangle + \langle \alpha^{21} , \Gamma^a_{\phantom{a}b} \Im \Phi_2 \rangle + \langle \alpha^{22} , \Gamma^a_{\phantom{a}b} \Re \Phi_2 \rangle = 0 \ ,
\end{equation}
where $a$ and $b$ are arbitrary $SO(4,4)$ indices. For the singlets
these equations imply additionally
\begin{equation}
 \langle \alpha^{11} , \Re \Phi_2 \rangle - \langle \alpha^{12} ,  \Im \Phi_2 \rangle + \langle \alpha^{21} , \Im \Phi_2 \rangle + \langle \alpha^{22} ,  \Re \Phi_2 \rangle = 0 \ .
\end{equation}
Together with the third equation in  \eqref{parameter} this implies
\begin{equation}
 \langle \alpha^{11}, \Re \Phi_2 \rangle = \langle \alpha^{12} , \Im \Phi_2 \rangle \ ,
\end{equation}
which reduces the number of singlets to two.

In Eq.~\eqref{stabilizer_constraint_6}
we can choose $\Gamma^a_{\phantom{a}b} \Re \Phi_2$ to be in
the ${\bf6}$ of $SU(2,2)_{\Phi_2}$ and
$\Gamma^a_{\phantom{a}b} \Im \Phi_2=0$ or the other way around.
This implies
\begin{equation}
 \alpha \equiv \operatorname{Proj}_{\bf6} \alpha^{11} = - \operatorname{Proj}_{\bf6} \alpha^{22} \ , \qquad \beta \equiv \operatorname{Proj}_{\bf6} \alpha^{12} = \operatorname{Proj}_{\bf6} \alpha^{21} \ ,
\end{equation}
where $\operatorname{Proj}_{\bf6}$ is the projection onto the ${\bf6}$ representations of $SU(2,2)_{\Phi_2}$.
This gives twelve conditions on the remaining $24$ degrees of freedom and eliminates two of the four ${\bf6}$ representations of $SU(2,2)_{\Phi_2}$.
Therefore, we can parameterize the $m^{i-}$ in the stabilizer by
\begin{equation}  \label{stabilizer6_representation}
 \admspin^{i-}_{\bf6} = \left( \begin{array}{c}
\Re K \wedge \alpha + \Im K \wedge \beta  \\
\Re K \wedge \beta  - \Im K \wedge \alpha
\end{array} \right) \ ,
\end{equation}
and
\begin{equation} \label{stabilizer1_representation_without_gaugings}
 \admspin^{i-}_{\bf1} = \left( \begin{array}{c}
\Re K \wedge (a \Re \Phi_2 + b \Im \Phi_2) - \Im K \wedge (b \Re \Phi_2 - a \Im \Phi_2) \\
 \Re K \wedge (b \Re \Phi_2 - a \Im \Phi_2) + \Im K \wedge (a \Re \Phi_2 + b \Im \Phi_2)
 \end{array} \right) \ ,
\end{equation}
where $a,b\in \mathbb{R}$.
We can reparameterize \eqref{stabilizer6_representation} and \eqref{stabilizer1_representation_without_gaugings} in a more elegant way by writing
\begin{equation}
  \label{stabilizer6_representation_elegant}
 \admspin^{i-}_{\bf6} =  \Re (u^i \bar{K} \wedge \alpha_\mathbb{C} ) \ ,\qquad
 \admspin^{i-}_{\bf1} = \Re ( c u^i K \wedge \bar{\Phi}_2 ) \ ,
\end{equation}
where $\alpha_\mathbb{C}$ is in the complexified ${\bf6}$ representation of $SU(2,2)_{\Phi_2}$ and $c \in \mathbb{C}$.
One can check that the algebra of the transformations in the ${\bf6}$ representation closes if one includes the $SO(2) \subset Sl(2,\mathbb{R})_S \times Sl(2,\mathbb{R})_U$ factor $R_{(+1,-1,0)}$ that is part of the stabilizer as well as the $SU(2,2)_{\Phi_2} \equiv SO(2,4) \subset SO(4,4)$. Together they form the group $SO(2,6)$.
The algebra of the $SU(2,2)_{\Phi_2}$ singlet transformations
$\admspin^{i-}_{\bf1}$ does close if one includes the transformation
$R_{(+1,+1,-2)}$ we introduced above. Together, they generate some
$SO(3)$ group and therefore the complete stabilizer is $SO(3) \times SO(2,6)$.

However, there are still some gauge transformations  of $\U_1$ that do
not have any physical meaning and must be removed.
In section \ref{section:moduliRRd4} we explain that $\U_1$ is part of some highest weight $SU(2)$ embedding $\U_a$ with an $SU(2)$ gauge freedom which is generated by the $\U_a$ themselves. The generator $R_{(+1,+1,+2)}$ coincides with $\U_3$, which is defined in \eqref{SU2embeddingU3}, and rotates $\U_1$ and $\U_2$ defined in \eqref{SU2embeddingU2} into each other. Furthermore, $\U_1$ and $\U_2$ are the other two $SU(2)$ generators $\admspin^{i-}_{\bf 1}$ that correspond to gauge freedom.
These three generators form the algebra of $su(2)$, and furthermore commute with the stabilizer $su(2) = so(3)$ on the singlets which we discussed above. Therefore, all of them together form the algebra of $so(4) = su(2) \oplus su(2)$, explicitly spanned by $R_{(+1,+1,+2)}$, $R_{(+1,+1,-2)}$ and
\begin{equation} \label{stabilizer1_representation}
\begin{aligned}
 \admspin^{i-}_{\bf1} & = \left( \begin{array}{c}
 \Re K \wedge (a \Re \Phi_2 + d \Im \Phi_2) - \Im K \wedge (b \Re \Phi_2 + c \Im \Phi_2)  \\
 \Re K \wedge (b \Re \Phi_2 + c \Im \Phi_2) + \Im K \wedge (a \Re \Phi_2 + d \Im \Phi_2)
\end{array} \right) \\
 & = \Re (c_1 u^i K \wedge \bar{\Phi}_2 + c_2 u^i K \wedge \Phi_2 ) \ .
\end{aligned}
\end{equation}
Here, $a,b,c$ and $d$ are real parameters that can be rewritten in
terms of the complex parameters $c_1$ and $c_2$. The generators
corresponding to $c_1$ are the elements in the stabilizer we already
had in
\eqref{stabilizer6_representation_elegant}.

As explained in section \ref{section:moduliRRd4}, the stabilizer of $\U_1$ coincides with the stabilizer of the other two $\U_a$.
Therefore, after removing pure gauge degrees of freedoms, we end up with the configuration space
\begin{equation}
\mathcal{M}_{\admu_a} =  \frac{SO(6,6)}{SO(4) \times SO(2,6)} \ .
\end{equation}
We see that $\U_1$ defines up to gauge equivalence a space-like four-dimensional subspace in a space of signature $(6,6)$. This can be understood as follows.
Under the breaking $SO(6,6)\to Sl(2,\mathbb{R})_S \times
Sl(2,\mathbb{R})_U \times SO(4,4)$, the vector representation of
$SO(6,6)$ 
decomposes analogously to  \eqref{Breaking_SO66_S044_SO22_vector} as
${\bf 12} \to ({\bf 2},{\bf 2},{\bf 1}) \oplus ({\bf 1},{\bf 1},{\bf
8}^s)$, where  we used the triality of $SO(4,4)$.
Under the same decomposition, $\admu$ sits in the  $({\bf 2},{\bf
2},{\bf 8}^s)$ representation which is just the tensor product $({\bf
2},{\bf 2},{\bf 1}) \otimes ({\bf 1},{\bf 1},{\bf 8}^s)$. The first
factor in this tensor product  is given by $u^i K$
while the second is given by $\Phi_2$. Both
define complex $SO(6,6)$ vectors which are orthogonal to each other.
From the normalization in \eqref{K_compatible} and \eqref{E7_normalization_u} we see that\footnote{Here, $\langle \cdot , \cdot \rangle_2$ is the usual spinor product on $\Lambda^\bullet T^*_2 \M_6$ that is related to the Mukai pairing by a two-dimensional volume factor.}
\begin{equation}
\epsilon_{ij} \langle u^i K , \bar{u}^j \bar{K} \rangle_2 = 2 \ .
\end{equation}
Thus, we can conclude that the real and imaginary part of $u^i K$
have positive norm. The same holds for the real and imaginary part of
$\Phi_2$. Hence, together they span the space-like four-plane defined by
$\admu$ (or equivalently, by the $\U_a$) in the space of signature $(6,6)$. Note that the $SO(4)$
transformations $R_{(+1,+1,+2)}$, $R_{(+1,+1,-2)}$ and the ones given
in \eqref{stabilizer1_representation} are really those that rotate the
space-like four-plane non-trivially into itself.

Now let us determine the most general form of the embedding \eqref{E7_embedding_spinor_adjoint}. For this, we consider the possible deformations of $\U_1$ under $G^{\textrm{IIA}} = Sl(2,\mathbb{R})_T \times SO(6,6)$ transformations. The transformations in $G^{\textrm{NS}} =
 Sl(2,\mathbb{R})_S \times
Sl(2,\mathbb{R})_T \times Sl(2,\mathbb{R})_U \times SO(4,4)$ act only on $u^i$ and $\Phi^-$. Thus, they do not change the embedding \eqref{E7_embedding_spinor_adjoint}. It remains to understand the transformation properties under $SO(6,6)$
deformations $\adnspin^{i-}$ which live in $\Lambda^\textrm{odd} T^*_2 \M_6 \wedge \Lambda^\textrm{even} T^*_4 \M_6$.
From~\eqref{stabilizer6_representation}
and~\eqref{stabilizer1_representation} we know the $\adnspin^{i-}$ that stabilize $\admu$. This means that those which deform $\U_1$ non-trivially are of the form
\begin{equation} \label{RR_deformations_adjoint_singlet} \begin{aligned}
  \adnspin^{i-}_{\bf1} & = \left( \begin{array}{c}
 \Re K \wedge (a \Re \Phi_2 + d \Im \Phi_2) + \Im K \wedge (b \Re \Phi_2 + c \Im \Phi_2)  \\
 \Re K \wedge (b \Re \Phi_2 + c \Im \Phi_2) - \Im K \wedge (a \Re \Phi_2 + d \Im \Phi_2)
\end{array} \right)  \\
 & = \Re (d_1 u^i \bar{K} \wedge \bar{\Phi}_2 + d_2 u^i \bar{K} \wedge \Phi_2 )
\end{aligned} \end{equation}
for the $SU(2,2)_{\Phi_2}$ singlets, and
\begin{equation} \label{RR_deformations_adjoint_6} \begin{aligned}
 \adnspin^{i-}_{\bf6} & =  \left( \begin{array}{c}
- \Re K \wedge \alpha + \Im K \wedge \beta  \\
- \Re K \wedge \beta  - \Im K \wedge \alpha
\end{array} \right)
 = \Re ( u^i K \wedge \beta_\mathbb{C} )
\end{aligned} \end{equation}
for the ${\bf6}$ representation of $SU(2,2)_{\Phi_2}$. The coefficients $a,b,c$ and $d$ again are real and can be expressed in the complex numbers $d_1$ and $d_2$. Analogously, $\alpha$ and $\beta$ are real $SO(4,4)$ spinors in the $\bf 6$ representation of $SU(2,2)$, while $\beta_\mathbb{C}$ is in the complexified $\bf 6$ representation.

In order to compute the transformations of $\admu$, we decompose
analogously to
\eqref{decomposition_SO66_subgroup_of_E7}
\begin{equation}
\delta \admu  = \left( (\delta \admui)^i_{\phantom{i}j} , (\delta \admuA)^p_{\phantom{p}q}, (\delta \admuA)^a_{\phantom{a}b} , (\delta \admuspin)^{i-} \right) \ .
\end{equation}
Using \eqref{commutator_SO66_subgroup} we find
\begin{equation}
\label{variation_RR_deformations_adjoint_6}
 (\delta_{\bf6} \admuA)^a_{\phantom{a}b} = 2 ( \langle \alpha, \Gamma^a_{\phantom{a}b} \Re \Phi_2 \rangle + \langle \beta, \Gamma^a_{\phantom{a}b} \Im \Phi_2 \rangle ) = 2 \Re ( \langle \beta_\mathbb{C} , \Gamma^a_{\phantom{a}b} \bar{\Phi}_2 \rangle )
\end{equation}
as the only non-vanishing deformation for the deformations in the ${\bf6}$ representation.
For the deformations in the ${\bf 1}$ representation the variation in the adjoint of $Sl(2,\mathbb{R})$ is
\begin{equation}
\label{variation_RR_deformations_adjoint_singlet1}
 (\delta_{\bf1} \admui)^i_{\phantom{i}j} =
 2 \left( \begin{array}{cc} a-c & b+d \\ b+d & c-a \end{array} \right) = 2 \Im (d_1 \epsilon_{jk} u^i u^k) \ .
\end{equation}
The deformations $(\delta_{\bf1} \admu)^p_{\phantom{p}q}$ that take values in the adjoint of $SO(2,2)$ do not fill it out completely but only the $sl(2,\mathbb{R})_U$ part. They read
\begin{equation}
 \label{variation_RR_deformations_adjoint_singlet2}
 (\delta_{\bf1} \admuA)^p_{\phantom{p}q} = \left( \begin{array}{cccc}
 a+c & d-b & 0 & 0 \\ d-b & -a-c & 0 & 0 \\ 0 & 0 & -a-c & b-d \\ 0 & 0 & b-d & a+c
 \end{array} \right) =  \left( \begin{array}{cc} \Im (d_2 \epsilon_{jk} \bar{u}^i \bar{u}^k) & 0 \\ 0 & - \Im (d_2 \epsilon_{jk} \bar{u}^i \bar{u}^k) \end{array} \right) \ ,
\end{equation}
with all other components being zero.

\subsubsection{Putting the parts together}\label{section:E7_stabilizer_both}
So far we just embedded the two pure $SO(6,6)$ spinors into appropriate $E_{7(7)}$ representations. In this section we finally discuss their purity and compatibility conditions as well as the moduli space spanned by both objects.

As we already mentioned in section \ref{section:moduliRRd4} a pure (complex) $SO(6,6)$ spinor $\Phi$ is equivalent to a stable (real) $SO(6,6)$ spinor $\chi= \Re \Phi$. Stability means in this context that such a spinor $\chi$ transforms in an open orbit under the action of $SO(6,6)$. Stability is measured by
the quartic $SO(6,6)$ invariant \cite{Hitchin:2004ut,Grana:2005ny}
\begin{equation}\label{quartic_invariant_SO66}
 q(\chi) = - \tfrac{1}{4} \, \bar{\chi} \Gamma_{AB} \chi \, \bar{\chi} \Gamma^{AB} \chi = \tfrac{1}{4} \operatorname{tr}(\mathcal{J}_\Phi^2) \ .
\end{equation}
For $\chi$ to be stable we need $q(\chi) <0$. Furthermore, $q$ measures the normalization of the generalized almost complex structure $\mathcal{J}_\Phi$. Thus, $\mathcal{J}_\Phi$ can be properly normalized if $q(\Re \Phi) <0$.
The quantity \eqref{quartic_invariant_SO66} is naturally embedded into the quartic invariants of the $E_{7(7)}$ representations \cite{Grana:2008tbp}
\begin{equation}\begin{aligned}
 q(\lambda) = & \epsilon_{ij} \epsilon_{kl} \mathcal{I}_AB \mathcal{I}_CD \lambda^{iA} \lambda^{kB}\lambda^{jC}\lambda^{lD} - \epsilon_{ij} \lambda^{iA}\lambda^{jB} \bar{\lambda}^+ \Gamma_{AB} \lambda^+ \\ &
- \tfrac{1}{24} \bar{\lambda}^+ \Gamma_{AB} \lambda^+ \bar{\lambda}^+ \Gamma^{AB} \lambda^+
\end{aligned}\end{equation}
and
\begin{equation}\begin{aligned}
 q(m) = & \operatorname{tr} m^4 + ( \operatorname{tr} m^2 )^2 + (\operatorname{det} \hat{m})^2 + \operatorname{tr} m^2 \epsilon_{ij} \bar{\tilde{m}}^{i-} \tilde{m}^{j-} + (\operatorname{det} \hat{m})^2  \epsilon_{ij} \bar{\tilde{m}}^{i-} \tilde{m}^{j-} \\ & + (\epsilon_{ij} \bar{\tilde{m}}^{i-} \tilde{m}^{j-} )^2  + \epsilon_{ij} \epsilon_{kl} \bar{\tilde{m}}^{i-} \Gamma^{AB} \tilde{m}^{k-}   \bar{\tilde{m}}^{j-} \Gamma_{AB} \tilde{m}^{l-} \ .
\end{aligned}\end{equation}
We see that in both expressions the last term just gives the Hitchin functional while the other terms vanish for the embeddings \eqref{E7_embedding_spinor_fund} and \eqref{E7_embedding_spinor_adjoint}. Hence, both quartic invariants just generalize the corresponding Hitchin functionals
and we impose the stability condition via
\begin{equation} \label{E7_purity_conditions}
 q(\lambda) < 0 \ , \qquad q(\U_1) < 0 \ .
\end{equation}
Now we come to the issue of compatibility. By studying \eqref{variation_fund_E7} one can check that \eqref{E7_compatibility_conditions}
is appropriate to reproduce the $SO(6,6)$ compatibility conditions given in \eqref{compatibility_conditions} for the embeddings \eqref{E7_embedding_spinor_fund} and \eqref{E7_embedding_spinor_adjoint}.
Furthermore, in principle both embeddings \eqref{E7_embedding_spinor_fund} and \eqref{E7_embedding_spinor_adjoint} could each be shifted by independent $E_{7(7)}$ transformations. Equation \eqref{E7_compatibility_conditions} ensures that the additional degrees of freedom which arise in the U-duality group are the same for both embeddings.

Finally, we want to determine the parameter space of compatible $\admu$
and $\lambda$. To do so, we first determine the intersection of the
two stabilizers, and afterwards eliminate all gauge redundancies.
Let us distinguish for the following analysis $SU(2) \times SU(2)$
singlets from the $({\bf 2}, {\bf 2})$ representation of $SU(2) \times
SU(2)$. The former one is spanned by the real and imaginary parts of
$\Phi_1$ and $\Phi_2$. The latter one is just the intersection of the
${\bf 6}$ representations of $SU(2,2)_{\Phi_1}$ and
$SU(2,2)_{\Phi_2}$.

Let us start with the $({\bf 2}, {\bf 2})$ representation.
The part of the common stabilizer that is in $so(4,4)$ is just $su(2)
\times su(2) \equiv so(4)$, as we know from the theory of pure
compatible spinors. It just describes rotations in the space
orthogonal to the real and imaginary parts of the two pure spinors and
therefore is in the $({\bf 2}, {\bf 2})$ representation. Furthermore, there are additional elements in the $({\bf 2}, {\bf 2})$ representation of the stabilizer which are of the form $\admspin^{i-}$. The $\admspin^{i-}$ in \eqref{stabilizer6_representation} split into those that are in the $({\bf 2}, {\bf 2})$ representation and those which are $SU(2)\times SU(2)$ singlets. Eq.~\eqref{stabilizer_Phi_plus} eliminates exactly the singlets. The algebra consisting of the $\admspin^{i-}$ in the $({\bf 2}, {\bf 2})$ representation and the $SU(2)
\times SU(2)$ generators closes if one includes the
$SO(2)$ factor $R_{(+1,-1,0)}$. Analogously to the analysis for the stabilizer of
$\admu$, these transformations form the group $SO(6)$.

Now let us consider the $SU(2) \times SU(2)$ singlets. These singlets are either $SU(2,2)_{\Phi_1}$ or $SU(2,2)_{\Phi_2}$ singlets. The singlets of $SU(2,2)_{\Phi_1}$ are removed from the stabilizer by \eqref{stabilizer_Phi_plus}. The $SU(2,2)_{\Phi_2}$ singlets all stabilize $\lambda$. Hence, the singlet component of the stabilizer is the same as the one for $\admu$. In section \ref{section:E7_stabilizer_adjoint}, we showed that it forms the group $SO(3)$.
Again there are
further gauge redundancies which have to be projected out.
Part of them we already discussed in the previous two subsections.
Furthermore, the gauge transformation that rotate the doublets $(\Re
\Phi_1 , \Im \Phi_1 )$ and $(\Re \Phi_2 , \Im \Phi_2 )$ are part of a
bigger group that rotates the vector $(\Re \Phi_1 , \Im \Phi_1, \Re
\Phi_2 , \Im \Phi_2)$.
In addition we have to mod out the gauge transformations generated by
$\admspin^{i-}$. On top of the transformations
in~\eqref{stabilizer1_representation},
this also includes
\begin{equation} \label{additional_gauging_dof} \begin{aligned}
 \admspin^{i-}_{\bf1} & = \left( \begin{array}{c}
 \Re K \wedge (\tilde{a} \Re \Phi_1 + \tilde{d} \Im \Phi_1) - \Im K \wedge (\tilde{b} \Re \Phi_1 + \tilde{c} \Im \Phi_1)  \\
 \Re K \wedge (\tilde{b} \Re \Phi_1 + \tilde{c} \Im \Phi_1) + \Im K \wedge (\tilde{a} \Re \Phi_1 + \tilde{d} \Im \Phi_1)
\end{array} \right) \\ & = \Re ( c_3 u^i K \wedge \bar{\Phi}_1 + c_4
u^i K \wedge \Phi_1 )\ .
\end{aligned} \end{equation}
Together with the $SO(4)$ rotations among $\Re \Phi_1$, $\Im \Phi_1$, $\Re \Phi_2$ and $\Im \Phi_2$ and the $SO(2)$ rotation of $u^i$, this forms an $SO(6)$ group which consists of elements that leave $\lambda$ and $\admu$ invariant up to gauge degrees of freedom.

So far we discussed only the transformations in $SO(6,6)$. The transformations in $Sl(2,\mathbb{R})_T$ stabilize $\mu$, but only an $SO(2)$ subgroup leaves $\lambda$ invariant up to gaugings.
Therefore, we end up with the parameter space
\begin{equation} \label{SO66_local_parameter_space}
\mathcal{M}_{\lambda,\admu} = \frac{SO(6,6)}{SO(6) \times SO(6)}  \times \frac{Sl(2,\mathbb{R})_T}{SO(2)} \ .
\end{equation}
The first factor describes the choice of a space-like six-dimensional subspace in a space of signature $(6,6)$.
Analogously to section \ref{section:E7_stabilizer_adjoint}, we can interpret this six-dimensional subspace as being spanned by the real and imaginary parts of $u^i K$, $\Phi_1$ and $\Phi_2$.
In fact, the $E_{7(7)}$ covariant formalism describes the procedure to descend from the flat superconformal cone over
\eqref{SO66_local_parameter_space} to the moduli space itself. First, one assumes that the
real and imaginary part of $u^i K$, $\Phi_1$ and $\Phi_2$, which are
space-like $SO(6,6)$ vectors, do not feel any constraints on their
position in the vector space of signature $(6,6)$ apart from being
space-like and linearly independent. This defines a $Gl(6)$ bundle over
\eqref{SO66_local_parameter_space}, which is an open set in $\bigoplus_{i=1}^6 \mathbb{R}^{(6,6)}$. The purity conditions
\eqref{E7_purity_conditions} and the compatibility condition
\eqref{E7_compatibility_conditions} together with gauge fixing then
project this flat cone to \eqref{SO66_local_parameter_space}. This
cone also survives the transition to a global moduli space by performing
the Kaluza-Klein truncation, as done in section
\ref{section:moduliRRd4}, where it becomes the $Gl(6)$ bundle over the
first factor in \eqref{E7_moduli_space_IIA} and is an open set in
$\bigoplus_{i=1}^6 \mathbb{R}^{(6,6+n)}$.

At the end of this section, we want to identify explicitly the scalar degrees of freedom coming from the RR-sector. For this we have to distinguish the deformations that are $SU(2) \times SU(2)$ singlets from those that are in the $({\bf2},{\bf2})$ representation.
The deformations in the $({\bf2},{\bf2})$ representation are in the
${\bf6}$ representation of both $SU(2,2)_{\Phi_1}$
and $SU(2,2)_{\Phi_2}$. Therefore, they are the deformations which are displayed
in~\eqref{RR_deformations_adjoint_6} but are not of the
form~\eqref{RR_deformations_fundamental_singlet}. This gives eight
degrees of freedom, corresponding to all RR-form fields
that are in the $({\bf2},{\bf2})$ representation of $SU(2) \times
SU(2)$. Their action on $\admu$ is given
by~\eqref{variation_RR_deformations_adjoint_6}, while $\lambda$ stays
invariant.

The situation for the singlets is a bit more involved, because some of
the deformations we presented
in \eqref{RR_deformations_fundamental_singlet}
and \eqref{RR_deformations_adjoint_singlet} might just refer to gauge transformations which we mod out. Indeed, exactly the deformations displayed in~\eqref{additional_gauging_dof} are pure gauge, and therefore, out of the eight deformations in~\eqref{RR_deformations_fundamental_singlet}, only those of the form
\begin{equation} \label{RR_deformations_fundamental_singlet_physical} \begin{aligned}
 \tilde{n}^{i-}_{\bf1} & = \left( \begin{array}{c}
 \Re K \wedge (\tilde{a} \Re \Phi_1 + \tilde{d} \Im \Phi_1) + \Im K \wedge (\tilde{b} \Re \Phi_1 + \tilde{c} \Im \Phi_1)  \\
 \Re K \wedge (\tilde{b} \Re \Phi_1 + \tilde{c} \Im \Phi_1) - \Im K \wedge (\tilde{a} \Re \Phi_1 + \tilde{d} \Im \Phi_1)
\end{array} \right) \\ & = \Re (d_3 u^i \bar{K} \wedge \bar{\Phi}_1 + d_4 u^i \bar{K} \wedge \Phi_1 )
\end{aligned} \end{equation}
are physical.\footnote{Note that these transformations deform $\lambda$ but leave $\admu$ invariant.}
Thus, in the singlets we have eight physical degrees of freedom, which are parameterized by \eqref{RR_deformations_adjoint_singlet} and \eqref{RR_deformations_fundamental_singlet_physical}.
These together with the degrees of freedom in the $({\bf2},{\bf2})$ representation form exactly one spinor $C \in \Lambda^\textrm{odd} T^*_2 \M_6 \wedge \Lambda^\textrm{even} T^*_4 \M_6 $ and represent the RR-scalars.
This spinor $C$ is composed out of the several RR-fields via the formal sum
\begin{equation}
 C= C_1 + C_3 + C_5 \in \Lambda^\textrm{odd} T^*_2 \M_6 \wedge \Lambda^\textrm{even} T^*_4 \M_6 \ ,
\end{equation}
and is identified with the deformations in \eqref{RR_deformations_adjoint_6}, \eqref{RR_deformations_adjoint_singlet} and \eqref{RR_deformations_fundamental_singlet_physical} via the decomposition
\begin{equation} \begin{aligned}
 C = & \Re K \wedge (\tilde{a} \Re \Phi_1 + \tilde{d} \Im \Phi_1 + a \Re \Phi_2 + d \Im \Phi_2 - \alpha) \\ & + \Im K \wedge (\tilde{b} \Re \Phi_1 + \tilde{c} \Im \Phi_1 + b \Re \Phi_2 + c \Im \Phi_2 + \beta) \\ = & \Re( d_1 \bar{K} \wedge \bar{\Phi}_2 + d_2 \bar{K} \wedge \Phi_2 +  d_3 \bar{K} \wedge \bar{\Phi}_1 + d_4 \bar{K} \wedge \Phi_1 - K \wedge \beta^{({\bf 2},{\bf 2})}_\mathbb{C}) \ ,
\end{aligned} \end{equation}
where $\alpha$ and $\beta$ are the real $SO(4,4)$ spinors that are in the $({\bf2},{\bf2})$ representation of $SU(2) \times SU(2)$, and $ \alpha^{({\bf 2},{\bf 2})}_\mathbb{C}$ is a complex $SO(4,4)$ spinor in the same representation. There is a simple way to display the RR-fields as $E_{7(7)}$ transformations that is
\begin{equation}
\label{RR_transformations}
 \admspin^{i-} = \Re (u^i  ( C - \iu \ast_B C) ) \ .
\end{equation}
If exponentiated, these transformations correspond to shifts in the RR-fields which read
\begin{equation}
 C_0 \to C_0 + C \ .
\end{equation}
Note that transformations of the form~\eqref{RR_transformations}
commute with each other up to gauge transformations and elements in the stabilizer of the $SU(2)\times SU(2)$ structure.

So far we did not discuss shifts in the complexified dilaton $B_6 + \iu \e^{-\phi}$.\footnote{Here, $B_6$ denotes the dualized $B$ field which is a scalar in four dimensions. Our notation refers to the democratic formulation and expresses the four-dimensional scalar in terms of the six-form $B_6$ which is dual to the ten-dimensional $B$ field and fills out all internal directions.} These are the  $Sl(2,\mathbb{R})_S$ transformations
\begin{equation}
\tilde{m}^i_{\phantom{i}j} =
\left( \begin{array}{cc}
 \e^\phi & 0 \\ - B_6 & \e^{-\phi}
\end{array} \right) \ ,
\end{equation}
which span $\frac{Sl(2,\mathbb{R})_S}{SO(2)} \subset \frac{SO(6,6)}{SO(6) \times SO(6)}$.

\section{$SU(2) \times SU(2)$ structures in five dimensions}\label{section:compd5}
In this appendix we study the missing case of backgrounds of the form
\eqref{stringbackground} with a five-dimensional Minkowskian space-time
($d=5$) times a five-dimensional compact manifold $\M_5$.
As before we focus on the situation where these backgrounds
preserve 16 supercharges which corresponds to $N =2$ in five dimensions.
More precisely the Lorentz group for these backgrounds
is $SO(1,4)\times SO(5)$. The ten-dimensional spinor representation
decomposes accordingly as\footnote{Note that there are no chiral spinor representations for $SO(1,4)$ and $SO(5)$.}
\begin{equation}
{\bf 16} \to ({\bf 4}, {\bf 4}) \ ,
\end{equation}
where the first ${\bf 4}$ denotes a spinor of $SO(1,4)$ while the second one denotes the spinor representation of $SO(5)$.
Preserving half of the supercharges amounts to choosing backgrounds
which admit one or two globally defined spinors which
corresponds to manifolds $\M_5$ with a reduced structure group $SU(2)$
or $SU(2)\times SU(2)$, respectively. We will start with a general analysis of the spectrum of type II supergravities in such backgrounds. We discuss first geometrical $SU(2)$ structure backgrounds, which are then generalized by the use of the pure spinor methods of generalized geometry and its analogues in exceptional generalized geometry.
Note that a similar, independently performed analysis of exceptional generalized geometry in $d=5$ has been presented in \cite{Sim:2008}.

\subsection{Field decomposition for $d=5$}\label{section:spectrumd5}
In this section we want to analyze the massless type II supergravity fields in terms of their representations under the five-dimensional Lorentz group and the $SU(2) \times SU(2)$ structure group and show analogously to section \ref{section:spectrumd6} how they assemble in $N =2, d=5$ multiplets, in the spirit of~\cite{Grana:2005ny}.

Again, we use the light-cone gauge where on-shell the fields form representations of $SO(3)$ instead of the whole $SO(1,4)$ Lorentz group. Since we treat the case of $SU(2) \times SU(2)$ structure group, we therefore examine the decomposition of massless type II supergravity fields under the group $SO(3) \times SU(2) \times SU(2)$. For this, let us recall the decomposition of the two Majorana-Weyl representations ${\bf 8}^s$ and ${\bf 8}^c$ and the vector representation ${\bf 8}^v$ under the breaking $SO(8) \to SO(3) \times SO(5) \to SO(3) \times SU(2)$. We get
\begin{equation}\begin{aligned} \label{decomposeSO8toSU2d5}
   {\bf 8}^s & \to {\bf 4}_{\bf 2} \to 2 \, {\bf 1}_{\bf \frac{1}{2}} \oplus {\bf 2}_{\bf \frac{1}{2}} \ , \\
   {\bf 8}^c & \to {\bf 4}_{\bf 2} \to 2 \, {\bf 1}_{\bf \frac{1}{2}} \oplus {\bf 2}_{\bf \frac{1}{2}}  \ , \\
   {\bf 8}^v & \to {\bf 1}_{\bf 1} \oplus {\bf 5}_{\bf 0} \to {\bf 1}_{\bf 1}  \oplus {\bf 1}_{\bf 0} \oplus 2 \, {\bf 2}_{\bf 0} \ ,
\end{aligned}\end{equation}
where the subscript denotes the spin $s$ under the $SO(3)$ Lorentz group, i.e.\ the representation has dimension $(2s+1)$. We note that both Majorana-Weyl representations decompose in the same way under a $SU(2)$ structure for $d=5$. Therefore, we expect compactifications of type IIA and type IIB to give the same low energy effective action in this dimension.

In type IIA the massless fermionic degrees of freedom originate from the $({\bf 8}^s,{\bf 8}^v)$ and $({\bf 8}^v,{\bf 8}^c)$ representation of $SO(8)_L \times SO(8)_R$, while in type IIB they form the $({\bf 8}^s,{\bf 8}^v)$ and $({\bf 8}^v,{\bf 8}^s)$ representation. However, since ${\bf 8}^s$ and ${\bf 8}^c$ decompose in the same way, we just investigate the former one.
Under the decomposition $SO(8)_L \times SO(8)_R \to SO(3) \times SU(2)_L \times SU(2)_R$ they behave as
\begin{equation}\begin{aligned}
  ({\bf 8}^s,{\bf 8}^v) & \to 2 ({\bf 1},{\bf 1})_{\bf \frac{3}{2}} \oplus 4 ({\bf 1},{\bf 1})_{\bf \frac{1}{2}} \oplus 4 ({\bf 1},{\bf 2})_{\bf \frac{1}{2}} \oplus ({\bf 2},{\bf 1})_{\bf \frac{3}{2}} \oplus 2 ({\bf 2},{\bf 1})_{\bf \frac{1}{2}} \oplus 2 ({\bf 2},{\bf 2})_{\bf \frac{1}{2}} \ , \\
  ({\bf 8}^v,{\bf 8}^s), ({\bf 8}^v,{\bf 8}^c) & \to 2 ({\bf 1},{\bf 1})_{\bf \frac{3}{2}} \oplus 4 ({\bf 1},{\bf 1})_{\bf \frac{1}{2}} \oplus ({\bf 1},{\bf 2})_{\bf \frac{3}{2}} \oplus 2 ({\bf 1},{\bf 2})_{\bf \frac{1}{2}} \oplus 4 ({\bf 2},{\bf 1})_{\bf \frac{1}{2}} \oplus 2 ({\bf 2},{\bf 2})_{\bf \frac{1}{2}} \ .
\end{aligned}\end{equation}
We see that half of the gravitinos come in the $({\bf 1},{\bf 1})$ representation while the other half is in the doublet representations $({\bf 1},{\bf 2})$ and $({\bf 2},{\bf 1})$ of $SU(2)_L \times SU(2)_R$. The latter ones again correspond to massive gravitino multiplets that must be projected out to end up with standard $N =2, d=5$ supergravity.
After this projection, the fermionic components in the $({\bf 1},{\bf 1})$ become part of the gravity multiplet and one vector multiplet, while the $({\bf 2},{\bf 2})$ components correspond to the fermionic degrees of freedom in the vector multiplets.\footnote{As we know from section \ref{section:spectrumd6}, the ${({\bf 1},{\bf 1})}$ representation corresponds to the gravity multiplet in $d=6$ (for type IIA). This decomposes into the gravity multiplet plus one vector multiplet in the ${({\bf 1},{\bf 1})}$ representation in $d=5$.}

The massless bosonic fields of type II supergravity can be decomposed in the same way. For the NS-NS-sector we consider the combination $E_{MN} = g_{MN} + B_{MN} + \phi \eta_{MN}$  which forms the $({\bf 8}^v,{\bf 8}^v)$ representation. It decomposes as
\begin{equation}\begin{aligned}
 E_{\mu \nu} & : ({\bf 1},{\bf 1})_{\bf 2}  \oplus({\bf 1},{\bf 1})_{\bf 0}  \oplus ({\bf 1},{\bf 1})_{\bf 1} \ , \\
 E_{\mu m} & : ({\bf 1},{\bf 1})_{\bf 1} \oplus 2 ({\bf 1},{\bf 2})_{\bf 1}  \ ,  \\
 E_{m \mu } & : ({\bf 1},{\bf 1})_{\bf 1} \oplus 2 ({\bf 2},{\bf 1})_{\bf 1}  \ ,  \\
 E_{mn} & : ({\bf 1},{\bf 1})_{\bf 0} \oplus 2 ({\bf 1},{\bf 2})_{\bf 0} \oplus 2 ({\bf 2},{\bf 1})_{\bf 0} \oplus 4 ({\bf 2},{\bf 2})_{\bf 0} \ ,
\end{aligned}\end{equation}
where in the first line the first component corresponds to the metric, the second one to the five-dimensional dilaton and the third one to the antisymmetric tensor field.
After the projection we are not only left with the four-dimensional metric and the antisymmetric two-tensor but also with two vectors of the form $E_{\mu m}$ and $E_{m \mu }$ and one scalar $E_{mn}^{({\bf 1},{\bf 1})}$ that become part of the gravity multiplet and the vector multiplet which was already mentioned. The other scalars $E_{mn}^{({\bf 2},{\bf 2})}$ sit in a vector multiplet. As in the $d=6$ case, they can be associated with deformations of the $SU(2) \times SU(2)$ background.

Finally, we decompose the RR-sector. This corresponds to the
decomposition of the $({\bf 8}^s,{\bf 8}^c)$ representation or the
$({\bf 8}^s,{\bf 8}^s)$, leading here to the same result. We find
\begin{equation} \begin{aligned}
  ({\bf 8}^s,{\bf 8}^c), ({\bf 8}^s,{\bf 8}^s) \to & 4 ({\bf 1},{\bf 1})_{\bf 1} \oplus 4 ({\bf 1},{\bf 1})_{\bf 0} \oplus 2 ({\bf 1},{\bf 2})_{\bf 1} \oplus 2 ({\bf 1},{\bf 2})_{\bf 0} \\
& \oplus 2 ({\bf 2},{\bf 1})_{\bf 1} \oplus 2 ({\bf 2},{\bf 1})_{\bf 0} \oplus ({\bf 2},{\bf 2})_{\bf 1} \oplus ({\bf 2},{\bf 2})_{\bf 0} \ .
\end{aligned}\end{equation}
Projecting out all $SU(2) \times SU(2)$ doublets leaves us with four
vectors and four scalars in the ${({\bf 1},{\bf 1})}$ representation
and one vector and one scalar in the ${({\bf 2},{\bf 2})}$ representation.

These fields can be assembled into a gravity multiplet and two vector
multiplets. The gravity multiplet contains the graviton, four
gravitini, six vector fields, four Weyl fermions and a real scalar
all in the ${({\bf 1},{\bf 1})}$ representation. The two vector
multiplets are in the ${({\bf 1},{\bf 1})}$ and ${({\bf 2},{\bf 2})}$
representation respectively, and
each contains one vector field, four gaugini and five scalars.

Analogously to section \ref{section:spectrumd6} we still deal with ten-dimensional fields which have been reordered in such a way that they form $N =2, d=5$ multiplets. The action corresponding to these multiplets only allows for manifest $SO(1,4)\times SO(5)$ symmetry and $N =2$ supersymmetry. Then we projected out the $SU(2)\times SU(2)$ doublets to achieve an action that is only $N =2$ supersymmetric.

\subsection{$SU(2)$ structures on manifolds of dimension $5$}\label{section:SU2d5}
For a five-dimensional manifold to admit a globally defined, nowhere vanishing $SO(5)$ spinor $\eta$ requires the structure group $G$ to be contained in $SU(2)$ since this is the largest subgroup that allows for a singlet in the spinor representation of $SO(5)$. This can be seen by the fact that the spin double cover of $SO(5)$ is $Sp(2)$. To allow for a spinor singlet $\eta$, this group has to be broken to $Sp(1) \equiv SU(2)$.
We will assume in the following that $\eta$ is normalized to one.
Considering the breaking $SO(5) \to SU(2)$, the spinor representation splits like
\begin{equation}
 {\bf 4} =  {\bf 2} \oplus {\bf 1} \oplus {\bf 1} \ .
\end{equation}
Here, $\eta$ and $\eta^c$ define the two singlets. From these two singlets, one can globally define the nowhere vanishing two-forms $J$ and $\Omega$ using \eqref{definition_two-forms}. Furthermore, one can define a real one-form $L$, given by
\begin{equation}
 L_{m} = \bar{\eta} \gamma_{m} \eta \ ,
\end{equation}
which defines an almost product structure
\begin{equation}
 P^m_{\phantom{m}n} = 2 L^m L_n - \delta^m_{\phantom{m}n} \ .
\end{equation}
It singles out one direction $L$ in $T \M_5$ at each point and breaks the structure group to $SO(4)$. Similarly to the discussion of appendix \ref{section:productstructured6}, all vectors orthogonal to $L$ are $SU(2)$ doublets. Therefore, in the $N=2$ theory only the prefactor of $L$ can be deformed and $P$ is rigid.
Analogously to the four-dimensional case, one can define two-forms via~\eqref{definition_two-forms} which reduce the structure group further to $SU(2)$.
For the $SU(2)$ structure part of the geometric moduli space, the discussion of section \ref{section:SU2d6} applies. We computed the geometric moduli space to be \eqref{moduli_space_geom}. Here, the prefactor of $L$ gives one additional degree of freedom. Thus, the result is
\begin{equation}
\mathcal{M}_{d=5}^\textrm{geom} \ =\ \frac{SO(3,n+3)}{SO(3) \times SO(n+3)} \times \mathbb{R}_+ \times \mathbb{R}_+  \ .
\end{equation}

\subsection{Pure spinors in five dimensions}\label{section:SU2SU2d5}
Now we want to apply the framework of generalized geometry to the case $d=5$.
However, one cannot define a valid generalized almost complex structure on the ten-dimensional bundle $T \M_5 \oplus T^* \M_5$ compatible with the canonical pairing $\mathcal{I}$. This generalizes the fact that one cannot define an almost complex structure on a manifold of odd dimension. Therefore, the techniques of generalized geometry seem not to apply for a manifold of dimension five. However, the language of pure spinors still makes sense, and therefore we can apply it even in this case.

In five dimensions, an $SO(5,5)$ spinor is defined to be pure if and only if (see for example~\cite{Berkovits:2000fe})
\begin{equation}
 \langle \Phi, \Gamma^M \Phi \rangle_5 = 0   \qquad \textrm{for all} \quad M = 1, \dots , 10 \ .
\label{purity_extended_spinor}
\end{equation}
Furthermore, we define a pure $SO(5,5)$ spinor to be normalizable if and only if
\begin{equation}
 \langle \Phi, \Gamma^M \bar{\Phi}\rangle_5 \ne 0 \ .
\end{equation}
For two pure spinors $\Phi^+$ and $\Phi^-$ of opposite chirality, we have to define appropriate compatibility conditions. Since we are not able to switch to the language of generalized almost complex structures, we cannot use the intuition which is usually gained there. However, in analogy to the case of a manifold of dimension four, we can define the compatibility conditions
\begin{equation}
 \langle \Phi^+, \Phi^-\rangle_5  =  \langle \Phi^+, \bar{\Phi}^-\rangle_5 = 0
 \label{SO5_comp}
\end{equation}
and the normalization
\begin{equation}
 \langle \Phi^+, \Gamma^M \bar{\Phi}^+\rangle_5 \ \mathcal{I}_{MN} \ \langle \Phi^-, \Gamma^N \bar{\Phi}^-\rangle_5 = c   \ ,
 \label{SO5_norm}
\end{equation}
where $\mathcal{I}_{MN}$ is the $SO(5,5)$ metric induced by the natural pairing of tangent and cotangent space.
Here it becomes clear why we chose $\Phi^\pm$ to be of opposite chirality. Only for such pairs, Eq.~\eqref{SO5_norm} can be different from zero. We choose in the following the gauge $c=1$.

A pair $\Phi^\pm$ with these properties already defines some generalized almost product structure $\mathcal{P}$ via
\begin{equation}
\label{SO55_product_structure}
 \mathcal{P}^M_{\phantom{M}N} = \big( \langle \Phi^+, \Gamma^M \bar{\Phi}^+\rangle_5 \langle \Phi^-, \Gamma^K \bar{\Phi}^-\rangle_5 + \langle \Phi^+, \Gamma^K \bar{\Phi}^+\rangle_5 \langle \Phi^-, \Gamma^M \bar{\Phi}^-\rangle_5 \big) \, \mathcal{I}_{KN} - \delta^M_{\phantom{M}N} \ .
\end{equation}
Since the pure spinors are globally defined, $\mathcal{P}$ is globally defined. It is symmetric with respect to $\mathcal{I}$ and divides $T \M_5 \oplus T^*\M_5$ into some two-dimensional eigenspace with eigenvalue $+1$ and an eight-dimensional eigenspace with eigenvalue $-1$. This splitting is compatible with $\mathcal{I}$ and therefore breaks $SO(5,5)$ to $SO(4,4) \times SO(1,1)$.
Note that in contrast to the case $d=4$, where we had to impose the existence of $\mathcal{P}$ in addition to the existence of a pure spinor pair, here it just results from the definition of $\Phi^\pm$.

We can use this splitting to decompose $\Phi^+$ and $\Phi^-$ into the corresponding pure $SO(4,4)$ spinors and the $SO(1,1)$ spinor parts. Projecting out all $SU(2)\times SU(2)$ doublets implies that the $SO(4,4)$ spinor components must be of even degree. Thus, we find
\begin{equation} \label{pure_spinors_5d}
 \Phi^+ = 1 \wedge \Phi_1 \ , \qquad \Phi^- = L \wedge \Phi_2 \ ,
\end{equation}
where $L$ is the one-form that is in the $+1$ eigenspace of $\mathcal{P}$.
Here, $\Phi_1$ and $\Phi_2$ are compatible pure $SO(4,4)$ spinors of even chirality that define some $SU(2) \times SU(2)$ structure, while $L$ is the one-form that corresponds to the fifth direction. Note that the ratio of volumes of the $SO(4,4)$ and the $SO(1,1)$ direction can be reparameterized by the $SO(1,1)$ vector
\begin{equation}
\label{SO55_fifth_vector}
U^K = \langle \Phi^+, \Gamma^K \bar{\Phi}^+\rangle_5 + \langle \Phi^-, \Gamma^K \bar{\Phi}^-\rangle_5 \ .
\end{equation}

Now we can again do a Kaluza-Klein truncation to obtain a finite-dimensional moduli space. By using the same methods as in section~\ref{section:moduliSU2d6} we can determine the moduli space of $SU(2) \times SU(2)$ structures. On top of \eqref{moduli_space_4d_NS} we get one additional $\mathbb{R}_+$ factor parameterized by $U^K$, corresponding to the length of $L$. Therefore, the moduli space is
\begin{equation} \label{moduli_space_5d_NS}
   \mathcal{M}^\textrm{NS}_{d=5} = \frac{SO(4,n+4)}{SO(4)\times SO(n+4)} \times \mathbb{R}_+ \times \mathbb{R}_+ \ .
\end{equation}

\subsection{Exceptional generalized geometry in five dimensions}\label{section:RRd5}
Analogously to $d=6$ we can use the U-duality covariant formalism to include the RR scalars into the moduli space.
The U-duality group in five dimensions is $E_{6(6)}$ with the T-duality subgroup being $SO(5,5)$ (for more details on the group $E_{6(6)}$ see~\cite{Cremmer:1979uq}). We discuss now the decomposition of the representations of $E_{6(6)}$ in terms of the maximal subgroup $\mathbb{R}_+ \times SO(5,5)$, where the $\mathbb{R}_+$ factor refers to shifts in the dilaton $\phi$.
The fundamental representation of $E_{6(6)}$ is of complex dimension $27$ and decomposes as
\begin{equation}
\label{E6_fundamental_decomposition}
 {\bf 27} \to {\bf 1}_{+4} + {\bf 10}_{-2} + {\bf 16}_{+1}  \ .
\end{equation}
$E_{6(6)}$ has also an anti-fundamental representation which decomposes analogously as
\begin{equation}
\label{E6_antifundamental_decomposition}
 {\bf \bar{27}} \to {\bf 1}_{-4} + {\bf 10}_{+2} + {\bf \bar{16}}_{-1}  \ .
\end{equation}
The adjoint of $E_{6(6)}$ decomposes as
\begin{equation}
\label{E6_decomposition}
 {\bf 78} \to {\bf 1}_0 + {\bf 16}_{-3} + {\bf \bar{16}}_{+3} + {\bf 45}_0 \ .
\end{equation}

To find the geometric realizations of these representations, we consider the charges in $d=5$, which form the fundamental representation of $E_{6(6)}$, analogously to section \ref{section:RRd4}. One finds~\cite{Hull:1994ys,Hull:2007zu}
\begin{equation}
\label{E6_fundamental_decomposition_geometrical_IIA}
  {\bf 27}^\textrm{IIA} \to (\mathbb{R})_{+4} + (T \M_5 \oplus T^* \M_5)_{-2} + (\Lambda^{\textrm{even}}T^* \M_5)_{+1}
\end{equation}
for type IIA, and
\begin{equation}
\label{E6_fundamental_decomposition_geometrical_IIB}
  {\bf 27}^\textrm{IIB} \to (\mathbb{R})_{+4} + (T \M_5 \oplus T^* \M_5)_{-2} + (\Lambda^{\textrm{odd}}T^* \M_5)_{+1}
\end{equation}
for type IIB.
From \eqref{E6_fundamental_decomposition_geometrical_IIA} we can derive the geometrical realization of the adjoint given in \eqref{E6_decomposition} for type IIA, which is
\begin{equation}
\label{E6_decomposition_geometrically}
 e_{6(6)}^{\textrm{IIA}} =  (\mathbb{R})_{0}  \oplus (\Lambda^{\textrm{odd}}T^* \M_5)_{-3} \oplus (\Lambda^{\textrm{even}}T^* \M_5)_{+3} \oplus (so(T \M_5 \oplus T^* \M_5))_{0} \ .
\end{equation}
For type IIB, only the charges of the dilaton are inverted. Therefore, we expect that both theories have the same moduli space.\footnote{ Actually, we just have to interchange $\Lambda^{\textrm{odd}}T^* \M_5$ and
  $\Lambda^{\textrm{even}}T^* \M_5$
  everywhere to get the result for type IIB. In terms of the pure spinors in \eqref{pure_spinors_5d}, we have to apply
  the operator $\tau_L \equiv \iota_L + L \wedge \in T \M_5 \oplus T^* \M_5$ which changes the chirality of $\Phi^+$ and
  $\Phi^-$. This symmetry is just the generalization of T-duality. Up to the gauging that interchanges $\Phi_1$ and $\Phi_2$
  in \eqref{pure_spinors_5d}, this just corresponds to the exchange of the two pure spinors $\Phi^+$ and $\Phi^-$, which is
  mirror symmetry.}
In \eqref{E6_decomposition_geometrically} the spinor representations of $SO(5,5)$ correspond to the additional generators of the group $E_{6(6)}$.

Following~\eqref{E6_fundamental_decomposition_geometrical_IIA}, we decompose an element $\lambda$ in the fundamental representation of $E_{6(6)}$ into its components
\begin{equation}
\label{E6_fundamental_splitting}
 \lambda = (\lambda_0 , \lambda^A , \lambda^+ ) \ ,\qquad A=1,\ldots,10 ,
\end{equation}
and similarly elements $\rho$ in the anti-fundamental representation
\begin{equation}
\label{E6_antifundamental_splitting}
 \rho = (\rho_0 , \rho^A , \rho^- ) \ .
\end{equation}
In the same way, we can write an element $M$ in the adjoint representation as
\begin{equation}
 M = (m_0, m^+ , m^- , m^{A}_{\phantom{A}B}) \ .
\end{equation}
The action of the adjoint representation on the fundamental is given by
\begin{align}
 M \cdot \lambda = & ( 4 m_0 \lambda_0 + \langle m^- , \lambda^+ \rangle_5, m^{A}_{\phantom{A}B} \lambda^B - 2 m_0 \lambda^A  + \langle m^+, \Gamma^A \lambda^+ \rangle_5 , \nonumber \\ &
 m_{AB} \Gamma^{AB} \lambda^+ +  \lambda^A \mathcal{I}_{AB} \Gamma^B m^- + m_0 \lambda^+ + \lambda_0 m^+ ) \ .
\end{align}
The product of two ${\bf 27}$s contains a $\bar{\bf 27}$ which is related to the existence of the cubic invariant of $E_{6(6)}$. More precisely, we have $ {\bf 27} \times  {\bf 27} \to \bar{\bf 27} \oplus \dots $ which by projection leads to the map
\begin{equation}\begin{aligned}
    {\bf 27} \times  {\bf 27} \longrightarrow \ & {\bf \bar{27}} \ ,  \\
     (\lambda,\lambda') \longmapsto \ & \lambda \times \lambda' \ ,
\end{aligned}\end{equation}
where
\begin{equation}\label{E6_cross_product}
\lambda \times \lambda'= (\lambda^A \mathcal{I}_{AB} \lambda'^B ,\ \lambda_0 \lambda'^A + \lambda'_0 \lambda^A + \langle \lambda^+, \Gamma^A \lambda'^+ \rangle_5 , \
      \lambda^A \mathcal{I}_{AB} \Gamma^B \lambda'^+ + \lambda'^A \mathcal{I}_{AB} \Gamma^B \lambda^+ ) \ .
\end{equation}
The scalar product
\begin{equation}\begin{aligned}
 {\bf 27} \times {\bf \bar{27}} \longrightarrow & {\bf 1} \ , \\
  (\lambda,\rho) \longmapsto & \lambda \cdot \rho \ ,
\end{aligned}\end{equation}
is defined as
\begin{equation}\label{E6_scalar_product}
\lambda \cdot \rho =  \lambda_0 \rho_0 + \lambda^A \mathcal{I}_{AB} \rho^B + \langle \lambda^+, \rho^- \rangle_5 \ .
\end{equation}

Next we embed the pair of compatible pure $SO(5,5)$ spinors $\Phi^+$ and $\Phi^-$ into $E_{6(6)}$ representations. It seems most natural to embed them into the complexified fundamental and anti-fundamental representation, respectively. Therefore, we define
\begin{equation} \label{E6_embedding_spinors}
 \lambda = ( 0 , 0 , \Phi^+ ) \ , \qquad \rho = ( 0 , 0 , \Phi^- ) \ .
\end{equation}
Then the purity condition~\eqref{purity_extended_spinor} is easily rephrased in
\begin{equation} \label{E6_purity}
 \lambda \times \lambda = 0 \ , \qquad \rho \times \rho = 0 \ ,
\end{equation}
and we can impose the compatibility conditions~\eqref{SO5_comp} in the form
\begin{equation} \label{E6_compatibility}
 \lambda \cdot \rho =  \bar{\lambda} \cdot \rho = 0 \ .
\end{equation}
The normalization condition~\eqref{SO5_norm} reads
\begin{equation}
\label{E6_normalization}
 ( \lambda \times \bar{\lambda} ) \cdot ( \rho \times \bar{\rho} ) = 1 \ .
\end{equation}\par
Note that $\lambda$ and $\rho$ define some isomorphism between the ${\bf27}$ and the ${\bf\bar{27}}$ representation, via
\begin{equation}�\label{E6_isomorphism}\begin{aligned}
 I \ : \  {\bf 27} \longrightarrow & {\bf \bar{27}} \ , \\
     \tilde{\lambda} \longmapsto & (\rho \times \bar{\rho}) \times \tilde{\lambda} \ ,
\end{aligned} \end{equation}�
and
\begin{equation}\label{E6_isomorphism_inverse} \begin{aligned}
  I^{-1} \ : \   {\bf \bar{27}} \longrightarrow & {\bf 27} \ ,  \\
     \tilde{\rho} \longmapsto & (\lambda \times \bar{\lambda}) \times \tilde{\rho} \ ,
\end{aligned}\end{equation}
where $I^{-1}$ is the inverse of $I$ due to \eqref{E6_normalization}.

However, so far we did not embed the dilaton degree of freedom into some $E_{6(6)}$ representation.
Analogously to~\eqref{SO55_embedding_SO11_vector}, we can embed some $SO(1,1)$ vector into the fundamental  representation of $E_{6(6)}$ which reflects the dilaton degree of freedom. First of all, one can either embed this doublet into the fundamental or into the anti-fundamental. However, both are related by the isomorphism \eqref{E6_isomorphism}.
Therefore, it does not make any difference whether we use the fundamental or the anti-fundamental representation.
The canonical $SO(1,1)$ doublet should correspond to the two singlets in~\eqref{E6_fundamental_decomposition} and~\eqref{E6_antifundamental_decomposition} because these do not transform under any geometric structure group and are of opposite charge. These two singlets are not part of the same representation, but we can use~\eqref{E6_isomorphism_inverse} to map one of the singlets into the other representation. Therefore, the embedded $SO(1,1)$ doublet is spanned by the two elements in the ${\bf 27}$ that are
\begin{equation}
 \mu^1 = (1, 0, 0 )_{\bf 27} \ ,
\end{equation}
and
\begin{equation}
 \mu^2 = (\lambda \times \bar{\lambda}) \times (1,0,0)_{\bf \bar{27}} \ .
\end{equation}
The embedding of the dilaton into the fundamental representation of $E_{6(6)}$ then reads
\begin{equation}\label{E6_mu_definition}
 \phi \longmapsto \mu(\phi) = \e^{\phi} \mu^1 + \e^{-\phi} \mu^2 \ .
\end{equation}
We choose the normalization
\begin{equation}
\label{E6_mu_normalization}
 \mu \times \mu = \tfrac{1}{2}\ \lambda \times \bar{\lambda} \ .
\end{equation}
We can always switch between $\mu$ and its `cousin' $\hat{\mu}$ in the anti-fundamental by
\begin{align}
 \hat{\mu} = (\rho \times \bar{\rho}) \times \mu \ , \\
 \mu = (\lambda \times \bar{\lambda}) \times \hat{\mu} \ .
\end{align}
Equation~\eqref{E6_mu_normalization} together with~\eqref{E6_normalization} then just states that
\begin{equation}
 \mu \cdot \hat{\mu} = 1 \ .
\end{equation}
Finally, we impose the conditions
\begin{equation} \label{E6_mu_compatibility1}�
 \mu \times \lambda = \mu \times \bar{\lambda} = 0
\end{equation}
and
\begin{equation} \label{E6_mu_compatibility2}
 \hat{\mu} \times \rho = \hat{\mu} \times \bar{\rho} = 0
\end{equation}
to make $\mu$ compatible with $\lambda$ and $\rho$.

Now we want to discuss the effect of the projection to $N=2$ which eliminates all $SU(2) \times SU(2)$ doublets. We have seen in section \ref{section:SU2SU2d5} that this fixes the generalized almost product structure $\mathcal{P}$, defined in~\eqref{SO55_product_structure}, and therefore reduces the T-duality group $SO(5,5)$ to $SO(4,4) \times \mathbb{R}_+$. Furthermore we know that modding out $SU(2)\times SU(2)$ doublets eliminates half of the degrees of freedom in $\Lambda^{\textrm{odd}}T^*\M_5$ and $\Lambda^{\textrm{even}}T^*\M_5$, while for $SO(5,5)$ vectors, only the components in the $+1$ eigenspace of $\mathcal{P}$ survive the projection.
We decompose~\eqref{E6_decomposition} further so that the adjoint in terms of $SO(4,4) \times \mathbb{R}_+ \times \mathbb{R}_+$ representations reads
\begin{equation} \label{E6_decomp_SO44}
 {\bf 78} \rightarrow {\bf 1}_{0,0} + {\bf 8}^f_{-3 , +1} + {\bf 8}^c_{-3 , -1} + {\bf 8}^f_{+3, -1}+ {\bf 8}^c_{+3, +1} + {\bf 1}_{0,0} + {\bf 8}^v_{0, +2} + {\bf 8}^v_{0, -2} + {\bf 28}_{0,0} \ .
\end{equation}
Here, we project out the representations ${\bf 8}^v$ and ${\bf 8}^c$ because they consist of $SU(2) \times SU(2)$ doublets.
Using the splitting of the adjoint of $SO(5,5)$ into $SO(4,4)$ representations, i.e.\
\begin{equation}
\label{SO55_decomposition}
 {\bf 45} \to {\bf 28}_0 + {\bf 8}_{+2} + {\bf 8}_{-2} + {\bf 1}_0  \ ,
\end{equation}
we identify the surviving pieces of \eqref{E6_decomp_SO44} with the Lie algebra of $SO(5,5) \times \mathbb{R}_+$. Here the extra $\mathbb{R}_+$ factor is a combination of the volume of the fifth direction and the dilaton degree of freedom. More precisely, to match the representations, we have to transform the charges under the two Abelian $\mathbb{R}_+$ factors as
\begin{equation}
 (p,q) \to \Big(\frac{p-q}2, - \frac{p+3q}2\Big) \ .
\end{equation}

Next we consider the decomposition of the fundamental representation of $E_{6(6)}$ in terms of $SO(4,4) \times \mathbb{R}_+ \times \mathbb{R}_+$ representations
\begin{equation}
 {\bf 27 } \to   {\bf 1}_{+4,0} + {\bf 8}_{-2,0}^v + {\bf 1}_{-2,+2} + {\bf 1}_{-2,-2} + {\bf 8}^f_{+1,+1} + {\bf 8}^c_{+1,-1} \ .
\end{equation}
Again, the representations ${\bf 8}^v$ and ${\bf 8}^c$ are projected out because they consist of $SU(2) \times SU(2)$ doublets. The surviving pieces form an $SO(5,5) \times \mathbb{R}_+$ vector
\begin{equation}
  {\bf 10}_{-2} \to {\bf 8}^f_{0,-2} + {\bf 1}_{+2,-2} + {\bf 1}_{-2,-2}
\end{equation}
and a singlet ${\bf 1}_{+4}$.
The same holds for the anti-fundamental representation, with the only difference that the Abelian charges have the opposite sign. The two singlets are spanned by $m_0 = (\bar{\lambda} \times \lambda)$ and $n_0 = (\bar{\rho} \times \rho)$, and both~\eqref{E6_cross_product} and~\eqref{E6_scalar_product} become the usual scalar product between $SO(5,5)$ vectors. $\lambda$, $\rho$ and $\mu$ are projected to vector representations of $SO(5,5)$, and $(\Re \lambda)$, $(\Im \lambda)$, $(n_0 \Re \rho)$, $(n_0 \Im \rho)$ and $\mu$ form a set of five orthonormal space-like $SO(5,5)$ vectors. Therefore, they span the coset $SO(5,5)/{SO(5)}$. From now on everything works analogously to section \ref{section:RRd6}.
After modding out the symmetry between these five vectors, we end up with
\begin{equation}\label{moduli_space_5d_local}
 \mathcal{M}_{\lambda,\rho,\mu} = \frac{SO(5,5)}{SO(5) \times SO(5)} \times \mathbb{R}_+ \ ,
\end{equation}
where the $\mathbb{R}_+$ factor is due to $m_0$ and $n_0$, which form an $SO(1,1)$ doublet parameterizing this degree of freedom. Using again the general argument that we presented in section \ref{section:moduliSU2d6}, one can argue that after Kaluza-Klein truncation the actual moduli space is of the form
\begin{equation} \label{moduli_space_5d}
   \mathcal{M}_{d=5} = \frac{SO(5,n+5)}{SO(5) \times SO(n+5)} \times \mathbb{R}_+ \ .
\end{equation}
The action of the RR fields can be determined analogously to section \ref{section:RRd6}, with the same result as \eqref{C_transforms}.

As we already discussed above, after the projection to a theory with $16$ supercharges the real and imaginary parts of $\lambda$ and $\rho$, which were introduced in \eqref{E6_embedding_spinors}, and $\mu(\phi)$ given in \eqref{E6_mu_definition} form a set of five space-like, linearly independent vectors in $\mathbb{R}^{(5,n+5)}$. As long as we do not constrain these objects by purity or compatibility conditions, they parameterize the flat cone over \eqref{moduli_space_5d} which is the moduli space of the corresponding superconformal supergravity.
Imposing the purity and compatibility conditions given in \eqref{E6_purity}, \eqref{E6_compatibility}, \eqref{E6_normalization}, \eqref{E6_mu_normalization}, \eqref{E6_mu_compatibility1} and \eqref{E6_mu_compatibility2} and the removal of gauge degrees of freedom reduce this to the moduli space \eqref{moduli_space_5d}.



\begin{thebibliography}{00}
\bibitem{Tsimpis:2005kj}
  D.~Tsimpis,
  ``M-theory on eight-manifolds revisited: N = 1 supersymmetry and generalized
  Spin(7) structures,''
  JHEP {\bf 0604} (2006) 027
  [arXiv:hep-th/0511047].

\bibitem{Hitchin}
N.~Hitchin, ``The geometry of three-forms in six and seven
dimensions,'' J.\ Diff.\ Geom.\ {\bf 55} (2000), no.3 547 [arXiv: math.DG/0010054].

\bibitem{Hitchin:2001rw}
N. Hitchin, ``Stable forms and special metrics,''
in ``Global Differential Geometry: The Mathematical Legacy of Alfred
Gray'', M.Fernandez and J.A.Wolf (eds.),
Contemporary Mathematics {\bf 288}, American Mathematical Society,
Providence (2001) [arXiv:math.DG/0107101].

\bibitem{waldram}
J.~P.~Gauntlett, N.~Kim, D.~Martelli and D.~Waldram,
  ``Fivebranes wrapped on SLAG three-cycles and related geometry,''
  JHEP {\bf 0111} (2001) 018
  [arXiv:hep-th/0110034];

J.~P.~Gauntlett, D.~Martelli, S.~Pakis and D.~Waldram,
  ``G-structures and wrapped NS5-branes,''
  Commun.\ Math.\ Phys.\  {\bf 247} (2004) 421
  [arXiv:hep-th/0205050].

\bibitem{CS}
S.\ Chiossi and S.\ Salamon, ``The Intrinsic Torsion of $SU(3)$ and $G_2$
Structures,'' in \emph{Differential geometry, Valencia, 2001}, pp. 115,
[arXiv: math.DG/0202282].

\bibitem{Hitchin:2004ut}
  N.~Hitchin,
  ``Generalized Calabi-Yau manifolds,''
  Quart.\ J.\ Math.\ Oxford Ser.\  {\bf 54}, 281 (2003)
  [arXiv:math/0209099].

\bibitem{Gauntlett:2003cy}
  J.~P.~Gauntlett, D.~Martelli and D.~Waldram,
  ``Superstrings with intrinsic torsion,''
  Phys.\ Rev.\  D {\bf 69}, 086002 (2004)
  [arXiv:hep-th/0302158].

\bibitem{Grana:2005jc}
  For reviews see, for example,

  M.~Grana,
  ``Flux compactifications in string theory: A comprehensive review,''
  Phys.\ Rept.\  {\bf 423} (2006) 91
  [arXiv:hep-th/0509003],

  M.~R.~Douglas and S.~Kachru,
  ``Flux compactification,''
  Rev.\ Mod.\ Phys.\  {\bf 79} (2007) 733
  [arXiv:hep-th/0610102],

  R.~Blumenhagen, B.~Kors, D.~Lust and S.~Stieberger,
  ``Four-dimensional String Compactifications with D-Branes, Orientifolds   and
  Fluxes,''
  Phys.\ Rept.\  {\bf 445} (2007) 1
  [arXiv:hep-th/0610327],

  B.~Wecht,
  ``Lectures on Nongeometric Flux Compactifications,''
  Class.\ Quant.\ Grav.\  {\bf 24}, S773 (2007)
  [arXiv:0708.3984 [hep-th]],

  H.~Samtleben,
  ``Lectures on Gauged Supergravity and Flux Compactifications,''
  Class.\ Quant.\ Grav.\  {\bf 25}, 214002 (2008)
  [arXiv:0808.4076 [hep-th]],

  and references therein.

\bibitem{Gualtieri:2003dx}
  M.~Gualtieri,
  ``Generalized complex geometry,''
  Ph.D. Thesis
  [arXiv:math/0401221].

\bibitem{Witt}
F.~Witt, ``Generalised $G_2$-manifolds'',
Commun.\ Math.\ Phys.\ {\bf 265} (2006) 275
[math.DG/0411642].

F.~Witt, ``Special metric structures and closed forms'',
Oxford University DPhil thesis (2004) [arXiv:math.DG/0502443].

\bibitem{Jeschek:2004wy}
  C.~Jeschek and F.~Witt,
  ``Generalised G(2)-structures and type IIB superstrings,''
  JHEP {\bf 0503}, 053 (2005)
  [arXiv:hep-th/0412280].

\bibitem{Grana:2005sn}
  M.~Grana, R.~Minasian, M.~Petrini and A.~Tomasiello,
  ``Generalized structures of N=1 vacua,''
  JHEP {\bf 0511}, 020 (2005)
  [arXiv:hep-th/0505212].

\bibitem{Grana:2005ny}
  M.~Grana, J.~Louis and D.~Waldram,
  ``Hitchin functionals in N = 2 supergravity,''
  JHEP {\bf 0601}, 008 (2006)
  [arXiv:hep-th/0505264].

  M.~Grana, J.~Louis and D.~Waldram,
  ``SU(3) x SU(3) compactification and mirror duals of magnetic fluxes,''
  JHEP {\bf 0704}, 101 (2007)
  [arXiv:hep-th/0612237].

\bibitem{Hull:2004in}
  A.~Dabholkar and C.~Hull,
  ``Duality twists, orbifolds, and fluxes,''
  JHEP {\bf 0309} (2003) 054
  [arXiv:hep-th/0210209];

C.~M.~Hull,
  ``A geometry for non-geometric string backgrounds,''
  JHEP {\bf 0510} (2005) 065
  [arXiv:hep-th/0406102];

 A.~Dabholkar and C.~Hull,
  ``Generalised T-duality and non-geometric backgrounds,''
  JHEP {\bf 0605}, 009 (2006)
  [arXiv:hep-th/0512005].

\bibitem{Kachru:2002sk}
  S.~Kachru, M.~B.~Schulz, P.~K.~Tripathy and S.~P.~Trivedi,
  ``New supersymmetric string compactifications,''
  JHEP {\bf 0303} (2003) 061
  [arXiv:hep-th/0211182].

\bibitem{Flournoy:2004vn}
  A.~Flournoy, B.~Wecht and B.~Williams,
  ``Constructing nongeometric vacua in string theory,''
  Nucl.\ Phys.\  B {\bf 706}, 127 (2005)
  [arXiv:hep-th/0404217].

\bibitem{Shelton:2005cf}
  J.~Shelton, W.~Taylor and B.~Wecht,
  ``Nongeometric flux compactifications,''
  JHEP {\bf 0510}, 085 (2005)
  [arXiv:hep-th/0508133].

\bibitem{Hull:2007zu}
  C.~M.~Hull,
  ``Generalised geometry for M-theory,''
  JHEP {\bf 0707}, 079 (2007)
  [arXiv:hep-th/0701203].

\bibitem{Pacheco:2008ps}
  P.~P.~Pacheco and D.~Waldram,
  ``M-theory, exceptional generalised geometry and superpotentials,''
  JHEP {\bf 0809}, 123 (2008)
  [arXiv:0804.1362 [hep-th]].

\bibitem{Sim:2008}
 A.~Sim,
   ``Exceptionally Generalised Geometry and Supergravity,''
   Imperial College London PhD Thesis, October 2008.

\bibitem{Grana:2008tbp}
   M.~Grana, J.~Louis, A.~Sim and D.~Waldram,
  ``$E_{7(7)}$ formulation of N=2 backgrounds,''
  arXiv:0904.2333 [hep-th].


\bibitem{Bovy:2005qq}
  J.~Bovy, D.~L\"ust and D.~Tsimpis,
  ``N = 1,2 supersymmetric vacua of IIA supergravity and SU(2) structures,''
  JHEP {\bf 0508}, 056 (2005)
  [arXiv:hep-th/0506160].


\bibitem{ReidEdwards:2008rd}
  R.~A.~Reid-Edwards and B.~Spanjaard,
  ``N=4 Gauged Supergravity from Duality-Twist Compactifications of String
  Theory,''
  arXiv:0810.4699 [hep-th].

  B.~Spanjaard,
  ``Compactifications of IIA supergravity on SU(2)-structure manifolds,'' 
  [http://www-library.desy.de/preparch/desy/thesis/desy-thesis-08-016.pdf].

\bibitem{Lust:2009zb}
  D.~L\"ust and D.~Tsimpis,
  ``Classes of AdS4 type IIA/IIB compactifications with SU(3)xSU(3)
  structure,''
  arXiv:0901.4474 [hep-th].

\bibitem{N4SUGRA}

M.~Haack, J.~Louis and H.~Singh,
  ``Massive type IIA theory on K3,''
  JHEP {\bf 0104} (2001) 040
  [arXiv:hep-th/0102110].

  R.~D'Auria, S.~Ferrara and S.~Vaula,
  ``N = 4 gauged supergravity and a IIB orientifold with fluxes,''
  New J.\ Phys.\  {\bf 4} (2002) 71
  [arXiv:hep-th/0206241].

  R.~D'Auria, S.~Ferrara, F.~Gargiulo, M.~Trigiante and S.~Vaula,
  ``N = 4 supergravity Lagrangian for type IIB on T**6/Z(2) in presence of
  fluxes and D3-branes,''
  JHEP {\bf 0306} (2003) 045
  [arXiv:hep-th/0303049].

  J.~P.~Derendinger, C.~Kounnas, P.~M.~Petropoulos and F.~Zwirner,
  ``Superpotentials in IIA compactifications with general fluxes,''
  Nucl.\ Phys.\  B {\bf 715} (2005) 211
  [arXiv:hep-th/0411276].

  J.~P.~Derendinger, C.~Kounnas, P.~M.~Petropoulos and F.~Zwirner,
  ``Fluxes and gaugings: N = 1 effective superpotentials,''
  Fortsch.\ Phys.\  {\bf 53} (2005) 926
  [arXiv:hep-th/0503229].

  J.~Sch\"on and M.~Weidner,
  ``Gauged N = 4 supergravities,''
  JHEP {\bf 0605} (2006) 034
  [arXiv:hep-th/0602024].

\bibitem{de Roo:1984gd}
  M.~de Roo,
  ``Matter Coupling In N=4 Supergravity,''
  Nucl.\ Phys.\  B {\bf 255}, 515 (1985).

  M.~de Roo and P.~Wagemans,
  ``Gauge Matter Coupling In N=4 Supergravity,''
  Nucl.\ Phys.\  B {\bf 262}, 644 (1985).

\bibitem{LST}
J.~Louis, D.~Martinez-Pedrera, B.~Spanjaard and H.~Triendl, in preparation.

\bibitem{LMM}
 J.~Louis, D.~Martinez-Pedrera and A.~Micu,
  ``Heterotic compactifications on SU(2)-structure backgrounds,''
  arXiv:0907.3799 [hep-th].

\bibitem{Ferrara:2008hwa}
See, for example,

S.~Ferrara, K.~Hayakawa and A.~Marrani,
  ``Lectures on Attractors and Black Holes,''
  Fortsch.\ Phys.\  {\bf 56} (2008) 993
  [arXiv:0805.2498 [hep-th]],

S.~Bellucci, S.~Ferrara and A.~Marrani,
  ``Attractors in Black,''
  Fortsch.\ Phys.\  {\bf 56} (2008) 761
  [arXiv:0805.1310 [hep-th]],

and references therein.

\bibitem{Swann:1990}
  A.~Swann,
  ``HyperKahler and quaternionic Kahler geometry,''
  Math.\ Ann.\  {\bf 289}, 421 (1991).

\bibitem{deWit:1999fp}
  B.~de Wit, B.~Kleijn and S.~Vandoren,
  ``Superconformal hypermultiplets,''
  Nucl.\ Phys.\  B {\bf 568} (2000) 475
  [arXiv:hep-th/9909228].

\bibitem{Andrianopoli:2001zh}
  L.~Andrianopoli, R.~D'Auria and S.~Ferrara,
  ``Supersymmetry reduction of N-extended supergravities in four dimensions,''
  JHEP {\bf 0203}, 025 (2002)
  [arXiv:hep-th/0110277].

  L.~Andrianopoli, R.~D'Auria and S.~Ferrara,
  ``Consistent reduction of N = 2 $\to$ N = 1 four dimensional supergravity
  coupled to matter,''
  Nucl.\ Phys.\  B {\bf 628}, 387 (2002)
  [arXiv:hep-th/0112192].

\bibitem{Romans:1986er}
  L.~J.~Romans,
  ``Selfduality For Interacting Fields: Covariant Field Equations For
  Six-Dimensional Chiral Supergravities,''
  Nucl.\ Phys.\  B {\bf 276} (1986) 71.

\bibitem{deWN}
  B.~de Wit and H.~Nicolai,
  ``$D = 11$ Supergravity With Local $SU(8)$ Invariance,''
  Nucl.\ Phys.\ B {\bf 274}, 363 (1986).


\bibitem{Koerber:2007xk}
  P.~Koerber and L.~Martucci,
  ``From ten to four and back again: how to generalize the geometry,''
  JHEP {\bf 0708} (2007) 059
  [arXiv:0707.1038 [hep-th]].

\bibitem{Aspinwall:1996mn}
  P.~S.~Aspinwall,
  ``K3 surfaces and string duality,''
  Boulder 1996, Fields, strings and duality 421-540
  [arXiv:hep-th/9611137].

\bibitem{Grana:2006kf}
  M.~Grana, R.~Minasian, M.~Petrini and A.~Tomasiello,
  ``A scan for new N=1 vacua on twisted tori,''
  JHEP {\bf 0705}, 031 (2007)
  [arXiv:hep-th/0609124].


\bibitem{Chevalley:1996}
  C.~Chevalley,
  ``The Algebraic Theory of Spinors and Clifford Algebras. Collected Works. Vol. 2,''
  {\it Berlin, Germany: Springer Verlag (1996) 227 p}

\bibitem{Charlton:1996PhD}
  P.~Charlton,
  ``The Geometry of Pure Spinors, with Applications,'' \\
  http://csusap.csu.edu.au/$\sim$pcharlto/charlton\_thesis.pdf

\bibitem{Cassani:2007pq}
  D.~Cassani and A.~Bilal,
  ``Effective actions and N=1 vacuum conditions from SU(3) x SU(3)
  compactifications,''
  JHEP {\bf 0709}, 076 (2007)
  [arXiv:0707.3125 [hep-th]].


\bibitem{Hitchin:1986ea}
  N.~J.~Hitchin, A.~Karlhede, U.~Lindstrom and M.~Rocek,
  ``Hyperkahler Metrics and Supersymmetry,''
  Commun.\ Math.\ Phys.\  {\bf 108}, 535 (1987).

\bibitem{Anguelova:2002kd}
  L.~Anguelova, M.~Rocek and S.~Vandoren,
  ``Hyperkaehler cones and orthogonal Wolf spaces,''
  JHEP {\bf 0205}, 064 (2002)
  [arXiv:hep-th/0202149].

\bibitem{Hull:1994ys}
  C.~M.~Hull and P.~K.~Townsend,
  ``Unity of superstring dualities,''
  Nucl.\ Phys.\  B {\bf 438}, 109 (1995)
  [arXiv:hep-th/9410167].

\bibitem{slansky}
See, for example, R.~Slansky,
  ``Group Theory For Unified Model Building,''
  Phys.\ Rept.\  {\bf 79} (1981) 1.



\bibitem{Cremmer:1978ds}
  E.~Cremmer and B.~Julia,
  ``The N=8 Supergravity Theory. 1. The Lagrangian,''
  Phys.\ Lett.\  B {\bf 80}, 48 (1978).

\bibitem{Duff:1995sm}
  M.~J.~Duff, J.~T.~Liu and J.~Rahmfeld,
  ``Four-Dimensional String-String-String Triality,''
  Nucl.\ Phys.\  B {\bf 459}, 125 (1996)
  [arXiv:hep-th/9508094].

\bibitem{VanProeyen:1999ni}
  A.~Van Proeyen,
  ``Tools for supersymmetry,''
  arXiv:hep-th/9910030.

\bibitem{Berkovits:2000fe}
  N.~Berkovits,
  ``Super-Poincare covariant quantization of the superstring,''
  JHEP {\bf 0004}, 018 (2000)
  [arXiv:hep-th/0001035].

\bibitem{Cremmer:1979uq}
  E.~Cremmer, J.~Scherk and J.~H.~Schwarz,
  ``Spontaneously Broken N=8 Supergravity,''
  Phys.\ Lett.\  B {\bf 84}, 83 (1979).


\end{thebibliography}
\end{document}